\documentclass[reprint,preprintnumbers,aps,prd,amsmath,amssymb,nobibnotes,nofootinbib,onecolumn]{revtex4-2}
\usepackage{xcolor}
\usepackage{subcaption}
\usepackage{amsmath}
\usepackage{amssymb}
\usepackage{graphicx}
\usepackage{hyperref}
\usepackage{slashed}
\usepackage[nolist]{acronym}
\usepackage{booktabs}

\usepackage{tikz}
\usepackage[compat=1.1.0]{tikz-feynman}

\newcommand{\astrolabe}{\textsc{astrolabe}}
\newcommand{\Nbody}{$N$-body}
\begin{document}

 \title{ASTROLABE: A 3D Particle-Level Surrogate for Cosmology Simulation\\ with Physical Flow Matching}
 
\author{Ibrahim Elsharkawy}
\affiliation{Department of Physics, University of Toronto and Vector Institute, Toronto, ON, M5S 1A7, Canada}
\affiliation{NERSC, Lawrence Berkeley National Laboratory, Berkeley, California, USA}
\author{David Curtin}
\affiliation{Department of Physics, University of Toronto, Toronto, ON, M5S 1A7, Canada}
\author{Yonatan Kahn}
\affiliation{Department of Physics, University of Toronto and Vector Institute, Toronto, ON, M5S 1A7, Canada}
\author{Neal Weiner}
\affiliation{Center for Cosmology and Particle Physics, Department of Physics, New York University, New York, NY 10003, USA}

\begin{abstract}
The microphysics of dark matter (DM) is imprinted on 
cosmological structure across an enormous dynamic range.  Nonlinear cosmological observables are most reliably modeled with $N$-body and hydrodynamic simulations. However, systematic $N$-body scans over DM theory space are prohibitively expensive. Machine learning emulators provide a possible solution, but are currently limited by either low resolution or only the ability to emulate specific observables. 
We introduce \astrolabe{}, a flow-matching-based generative emulator designed to be free of both constraints. \astrolabe{} acts directly on the $6N$ phase-space coordinates of $N$ individual particles.
With no voxel grid in the output, there is no voxel-resolution ceiling, and in principle any observable can be measured from inference. 
\astrolabe{} is computationally tractable via a field-particle hybrid transformer in which every particle cross-attends to multiscale field tokens, so compute and memory are linear in $N$. \astrolabe{} incorporates physical flow matching by taking the prior distribution for the flow to be approximate COmoving Lagrangian Acceleration (COLA)-based coordinates rather than unphysical noise, with momentum and mass conservation enforced at all points along flow time. In the $\Lambda$CDM cases we test, \astrolabe{} produces realistic fields with accurate power spectra, halo mass functions, velocity profiles, and halo density profiles. We also observe out-of-distribution generalization, with accurate emulation of dark-acoustic oscillation initial conditions. \astrolabe{} is also capable of being fine-tuned to simulate very large $N$, and we demonstrate this by emulating a \texttt{TNG50-3-dark} simulation at a fraction of the cost. 
This method represents a first step toward a general and powerful cosmology simulation surrogate, which may eventually permit the inclusion of baryonic physics and a systematic exploration of richer dark sector interactions.
\end{abstract}

\maketitle
\tableofcontents

\section{Introduction}
\label{sec:intro}

The nature of dark matter (DM) remains one of the central open problems in physics, and cosmological structure formation is a critical probe. The properties of a dark sector, and in particular the ways it might deviate from standard collisionless Cold Dark Matter (CDM), 
imprint themselves on the clustering of matter across an enormous range of scales, from formation rates for 
the smallest halos to the shape of the cosmic web. The most reliable tools for mapping a given dark matter model onto nonlinear observables are \Nbody{} and hydrodynamical simulations, which act as the
forward model from Lagrangian microphysics to late-time
matter distribution. This forward model is, however, extraordinarily expensive.
A single full-resolution DM-only (DMO) \Nbody{} simulation can consume millions of CPU-hours; adding even basic gas physics and star formation for baryons multiplies that cost by a large factor; and exploring the effect of cosmic variance or the many nuisance parameters that parameterize sub-grid physics becomes literally astronomically expensive~\cite{Rose:2024xcb,Roy:2023zar, Hopkins:2014qka,Potter:2016ttn,Springel:2009aa,Springel:2017tpz,Springel:2020plp,Pillepich:2017jle,Nelson:2018uso,Schaye:2014tpa,Hopkins:2017ycn,Villaescusa-Navarro:2019bje,CAMELS:2020cof,Dai:2026jme}. Systematic scans over the parameter space of any particular dark matter model with non-minimal cosmological or astrophysical effects are computationally prohibitive. This represents a significant bottleneck in our understanding of dark matter, in particular at cosmologically small scales.

A fast and reasonably accurate surrogate of a full simulation, which generalizes across dark matter physics, expands the scope of possible probes of the dark sector. Emulation has a long history in cosmology, and existing approaches fall into two broad regimes. The first emulates a \textit{fixed observable}. For example, Gaussian processes have been used to emulate the nonlinear matter power spectrum \cite{Heitmann:2013bra, Euclid:2018mlb,Nishimichi:2018etk, Angulo:2020vky}. The same strategy has been successful for the
Lyman-$\alpha$ forest where emulators trained on datasets of
hydrodynamical simulations \cite{Bird:2018efe,Pedersen:2020kaw} now set constraints on models that produce dark acoustic oscillations \cite{Rogers:2020ltq,Yuan:2026zlr}. These emulators are extremely useful, but they are limited by being bound to a few bespoke observables. A simulation surrogate is most useful if it is \textit{observable-independent}, so that any dark matter signature, including ones not yet well-studied, can be probed by the output of the simulation.

The second regime targets the density \textit{field} itself but
at generally limited resolution. Convolutional networks were first shown to predict
nonlinear displacements from linear initial conditions on coarse particle grids
\cite{He:2018ggn} and generative adversarial networks (GANs) have produced two-dimensional slices
and small three-dimensional volumes of the cosmic web
\cite{Rodriguez:2018mjb, Perraudin:2019bxl}. More recently, super-resolution networks have been found capable of adding small-scale power to a low-resolution run \cite{Ramanah:2020vyl, Li:2020vor} using voxelized fields as input. A major limitation of these field/voxel based models, when trained on some fixed voxel spacing $L$, is that one must pay $\mathcal{O}(1/L^3)$ compute for increasing resolution in three dimensions. Currently, no existing surrogate resolves the full dynamic range, from individual collapsed halos to the largest modes of the box, likely due to the $\mathcal{O}(1/L^3)$ scaling.

We thus argue that the path to near-full resolution is a generative model that operates directly on point clouds sampled from a phase space density. Point-cloud native models are naturally resolution-adaptive: regions with more structure contain more particles, and regions with less contain fewer. There is no wasted capacity on voxels in empty regions. Point clouds may also offer a better inductive bias. \Nbody{} simulations are simulations of gravitational pairwise interactions acting on $N$ simulation particles taken as the evolution of a collisionless fluid. Such interactions are more naturally captured by models that act directly on the simulation particles and the edges between them rather than ones that are trained on summary rasterizations.\footnote{It is important to note that a point-cloud sampling of phase space is simulation specific. Different software handle it differently, and thus a model trained on such a point cloud implicitly learns a given simulation's specific representation.} In the past, cosmological point-cloud models have been hard to scale to large particle numbers, and consequently have been demonstrated only on smaller halo catalogs \cite{Cuesta-Lazaro:2023zuk, Mikuni:2025ocp, Huang:2025dmm,Nguyen:2024ndo}.

\begin{figure}
\centering
    \includegraphics[width=.65\linewidth]{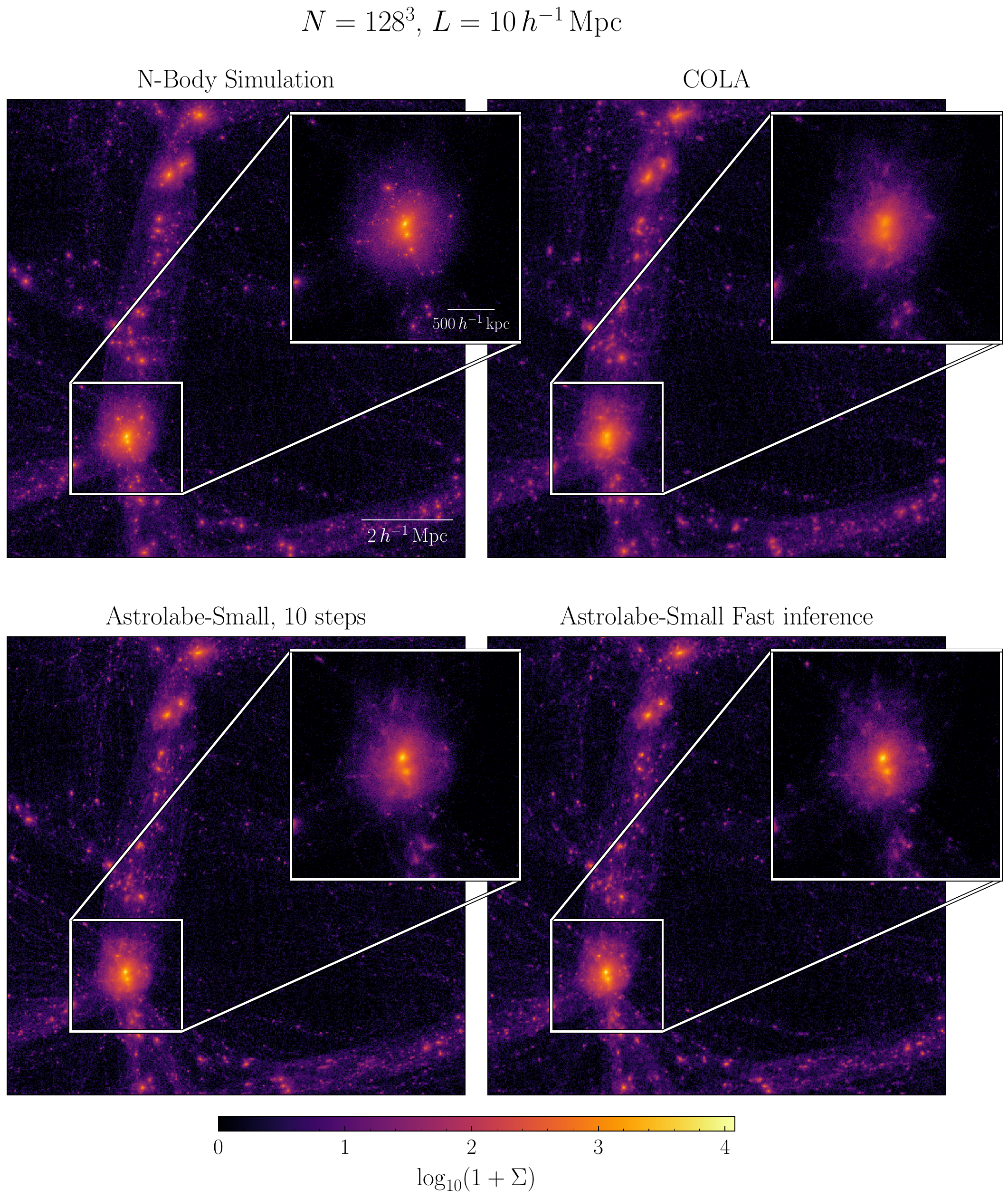}
    \caption{Example $\Lambda$CDM Field for $128^3\approx2\rm M$ particles in a 10 Mpc/$h$ box. We plot surface density $\Sigma$ in units of particles per pixel on a log scale post-rasterization and zoom into the largest halo. \textbf{Top left:} A \textsc{GIZMO} \cite{Hopkins:2014qka} $N$-body simulation, which serves as the truth-level phase space density for this example. \textbf{Top right:} Coarse approximation to the truth-level density given by COLA with a mesh of $256^3$ and 40 timesteps \cite{Tassev:2013pn}. \textbf{Bottom left:} \astrolabe-Small evaluated with integration steps and no subsampling (see Section \ref{sec:training-inference}). \textbf{Bottom right:} \astrolabe-Small Fast Inference evaluated with 5 integration steps and $c=4$ subsampling (see Section \ref{sec:compute-res} and Appendix \ref{app:inference-ablation}). }
    \label{fig:fast-inference}
\end{figure}
In this paper, we introduce \astrolabe{} (\textbf{A} \textbf{S}calable \textbf{T}ransport for \textbf{R}esolved
c\textbf{O}smological \textbf{LA}rge-\textbf{B}ody \textbf{E}mulation)\footnote{The name backronyms the astrolabe (from the Greek \textit{astrolabos}, ``star-taker''), the analog computer that compressed the hours of trigonometry needed to compute positions of the stars \cite{Evans:1998,Morrison:2007}, a parallel with \astrolabe{}'s purpose of speeding up cosmology simulation. Further, the instrument began as a large, single-latitude, narrow-purpose device built on the work of Hipparchus and Ptolemy \cite{Evans:1998} and was later transformed by medieval astronomers such as al-Fazari, al-Sufi, and al-Zarqali \cite{Hockey:2007,King:2005}, who shrank the device to fit in a pocket \cite{King:2005}, and increased its use far beyond reading the sky \cite{Calvo:2006,Hockey:2007,King:2004,Charette:2003}. Similarly, where previous emulators often are single-observable and/or impractical to scale, we intend for \astrolabe{} to be a compact model, useful for extracting many observables across redshifts, dark matter models, and box sizes while remaining scalable to large particle counts.}, a generative emulator which is trained on 1500 distinct DMO \Nbody{} simulations, 100,000 redshift snapshots, and 200 billion individual particles. The initial conditions for \astrolabe{} are point clouds sampled from the phase space density of COmoving Lagrangian Acceleration (COLA) \cite{Tassev:2013pn}, a fast, coarse-grained approximation to the full \Nbody{} dynamics which successfully reproduces large-scale evolution. When conditioned on a desired redshift, \astrolabe{} transports the COLA point clouds to an approximation of the phase space density of the full \Nbody{} simulation which recovers small-scale structure. The division of labor by hierarchies of scale (COLA for large scale, \astrolabe{} for small scale) is similar in structure to the hierarchical world models in \cite{zhang2026h} (H-JEPA), with the largest scales here given by an analytic approximation rather than learned latent prediction. 

At the image level, the fidelity of \astrolabe{} is immediately obvious, producing a 3-dimensional density field that is in close visual agreement with the \Nbody{} simulation truth; see Fig.~\ref{fig:fast-inference} for an example. \astrolabe{} improves on COLA in several quantitative metrics: in a 10 Mpc box, \astrolabe{} reproduces the full truth-level comoving power spectrum at wavenumbers smaller than $k=10^{-2}$ $h/{\rm kpc}$, matches the true halo mass function for halos as small as $10^9$ $M_\odot/h$, recovers Navarro-Frenk-White (NFW) \cite{Navarro:1996gj} halo mass profiles (including approximately matching the slope well below the virial radius) with near-correct concentrations, and improves the correlation of the position and velocity coordinates in phase space as measured by consistency with the Jeans equation. Critically, we observe \astrolabe{} is capable of transforming a broad range of initial conditions, both in its $\Lambda$CDM training distribution and from different dark matter models not in its training set, with essentially the same fidelity as $\Lambda$CDM initial conditions. The generalization across initial conditions separates \astrolabe{} from previous approaches to \Nbody{} emulation. \astrolabe{} is also easily fine-tuned to recover most of its performance on box sizes and particle numbers in the box that are different from its training set.

The novel technical aspects of our emulator are as follows: 
\begin{enumerate}
    \item We introduce a point-field
    hybrid transformer architecture in which individual particles attend to a coarse grid of
     vision transformer-style field tokens \cite{vaswani2017attention,dosovitskiy2020image}. Acting on individual dark matter simulation particles allows for full resolution, while attending to field tokens leads to compute and memory that scale linearly in particle number and permit embarrassingly-parallel training and inference for large particle counts.
    
    \item We introduce \textit{physical flow matching} as a transport of the dark matter phase space density. A single model learns to transport samples from the COLA phase space density to samples from the true \Nbody{} distribution for any redshift. This flow of phase space densities conserves total momentum and mass by construction, is independent of the initial conditions and dark matter parameters that produced the starting distribution for the flow, and produces temporally-coherent movies of structure formation when the conditioning redshift is rolled forward.
\end{enumerate}  
At inference, running \astrolabe{} compared to other GPU-utilizing \Nbody{} simulations for an apples-to-apples comparison represents a computational speedup by at least a factor of 10 to, in some cases, a factor of 500. However, we emphasize that the version of \astrolabe{} presented in this paper should be regarded as a first iteration of this kind of architecture which can almost certainly be made more efficient. Furthermore, given the holistic nature in which \astrolabe{} learns how to generate accurate \Nbody{} configurations, we expect that our methodology may be generalized to include baryonic physics (and potentially dark matter microphysics away from the $\Lambda$CDM paradigm, for example, various forms of self-interaction) with similar, if not much greater gains in computational efficiency compared to conventional simulations. 

The remainder of the paper is organized as follows.
Section \ref{sec:model} presents the \astrolabe{} architecture, including the
point-field hybrid attention and the conditioning scheme, and the physical flow matching objective,.
Section \ref{sec:training} describes the training set, and the inference procedure. Section \ref{sec:results} presents \astrolabe{} results on in-distribution $\Lambda$CDM and an out-of-distribution non-CDM dark acoustic oscillation case study. In the section, we also investigate \astrolabe{}'s ability to emulate larger particle number with fine-tuning, emulating a 150M particle TNG sim, and do a comparison of compute cost. We conclude and discuss the outlook for observable-independent emulation across dark-sector physics in Section \ref{sec:conclusion}.

\section{\astrolabe{}}
\label{sec:model}

The task of a cosmological machine-learning emulator is to efficiently and effectively learn and sample from the 6-dimensional phase-space density of particles, conditional on redshift $z$ and cosmological parameters $\boldsymbol{\theta}$:
\begin{equation}
p\big(\{\mathbf{x}_i\}_{i=1}^{N} \big| z, \boldsymbol{\theta}\big),
\qquad
\mathbf{x}_i = (\mathbf{r}_i, \mathbf{v}_i) \in \mathbb{R}^{6},
\label{eq:target}
\end{equation}
where $\mathbf{r}_i$ is the co-moving position of particle $i$ inside a periodic box of
size $L$ and $\mathbf{v}_i$ is its peculiar velocity. A current state-of-the-art technique for learning and sampling from high-dimensional distributions conditioned on latent variables is flow matching \cite{albergo2023buildingnormalizingflowsstochastic,lipman2023flowmatchinggenerativemodeling,liu2022flowstraightfastlearning}. We first review the standard flow matching procedure in Sec.~\ref{sec:FMreview} and then describe our modifications to incorporate important physical constraints in our training objective in Sec.~\ref{sec:physFM}, followed by our specific architecture choices in Sec.~\ref{sec:arch}.

\subsection{Review of standard flow matching}
\label{sec:FMreview}
\begin{figure}
    \centering
    \includegraphics[width=\linewidth]{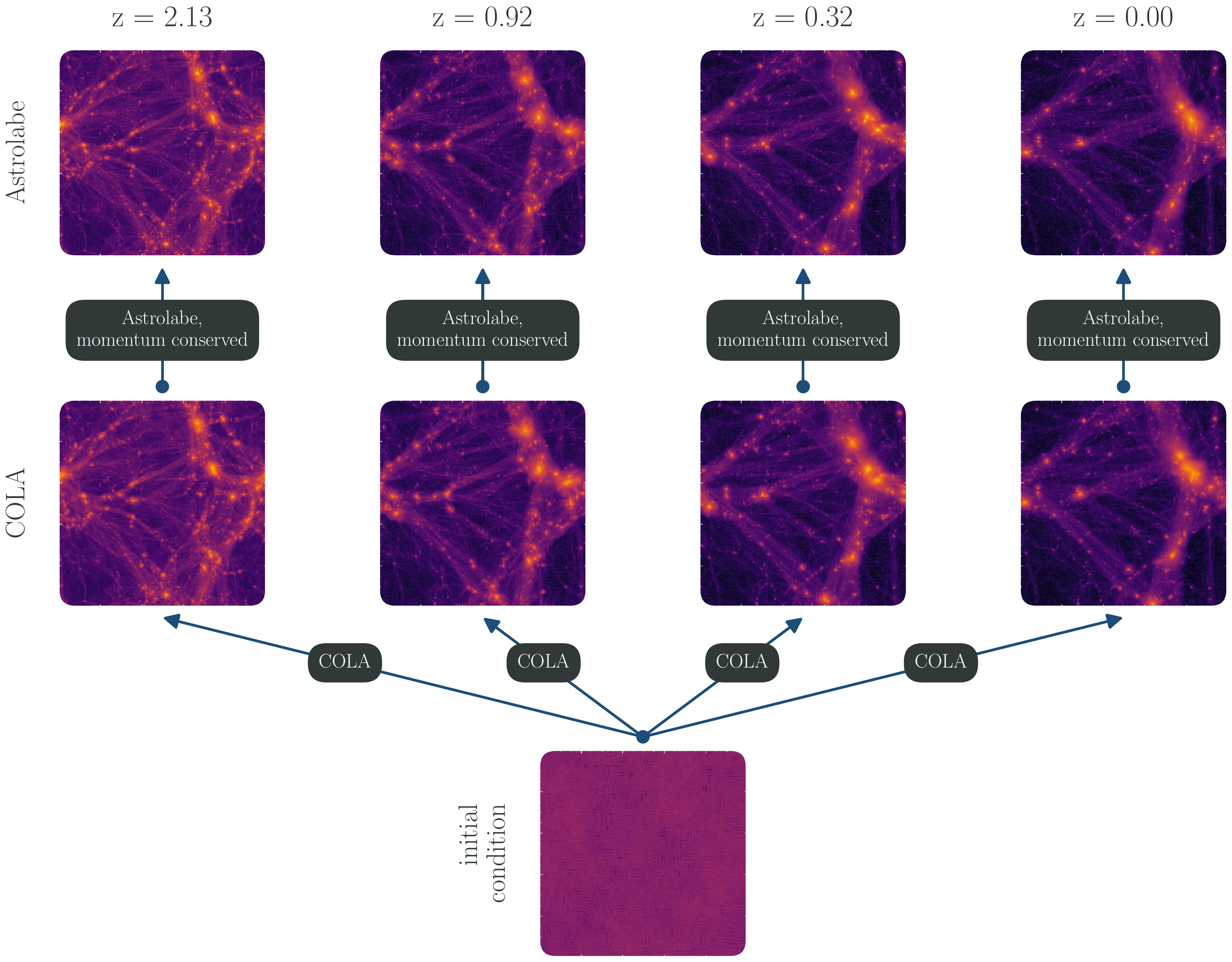}
    \caption{Illustration of physical flow matching, with flow time decoupled from redshift. \textbf{Bottom:} Simulation initial conditions. \textbf{Middle row:} COLA fields, evolved under cosmic time from the initial conditions. \textbf{Top row:} \astrolabe{} density field, evolved under flow time from the COLA distributions in the middle row.}
    \label{fig:flowtime}
\end{figure}

Flow matching models learn a continuous transport from a prior distribution $p_0(\mathbf{x})$ to a desired data distribution $p\big(\mathbf{x} \big|\boldsymbol{\theta}\big)$ via a learned velocity field $\mathbf{V}_\phi(\mathbf{x}, t|\boldsymbol{\theta})$, where the subscript $\phi$ refers to the parameters of the neural network used to parameterize $\mathbf{V}$. Given a sample from the prior $\mathbf{x}_0\sim p_0(\mathbf{x})$, samples from the target distribution are obtained by integrating the ordinary differential equation (ODE)
\begin{equation}
\frac{\mathrm{d}\mathbf{x}}{\mathrm{d}t} = \mathbf{V}_\phi(\mathbf{x}, t | \boldsymbol{\theta}),
\qquad \mathbf{x}(0) = \mathbf{x}_{0},
\label{eq:FM_ode_standard}
\end{equation}
up to flow time $t = 1$. The goal of flow matching is to find a velocity field $\mathbf{V}_\phi$ such that $\mathbf{x}(t = 1) \sim p(\mathbf{x}|\boldsymbol{\theta} )$ after integrating the ODE.

To obtain an approximation to this velocity field, flow matching takes pairs of samples from the training data distribution $\mathbf{x}_1(\boldsymbol{\theta})\sim p\big(\mathbf{x} \big|\boldsymbol{\theta}\big)$ and the prior $\mathbf{x}_0\sim p_0(\mathbf{x})$ and constructs a conditional path $\mathbf{x}_t(\mathbf{x}_1,\mathbf{x}_0,t)$ between each pair. A common choice for this path is linear \cite{liu2022flowstraightfastlearning,lipman2023flowmatchinggenerativemodeling},
\begin{equation}
\mathbf{x}_t(\mathbf{x}_1,\mathbf{x}_0,t)=t \mathbf{x}_1(\boldsymbol{\theta})+(1-t)\mathbf{x}_{0},
    \label{eq:FMlinpath}
\end{equation}
which satisfies the boundary conditions
\begin{equation}
    \mathbf{x}_t(\mathbf{x}_1,\mathbf{x}_0,t=0) = \mathbf{x}_0, \qquad \mathbf{x}_t(\mathbf{x}_1,\mathbf{x}_0,t=1) = \mathbf{x}_1(\boldsymbol{\theta}).
\end{equation}
Taking the derivative of Eq.~(\ref{eq:FMlinpath}) with respect to flow time gives a conditional velocity field
\begin{equation}
    \dot{\mathbf{x}}_t = \mathbf{x}_1(\boldsymbol{\theta}) - \mathbf{x}_{0}.
\end{equation}
Averaging over draws from the prior and training distributions, as well as flow times, gives the flow matching loss\footnote{Note that even though each conditional path $\mathbf{x}_t$ is linear, the learned velocity $\mathbf{V}_\phi(\mathbf{x}, t)$, which approximates the population-optimal marginal velocity $\mathbf{V}^*(\mathbf{x},t) = \mathbb{E}[\mathbf{x}_1 - \mathbf{x}_0 | \mathbf{x}_t = \mathbf{x}]$, need not be constant as it averages over many endpoint pairs $(\mathbf{x}_0, \mathbf{x}_1)$; see Ref.~\cite{lai2025principlesdiffusionmodels} for a pedagogical review including an illustrative example with 1-dimensional Gaussians.}
\begin{equation}
\mathcal{L}_{\rm FM}
=  \mathbb{E}_{t \sim \mathcal{U}[0,1]} \mathbb{E}_{\mathbf{x}_0 \sim p_0,  \mathbf{x}_1 \sim p} \left [\big\| \mathbf{V}_\phi(\mathbf{x}, t|\boldsymbol{\theta})
- \dot{\mathbf{x}}_t  \big\|^2 \right] = \mathbb{E}_{t \sim \mathcal{U}[0,1]} \mathbb{E}_{\mathbf{x}_0 \sim p_0,  \mathbf{x}_1 \sim p} \left [\big\| \mathbf{V}_\phi(\mathbf{x}, t|\boldsymbol{\theta})
- (\mathbf{x}_1(\boldsymbol{\theta}) - \mathbf{x}_{0})  \big\|^2 \right].
\label{eq:loss}
\end{equation}
Minimizing $\mathcal{L}_{\rm FM}$ over neural network parameters $\phi$ gives the desired $\mathbf{V}_\phi$. Unlike normalizing flows \cite{dinh2015nicenonlinearindependentcomponents,rezende2016variationalinferencenormalizingflows,dinh2017densityestimationusingreal,Elsharkawy:2025six}, flow matching never changes variables in order to transport distributions, and unlike diffusion-based transports \cite{song2021scorebasedgenerativemodelingstochastic,ho2020denoisingdiffusionprobabilisticmodels,karras2022elucidatingdesignspacediffusionbased}, flow matching is deterministic and invertible, although there exists correspondence between the two \cite{albergo2025stochastic,albergo2022building}.   

\subsection{Physical flow matching with COLA}
\label{sec:physFM}

Our modification to the standard flow matching procedure, physical flow matching, uses a \emph{physical prior} distribution furnished by COLA, and \emph{explicitly enforces conservation laws along the flow}. We describe each in turn.  

\textbf{Physical priors.} In flow matching, the only requirement on the prior is that it be computationally cheap to sample from. One choice is simply to take $p_0$ to be the initial conditions of the \Nbody{} simulation itself. We attempted this choice, but the resulting model had trouble achieving the visual fidelity we desired. In a sense, a generative model learning a continuous flow where flow time is in 1-to-1 correspondence with cosmic time must be approximating the solution of the differential equations simulated by \Nbody{} dynamics, and is thus subject to accumulation of discretization errors which eventually overwhelm any small-scale structure.  We saw some success with using a chain of two flows, where the first flow learns a transport from the \Nbody{} initial conditions to a coarse-grained phase space density which captures large-scale structure, and the second flow takes the endpoint of the first flow as its prior and learns to add small-scale structure. The necessity of a decomposition into large- and small-scale emulators for physical systems with structure across a wide range of scales is well-studied in the climate~\cite{Kochkov_2024,kim2026mindresidualgapprobabilistic,sundar2025taudiffhighlyefficientkilometerscale,pathak2024kilometerscaleconvectionallowingmodel,mardani2024residualcorrectivediffusionmodeling} and fluids/partial differential equation ~\cite{Kochkov_2021,um2021solverinthelooplearningdifferentiablephysics,Bar_Sinai_2019} emulation communities. Without such a decomposition, the model seems to converge on a smeared-out mean transport of the desired field, with potential explanations ranging from gradient starvation \cite{NEURIPS2021_0987b8b3} to the spectral bias \cite{rahaman2019spectralbiasneuralnetworks}. 

However, a much more computationally-efficient large-scale transport is provided by COLA~\cite{Tassev:2013pn}, which obviates the need for training two separate generative models. COLA acts as an approximate, fast mesh-based solver of \Nbody{} dynamics that is easy to run on a GPU. It is accurate on large-scale structure \cite{Tassev:2013pn,Ding:2023uro}, as can be seen visually in Fig.~\ref{fig:fast-inference} (top right), but it fails to reproduce substructure, small-scale power, and velocity distributions. We thus cast our flow matching model as transport from an approximate COLA field $\mathbf{x}_{\rm COLA} \sim p_{\rm COLA}(\mathbf{x})$ to the full \Nbody{} distribution $\mathbf{x}_1 \sim p(\mathbf{x} | \boldsymbol{\theta})$ via the conditional path
\begin{equation}
     \boldsymbol{\varphi}=t\cdot\mathbf{x}_1(z, \boldsymbol{\theta})+(1-t)\cdot\mathbf{x}_{\rm COLA}( \boldsymbol{\theta}) \implies \dot{\boldsymbol{\varphi}} = \mathbf{x}_1(z, \boldsymbol{\theta}) - \mathbf{x}_{\rm COLA }( \boldsymbol{\theta}).
\end{equation}
$\dot{\boldsymbol{\varphi}}(t,z,\boldsymbol{\theta})$ becomes our target for training, where the loss is\footnote{In practice, we find that sampling in flow time from a distribution $p_t$ which is linear in $t$ rather than uniform gives better performance, see Sec.~\ref{sec:training}.}
\begin{equation}
\mathcal{L}_{\rm PHYS-FM}
=  \mathbb{E}_{t \sim p_t}  \mathbb{E}_{\mathbf{x}_{\rm COLA} \sim p_{\rm COLA},  \mathbf{x}_1 \sim p} \left [\big\| \mathbf{V}_\phi(\mathbf{x},t|z,\boldsymbol{\theta})
- \dot{\boldsymbol{\varphi}}  \big\|^2 \right ],
\label{eq:loss_phys}
\end{equation}
This approach is similar in spirit to \textsc{COCA} \cite{Bartlett:2024cra} and \textsc{PGD} \cite{Dai:2018vvv}, which also learn a correction to a fast but approximate \Nbody{} solver (indeed, COCA is designed to correct COLA itself). However, both methods work as a correction to the partial differential equation (PDE) solved on the mesh. Their corrections are thus deterministic, and bound to the mesh resolution used in the fast solver. \astrolabe{}, on the other hand, acts directly on particles. This allows sub-mesh corrections, stochasticity (for example, by producing many realizations per low-mesh COLA prior), and the possibility of incorporating hydrodynamic emulation of baryons with distinct particle types, which in principle requires minimal change to the method other than training data.

\textbf{Conservation laws.} In physical flow matching, we constrain all physical data along the flow to satisfy certain conservation laws for all flow time. In this work, we focus on the conservation of momentum of the dark matter particles along the flow, and we do so by defining the projection operator $\mathcal{P}$ which removes any component of flow velocity that would modify the momentum of a given point $\mathbf{x}_{\rm COLA}$ sampled from the COLA prior, where COLA also conserves momentum. In the particular case where points sampled from the prior consist of $N$ equal-mass particles, this projection can be accomplished by removing the mean change of particle phase-space velocity from the flow velocity. From the ODE formulation Eq.~(\ref{eq:FM_ode_standard}), with $\mathbf{x} = \{\mathbf{x}_i\}_{i=1}^N \in \mathbb{R}^{6N}$, the flow velocity $\mathbf{V}_\phi$ is a tangent vector to $6N$-dimensional phase space. Let $\mathbf{V}_\phi^{(\mathbf{r}_i)}$ and $\mathbf{V}_\phi^{(\mathbf{v}_i)}$ denote the components of flow velocity tangent to the position coordinates $\mathbf{r}_i$ and velocity coordinates $\mathbf{v}_i$ of particle $i$, respectively. We define the momentum-conserving projection operator $\mathcal{P}$ as
\begin{equation}
\mathcal{P}[{\mathbf{V}_\phi(\mathbf{x}_t,t|z,\boldsymbol{\theta})}^{(\mathbf{v}_i)}] = \mathbf{V}_\phi(\mathbf{x}_t,t|z,\boldsymbol{\theta})^{(\mathbf{v}_i)}-
\frac{1}{N}\sum_{j=1}^N \mathbf{V}_\phi(\mathbf{x}_t,t|z,\boldsymbol{\theta})^{(\mathbf{v}_j)}, \qquad \mathcal{P}[{\mathbf{V}_\phi(\mathbf{x}_t,t|z,\boldsymbol{\theta})}^{(\mathbf{r}_i)}] = {\mathbf{V}_\phi(\mathbf{x}_t,t|z,\boldsymbol{\theta})}^{(\mathbf{r}_i)},
\end{equation}
and we modify our objective in Eq.~(\ref{eq:loss_phys}) to 
\begin{equation}
\mathcal{L}'_{\rm PHYS-FM}
=  \big\| \mathcal{P}[\mathbf{V}_\phi(\mathbf{x}_t,t|z,\boldsymbol{\theta})]
- \dot{\boldsymbol{\varphi}}  \big\|^2 .
\label{eq:loss_proj}
\end{equation}
The projection operators can be applied to every velocity evaluation during inference. The projected velocity field which sources the ODE evolution therefore lives in the tangent space to the zero-net-momentum submanifold of possible phase space flows, ensuring that the integrated trajectory \emph{exactly} conserves the total momentum of the box at every step, rather than approximately enforcing conservation by a loss penalty.\footnote{For a similar approach applied to distributions of relativistic particles, see Ref.~\cite{Bogorad:2026oxa}.} Conservation of mass is trivially enforced by keeping the number of particles constant when the flow transports equal-mass particles. In a box with periodic boundary conditions, there are no independent symmetry generators associated with angular momentum, so conservation of mass and momentum exhaust the physical conservation laws of this system.

\subsection{Field-Particle Hybrid Architecture}
\label{sec:arch}

\begin{figure}[t]
    \centering
    \includegraphics[width=\linewidth]{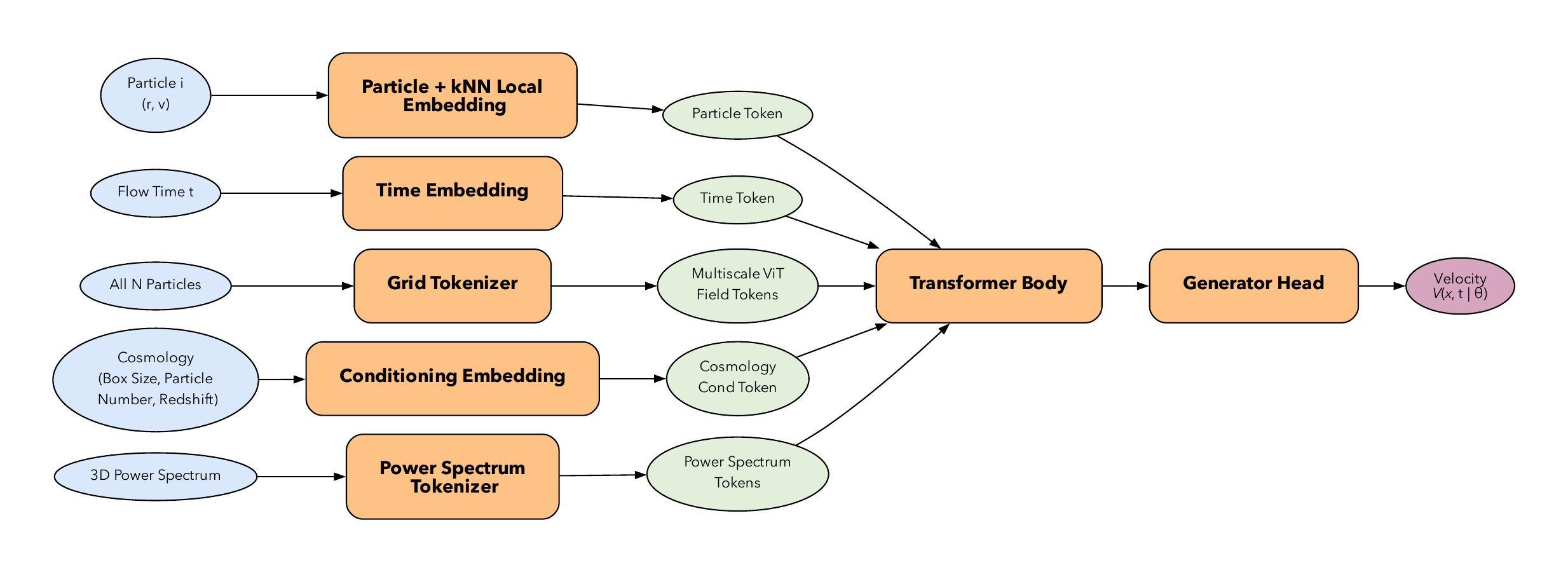}
    \caption{The \astrolabe{} field$+$particle hybrid network. Inputs
    (blue) are the flowing particle, the full particle set, the flow time, and any conditioning
    variable. Model blocks are orange and token types green. The body alternates token and particle
    cross-attention updates at cost linear in $N$, and the generator head outputs the six-channel
    velocity $\mathbf{V}_\phi(\mathbf{x},t|\boldsymbol{\theta})$ for conditioning variables $\boldsymbol{\theta}$.}
    \label{fig:arch}
\end{figure}

The learned flow velocity field must be sensitive to two very different regimes: (1) short-range, highly nonlinear physics, and (2) long-range
field interactions due to the universally attractive nature of gravity. All-to-all attention over
particles captures both but costs $\mathcal{O}(N^2)$, which is prohibitive at the particle counts of interest. \astrolabe{} instead splits the interaction into a \textit{particle} channel and a \textit{field} channel, each with linear complexity in particle number $N$. See Figure \ref{fig:arch} for an overview diagram of the architecture, which we will describe in more detail below. Hyperparameters are given in detail in Appendix \ref{app:hparams}.
\begin{figure}[h]
    \centering
    \includegraphics[width=\linewidth]{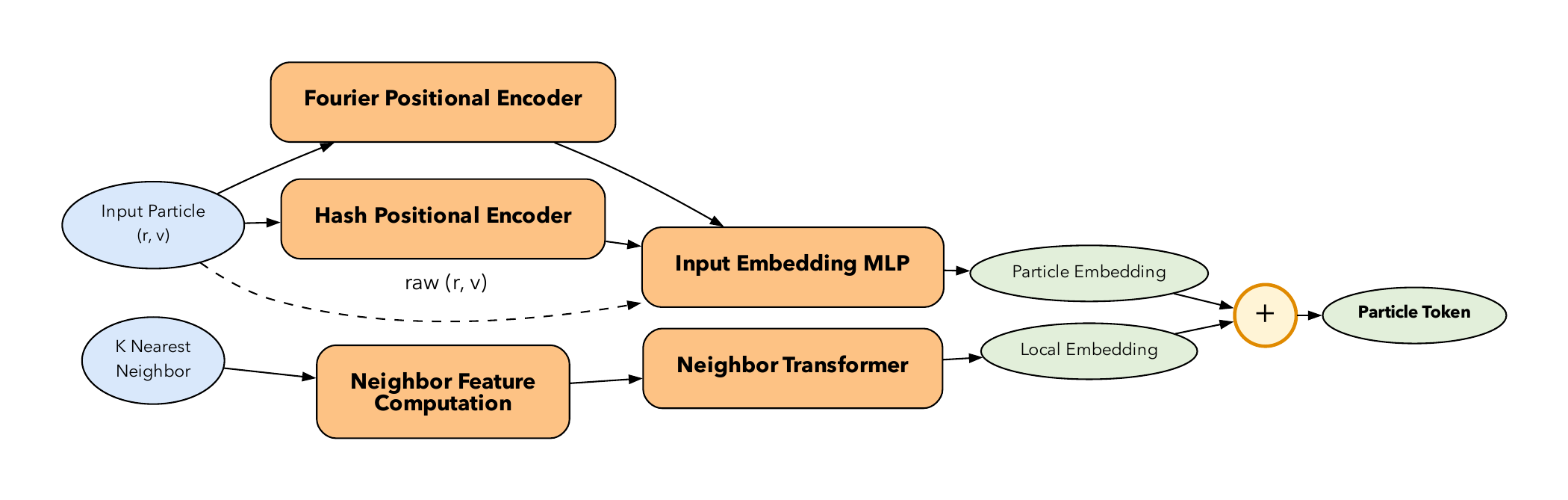}
    \caption{The particle and local encoder. Each particle's position is hash-position-encoded, Fourier position-encoded, and embedded by an MLP. In parallel, its $K$ nearest neighbors are turned into pair features, passed through a small transformer, and pooled into a local embedding. The two are summed into the particle token.}
    \label{fig:localknn}
\end{figure}

\begin{figure}[t]
    \centering
    \includegraphics[width=\linewidth]{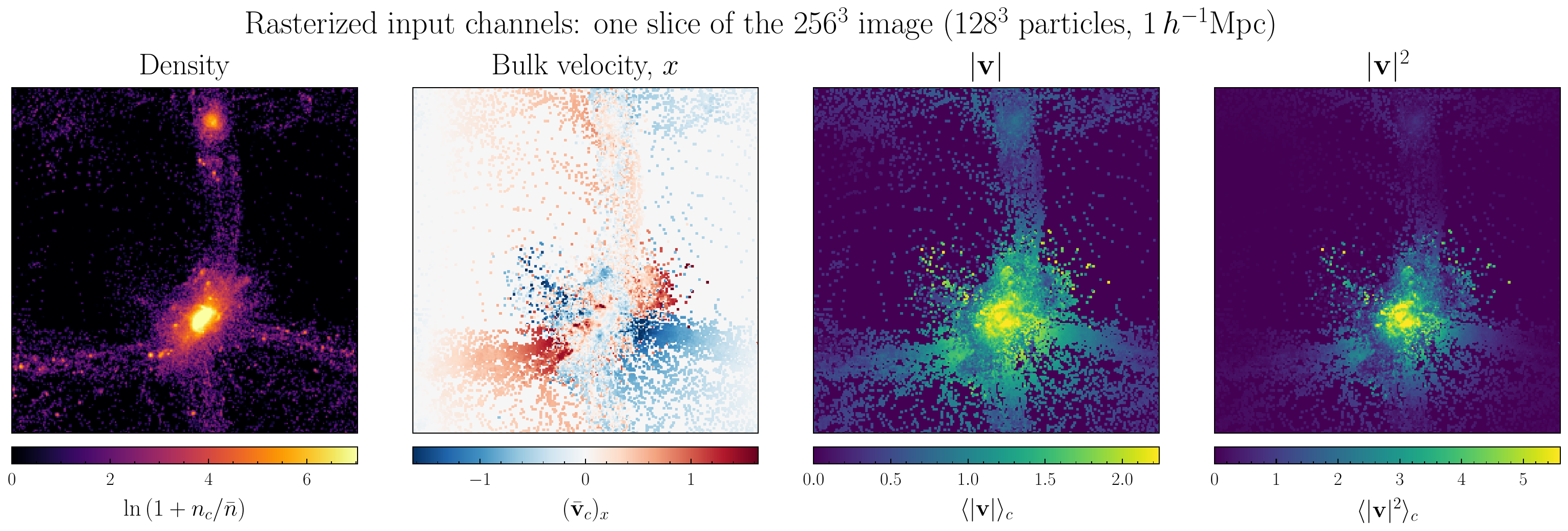}
   \caption{2-dimensional slice of example rasterized field fed after normalization to 3-dimensional convolutions for four of the six input channels (the remaining two channels are momenta $p_y$ and $p_z$).}   
    \label{fig:field-tokens}
\end{figure}

\begin{figure}
    \centering
    \includegraphics[width=\linewidth]{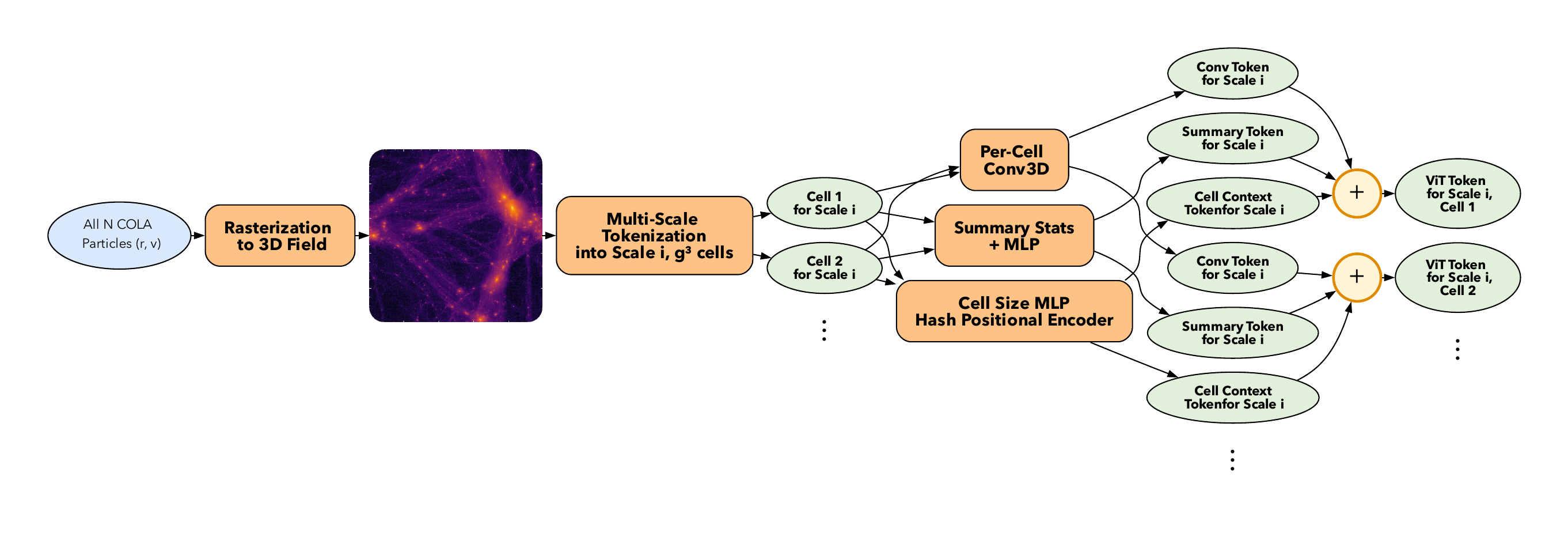}
    \caption{The multi-scale grid tokenizer. All $N$ particles are rasterized into a 3-dimensional field and split into $g^3$ cells at many scales. Each cell's ViT token is the sum of three embeddings of the cell: 1) a pooled 3-dimensional convolution, 2) an MLP over summary statistics, and 3) a cell context MLP over the (hash-encoded) cell center positions and size.}
    \label{fig:gridtokenizer}
\end{figure}

\textbf{Particle channel and local embedding kNN.} Each particle's COLA state $\mathbf{x}_{\rm COLA}$ is concatenated with its initial phase space coordinates $\mathbf{x}_0$ and the resulting 12-dimensional vector is embedded by a hash encoder \cite{takikawa2021neuralgeometricleveldetail,Huang_2024,muller2022instant} and a Fourier positional encoding \cite{rahaman2019spectralbiasneuralnetworks,sitzmann2020implicitneuralrepresentationsperiodic,tancik2020fourierfeaturesletnetworks,mildenhall2020nerfrepresentingscenesneural}. This ensures that the network can refine fine-scale detail, as discussed further in Appendix~\ref{app:hash}. The output of the hash encoder and the Fourier encoder, along with raw particle positions and velocities, is fed to a final embedding multi-layer perceptron (MLP) which creates a per-particle embedding token of dimension $D$. 

For every particle $i$, we then gather its $K=256$ nearest neighbors as measured by Euclidean distance in position space, and build, per pair $(i,j)$ for $j = 1, \dots, 256$, a vector of $N_{\rm pair}=49$ physically-motivated scalars given by,  
\begin{equation}
    \vec x_{\rm pair} = \Big(\mathbf{r}_{ij}/\bar d,~ r_{ij},~ \log r_{ij},~ 1/r_{ij},~ 1/r_{ij}^2,~\Delta\mathbf{v}_{ij},~|\Delta \mathbf{v}_{ij}|,~ \Delta \mathbf{v}_{ij} \cdot\hat{\mathbf{r}}_{ij},~ \hat{\mathbf{r}}_{ij},~ \cos\theta_{ij},~ v^{\rm tan}_{ij},~ \big\{{\rm RBF}(\mathbf{r}_i,\mathbf{r}_j)_m\big\}\Big)
\end{equation}
Here, $\mathbf{r}_{ij}$ is the pairwise separation vector, $\bar d$ is the mean interparticle spacing, $r_{ij} = |\mathbf{r}_{ij}|$, $\hat{\mathbf{r}}_{ij} = \mathbf{r}_{ij}/r_{ij}$, $\Delta\mathbf{v}_{ij} = \mathbf{v}_i - \mathbf{v}_j$ is the relative velocity, $\cos\theta_{ij} = \Delta\mathbf{v}_{ij}\cdot\hat{\mathbf{r}}_{ij}/|\Delta\mathbf{v}_{ij}|$, and $v^{\rm tan}_{ij} = \sqrt{|\Delta\mathbf{v}_{ij}|^2 - (\Delta\mathbf{v}_{ij}\cdot\hat{\mathbf{r}}_{ij})^2}$ is the relative tangential speed. Finally, 
\begin{equation}
    {\rm RBF}(\mathbf{r}_i,\mathbf{r}_j)_m =
    \exp\left(-\frac{\left(r_{ij}/\bar d-\mu_m\right)^2}{\delta^2}\right)
    \frac{1}{2}\left[\cos\left(\frac{\pi r_{ij}}{6\bar d}\right)+1\right],
\end{equation}
for $m = 1, \dots, 32$ is a set of 32 radial basis functions that act as a flexible basis for a network to transform, where $\mu_m$ is sampled uniformly from $[0,6]$, and we choose $\delta=6/31\sim 0.2$. This strategy of encoding physically-motivated features and the use of RBFs is inspired by Refs.~\cite{Mikuni:2025ocp,Bhimji:2025isp,Elsharkawy:2026kwp}, which showed success for learning point-cloud distributions; our choices represent an attempt to build an inductive bias into the network of the $1/r^2$ central force nature of gravity. The resulting set of $K=256$ vectors are fed into $L_{\rm loc}$ small self-attention blocks. We take the mean of the transformed $K=256$ vector output and this gives us a local embedding: a token encoding a particle's relevant local interactions. This local embedding is summed with the particle embedding to give us our final per-particle token. See Figure \ref{fig:localknn} for a diagrammatic representation. 

\begin{figure}[t]
    \centering
    \includegraphics[width=\linewidth]{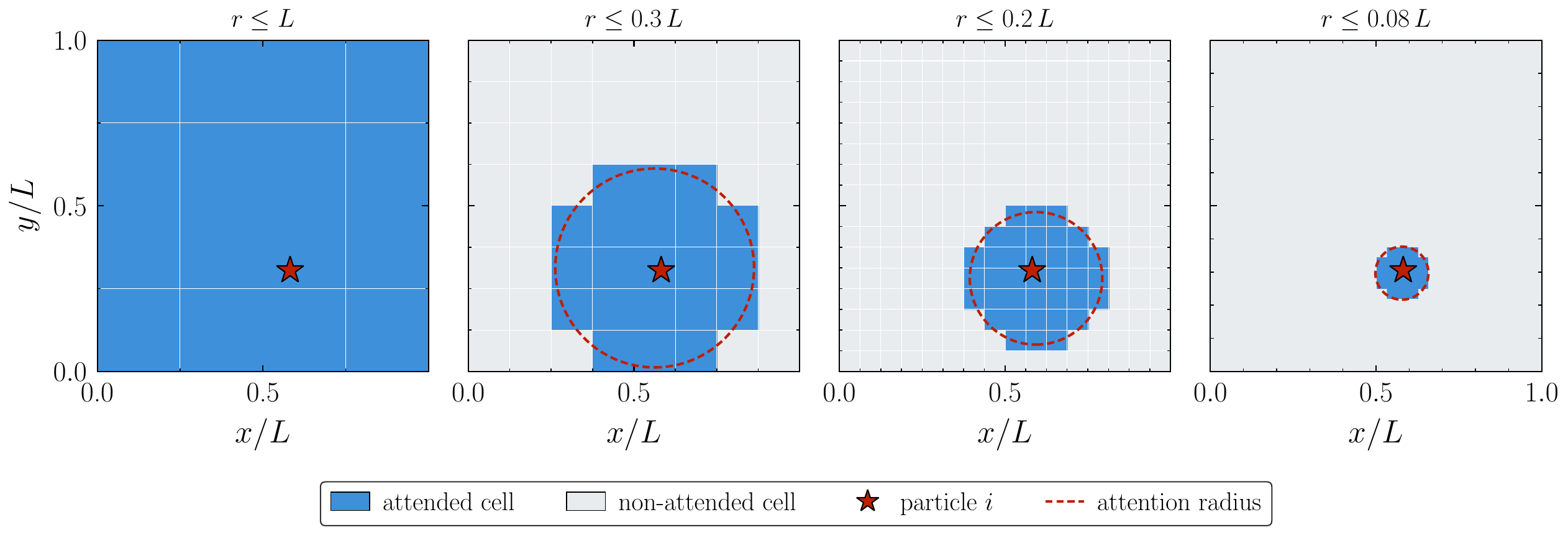}
   \caption{Distance-gated field attention for one particle (red star). Here we plot one panel per scale $g$. For this particle, blue cells are attended to, while grey cells are excluded. The dashed circle is the radius of attention for the networks $f_g L$ about the particle's cell center: from left to right, $f_g = 1,  0.3,  0.2,  0.1$. Hyperparameters are chosen so that coarse scales stay global, while finer scales are confined to smaller local neighborhoods. The number of attended tokens per particle stays small and independent of the total number of cells across all scales.}   
    \label{fig:ViT-atten}
\end{figure}

\textbf{Multi-scale field tokens.} Long-range correlations are fed to the network through a 3-dimensional vision-transformer (ViT)-style tokenization of the particle field. We first rasterize the particles into a high-resolution 3-dimensional image with 6 channels (number density, momentum, $|\mathbf{v}|$, $|\mathbf{v}|^2$). We take the log and normalize the number density and mean-normalize the velocity channels by,
\begin{equation}
    n_{c,\rm input}= \ln(1 +\frac{n_c}{\bar{n}}),\quad \boldsymbol{v}_{c,\rm input}= \frac{\boldsymbol{v}_{c}}{\bar{\boldsymbol{v}}}
\end{equation}
where $\bar{n}$ is the average number density of particles and $\bar{\boldsymbol{v}}$ is the average particle velocity in voxels of the cell size $c$. Example inputs can be found in Figure \ref{fig:field-tokens}. For each grid size in $g\in\{2,4,8,16,32\}$ the box is subdivided into $g^3$ cells, each of which is then subsampled to $m^3$ voxels.

Each cell is encoded via three independent blocks, with the final field token constructed by summing their contributions. The first block is a convolutional neural network, where each cell passes through two $3\times3\times3$ learned convolutions to output $C$ channels over the cell's $m^3$ voxels. The output of the convolution is then flattened and projected via a learned linear transformation to width $D$. 

The second is a summary path, which first computes per-cell (rather than per-voxel) a cell summary vector given by,
\begin{equation}
    \vec{s}_c = \Big(\ln\big(1 + n_c/\bar n\big),~
\langle |\mathbf{v}| \rangle_c,~
\sigma_{v,c},~
|\bar{\mathbf{v}}_c|,~
\delta_c,~
\lambda_1/\lambda_3,~
\lambda_2/\lambda_3,~
\sqrt{\lambda_1+\lambda_2+\lambda_3}/\ell,~
\rho_{r,c},~
f_c~
\Big),
\end{equation}
where $n_c$ is the number of particles in the cell, $\bar n$ is the mean number of particles in cells of the same size. $\langle |\mathbf{v}| \rangle_c$ is the mean particle speed in the cell, $\sigma_{v,c}$ is the velocity dispersion, $|\bar{\mathbf{v}}_c|$ is the bulk velocity vector of the cell, $\delta_c = \frac{|\langle \mathbf{r} \rangle_c|}{\ell/2}$ is the distance between the cell's center of mass and the cell's geometric center. The shape parameters $\lambda_1/\lambda_3,,
\lambda_2/\lambda_3,~$ and $\sqrt{\lambda_1+\lambda_2+\lambda_3}/\ell$, constructed from the eigenvalues $\lambda_{1,2,3}$ of the cell's moment of inertia tensor, roughly quantify how the mass is distributed in the cell. Finally, $\rho_{r,c} = \frac{\mathbf{R}\cdot\mathbf{V}}{|\mathbf{R}|\,|\mathbf{V}|}$, with $\mathbf{R},\mathbf{V}\in \mathbb{R}^{3N}$ captures the corelation between a particle's radial position from a cell's center and its velocity as a measure for radial flow, and $f_c$ is the occupancy fraction, the number of voxels in the cell that contain a particle. The summary vector $\vec{s}_c $ is passed through an MLP that provides a summary encoding.

Finally, we encode each cell's location and size through another MLP and the same hash encoder as was used in the per-particle embedding, see Appendix \ref{app:hash}. We then add all three cell embeddings (the field, the summary, and the location/size embedding) to find a final per-cell embedding. We do this for all cells for all grid sizes $g\in\{2,4,8,16,32\}$, giving $T_{\rm field} =\sum_g g^3 = 37,448$ cell embedding tokens of dimension $D$ as the output of the field tokenizer. See Fig.~\ref{fig:gridtokenizer} for a diagram of the architecture.

\textbf{Time embedding and additional conditioning.} In addition to particle-level features, we also embed global cosmological and flow time as additional context. Flow time is embedded with Fourier features followed by an MLP.  The conditioning variables $\boldsymbol{\theta}$ (the redshift $z$ transformed to scale factor with $a=1/(1+z)$, the box size given in $\log({\rm Mpc}/h)$, the particle count, and any additional global parameters) are embedded by the same mechanism and summed into that token.

We also tokenize the comoving power spectrum of the field at the current flow time $t$. Importantly, we do \emph{not} encode the truth-level power spectrum, only the power spectrum of the state of the field at flow time $t$ at inference as additional conditioning. We compute the 3D power spectrum and bin to $16$ mean 1D values corresponding to 16 values of $k$. The power spectrum along with the value of $k$ is embed via an MLP to 16 tokens used to condition the output.

\textbf{Distance-gated cross-attention.} The transformer body learns particle correlations via cross-attention \cite{vaswani2017attention} applied to tokens. Attending to all $T_{\rm field}$ tokens from every particle is potentially prohibitively expensive given that the computational cost of attention is quadratic in tokens. As a solution, for a given particle and for each scale $g$, we restrict attention to a fraction $f_g$ of the box where the distance between the two cell centers is $\le f_g L$. Figure~\ref{fig:ViT-atten} shows what one particle would attend to given its location and a set of token scales $\{g\}$. This makes the finest grids affordable while allowing coarse scales to stay global for all particles, as well as providing an inductive bias for locality in position space, with more information provided for objects near a given particle.

\textbf{Attention bias.} Following Refs.~\cite{Mikuni:2025ocp,Bhimji:2025isp,Elsharkawy:2026kwp,Elsharkawy:2026dte}, for each attended particle-cell pair $(i,j)$ we add a learned bias. For each distance-gated pair, we take the 3-vector from the particle to the cell center and its eight corners, append the cell size,
and pass the resulting $28$ features through a narrow MLP that outputs a scalar bias $B_{ij}$ which we use to bias the transformer body's attention logits $A_{ij}$ as
\begin{equation}
A_{ij}^{*} = A_{ij} + B_{ij}.
\label{eq:attn-bias}
\end{equation}
This bias gives the network an additional explicit physical prior on locality in position space.

\textbf{Head and conditioning.} The final architectural component of \astrolabe{} is the generator head, which applies $L_{\rm h}$ further cross-attention blocks, followed by an MLP that reshapes the data into our six output channels $(\mathbf{V}_\phi^{(\mathbf{r}_i)}, \mathbf{V}_\phi^{(\mathbf{v}_i)})$ representing the flow velocity field in the tangent space to 6-dimensional phase space for each particle $i$.

\section{Training and Inference}
\label{sec:training-inference}

In this section, we describe our training and inference procedure. In Sec.~\ref{sec:dataset} we describe the composition and generation of our training set, in Sec~\ref{sec:training} we describe our training procedure, and in Sec.~\ref{sec:inference} we explain how we perform inference at scale.

\subsection{The \Nbody{} and COLA dataset}
\label{sec:dataset}
 
We generate a training set consisting of matched pairs of \Nbody{} and COLA simulations that share the same initial conditions (that is, $z=99$ particle positions). We compute the initial conditions with a $\Lambda$CDM transfer function computed with \textsc{CLASS} \cite{Blas:2011rf}, fed to \textsc{MUSIC} \cite{Hahn:2011uy} to generate many different particle-level realizations (e.g. different particle point-clouds from the same initial distribution functions) at redshift $z=99$, which serves as our starting point for cosmic time evolution. For truth-level \Nbody{} simulations, we take each initial condition and use \textsc{GIZMO} \cite{Hopkins:2014qka} as a collisionless gravity solver. We use periodic boundary conditions, a $\Lambda$CDM cosmology, and save $100$ snapshots per run at redshifts uniform in scale factor. For training, we fix the number of particles per simulation to $128^3=2{,}097{,}152$; for brevity, we refer to this density as the 2M simulation.  We use $\Omega_m = 0.3111$, $\Omega_b = 0$, $h = 0.6766$, $\sigma_8 = 0.8102$. The 2M training runs in boxes of 1 and 10 Mpc/$h$ have interparticle spacings of 7.8 and 78 kpc/$h$, particle masses of $4.1\times10^{4}$ and $4.1\times10^{7}~M_\odot/h$, and comoving softening lengths of 0.195 and 1.95 kpc/$h$ respectively \cite{Planck:2018vyg}.

We then use COLA as a solver for the same initial conditions. COLA's major advantage is a drastically reduced number of timesteps compared to the true \Nbody{} simulation, but its accuracy is still strongly correlated with the number of timesteps used. Using intermediate snapshots of one COLA run to $z=0$ would therefore result in different levels of convergence at each redshift snapshot~\cite{Tassev:2013pn}. To account for this, we run COLA independently for each initial condition GIZMO-snapshot pair (snapshots at varying $z$), and each run is done with 40 timesteps and a particle mesh of $256^3$. 

The final dataset consists of $N_{\rm runs} = 1500$ independent $N$-body runs, each
with $N_{\rm snap} = 100$ uniform-in-scale-factor snapshots between the initial condition ($z = 99$) and $z =0$. Of these $1500$, half use a box size of $L=1$ Mpc/$h$, with the other half at a box size of $L=10$ Mpc/$h$. We run COLA for $104,813$ of these initial-condition GIZMO-snapshot pairs ($\sim$50,000 per box size, random snapshots drawn uniformly in scale factor from the 100 saved snapshots). At $\sim 2$M particles per snapshot, the dataset totals $22 \,\rm TB$ and contains roughly $200$ billion particles. Since each particle is treated individually in training (see Sec.~\ref{sec:training}), our models are thus trained on $\sim200$ billion individual tokens. We hold out eight GIZMO runs (i.e.\ hold out all snapshots associated with eight sets of initial conditions) for validation and eight GIZMO runs for test, and train on the rest, a $\sim 97\%/2\%/2\%$ split. 

\subsection{Training procedure}
\label{sec:training}
\begin{figure}
    \centering
    \includegraphics[width=.8\linewidth]{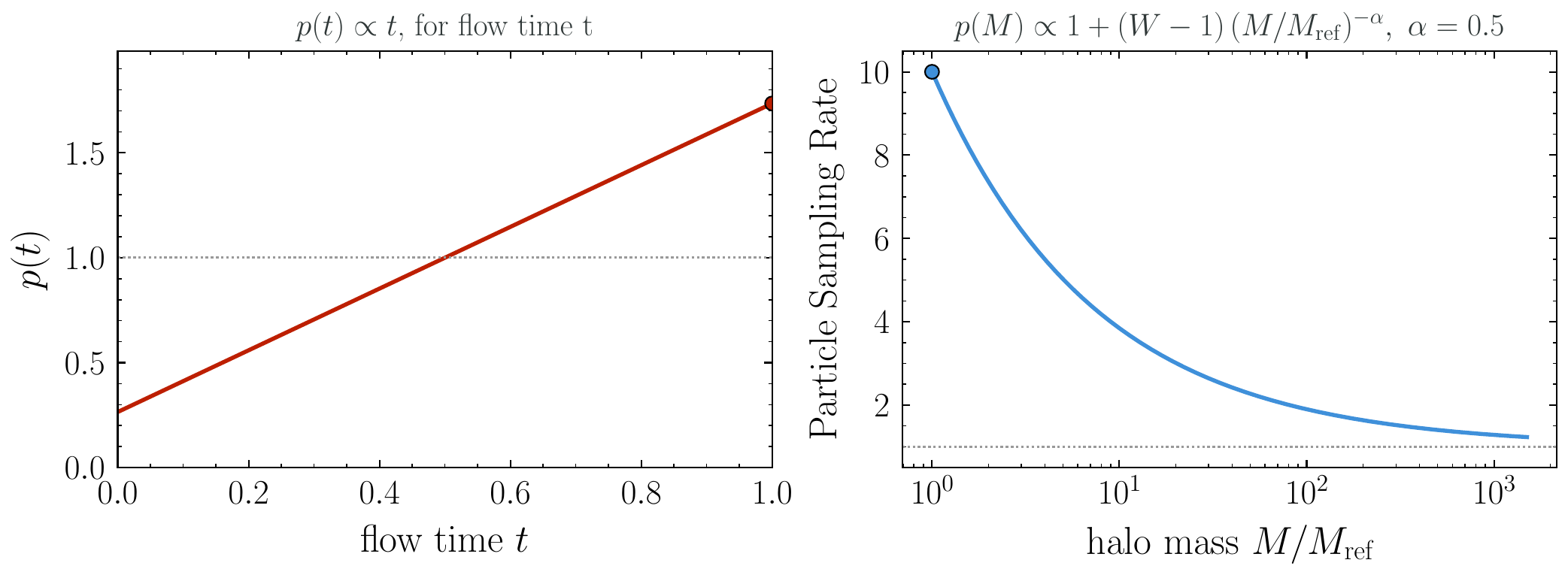}
    \caption{Two schedules used during training and inference. \textbf{Left:} the linear flow time sampling distribution, which prefers later flow time. \textbf{Right:} the halo-size sampling weight, which up-weights the likelihood of particles in small halos to be drawn and tapers to $1$ for the most massive halos. Here, we have converted $p(N)$ to $p(M)$ so that the mass dependence is in terms of a physical halo mass, in units of $M_{\rm ref} = 6.59\times10^9~M_\odot/h$.  }
    \label{fig:train-inf-weight}
\end{figure}
Each training step draws one paired COLA field $\mathbf{x}_{\rm COLA}$ and associated GIZMO snapshot $\mathbf{x}_1$ at the same redshift. We then sample the flow time linearly, $p_t \propto t$ (see Fig~\ref{fig:train-inf-weight}, left), construct the conditional path, and regress to the flow velocity using the loss function in Eq.~\eqref{eq:loss_proj}, with the target redshift transformed to a scale factor $a=1/(1+z)$ and box size supplied as conditioning. Naively this process would be intractable as $N$ increases, eventually far surpassing reasonable GPU memory budgets. However, large-$N$ training is made practical by \textit{particle subsampling}, which is uniquely possible due to our hybrid network's lack of all-to-all attention. Specifically, the grid tokens and neighbor candidates must always be built from all $N$ particles, but only a random subset of $n\ll N$ particles need to be embedded and supervised for each training batch for the network to converge to the true per-particle flow velocity $\dot{\boldsymbol{\varphi}}$. That is, uniform subsampling of $N$ and a per-particle-mean loss allow each step to use an unbiased estimator of the full loss gradient, cutting the per-step cost from $\mathcal{O}(N)$ to $\mathcal{O}(n)$. This construction permits training on simulations with any particle number $N$. A batch in this setup is thus some randomly drawn number of particles from the same COLA/GIZMO snapshot pair.

However, because the flow matching loss is a per-particle mean, it is dominated by the field's bulk motion. Particles within single large halos or living in the non-halo bulk are by far the most common, so the rare particles that end up in small, low-mass halos contribute negligibly to the gradient. Of course, a critical goal of our emulator is to reproduce the small-scale halo population. To remedy this, we up-weight the probability of a particle to be drawn in the sub-sampling if at $z = 0$ it ends up in a halo, with a weight depending on the $z = 0$ halo mass.\footnote{All halos in this work are found via a friend-of-friends halo finder. We use a minimum halo size of 20 particles, and the standard linking length parameter of $b=0.2$ \cite{More:2011dc,Davis:1985rj}.\label{ft:halo-finder}} In our training setup where each particle has equal mass, particle number is a proxy for halo mass, so our particle-number-dependent weighting is
\begin{equation}
    w(N)=1+(W-1)(N/N_{\rm ref})^{-\alpha}
\end{equation}
as illustrated in Fig.~\ref{fig:train-inf-weight} (right). With $W>1$ and
$\alpha>0$, small halos get a large weight relative to larger halos, encouraging the network to learn small-scale structure. In our training, we use $W=10$, $\alpha = 0.5$, and $N_{\rm ref}=20$, corresponding to a reference halo mass $M_{\rm ref} = 6.59\times10^9~M_\odot/h$ . Critically, such weighting is training-only and thus introduces no direct prior in inference, and simply acts as a bias on training gradients to encourage more network capacity spent on small halos.

Training uses \texttt{bf16} mixed precision, gradient clipping, and the \texttt{Ranger} optimizer \cite{wright2021ranger21synergisticdeeplearning,zhang2019lookaheadoptimizerksteps,liu2021varianceadaptivelearningrate}, with a linear warm-up of $n_{\rm warm}$ steps followed by cosine decay over $n_{\rm total}$ steps. Redshift, per-scale field tokens, and box size are droped out independently per sample. We train two model sizes as shown in Table~\ref{tab:sizes}. Both models are trained with 4x4 A100 40 GB nodes for 48 hours with early stopping. More details on training and model hyperparameters can be found in Appendix \ref{app:hparams}.

\begin{table}
    \centering
    \begin{tabular}{lccccccc}
        \hline
        Preset & $D$ & $L_{\rm b}$ & $L_{\rm h}$ & $H$ & $D_{\rm MLP}$ & Params \\
        \hline
        small  & 128  & 8  & 2 & 8  & 256 & 38.50 M \\
        medium & 512  & 12 & 2 & 16 & 1028 & 195.18 M \\
        \hline
    \end{tabular}
        \caption{Model-size presets. $D$ is the token dimension, $L_{\rm b}$ the number of body attention 
    blocks, $L_{\rm h}$ the number of generator-head blocks, $H$ the number of attention
    heads, and $D_{\rm MLP}$  the MLP hidden dimension.}
        \label{tab:sizes}
\end{table}

\subsection{Inference}
\label{sec:inference}

As described in Secs.~\ref{sec:FMreview}--\ref{sec:physFM}, we generate samples by solving the flow matching ODE with the learned flow velocity field. Given the COLA phase space distribution of a held-out run, we solve,
\begin{equation}
\frac{\mathrm{d}\mathbf{x}}{\mathrm{d}t} = \mathcal{P}[\mathbf{V}_\phi(\mathbf{x}, t | \theta)],
\qquad \mathbf{x}(t=0) = \mathbf{x}_{\rm COLA}.
\end{equation}
 We use the Euler integration scheme with $n_{\rm steps}$ network evaluations per output frame, and we apply the momentum projection at every integration step. Because flow time is a pure integration variable, Euler steps are placed uniformly in flow time on $[0,1]$. The desired generated phase space density is $\mathbf{x}(t=1)$.
 
 Our integration takes each COLA field and performs independent inference per desired redshift conditioned on both redshift and the box size. The field-particle hybrid architecture makes inference embarrassingly parallel. Per velocity evaluation, the rasterized 3-dimensional grid tokens are built \textit{once} from all $N$ particles. Since particle-to-particle communication only happens through the field tokens and local neighbors, the particle evaluation can be split into separate batches, one per device, and each batch runs its own particle forward. Consequently, the per-device compute and memory scales as $\mathcal{O}(N/n_{\rm device})$. Thus, adding GPUs increases the possible particle count essentially without limit, assuming the model generalizes to different particle number and different resolution. This allows us to probe the model's generalization beyond the particle number it was trained on.

\section{Results}
\label{sec:results}

\begin{figure}[t!]
    \centering
    \includegraphics[width=.9\linewidth]{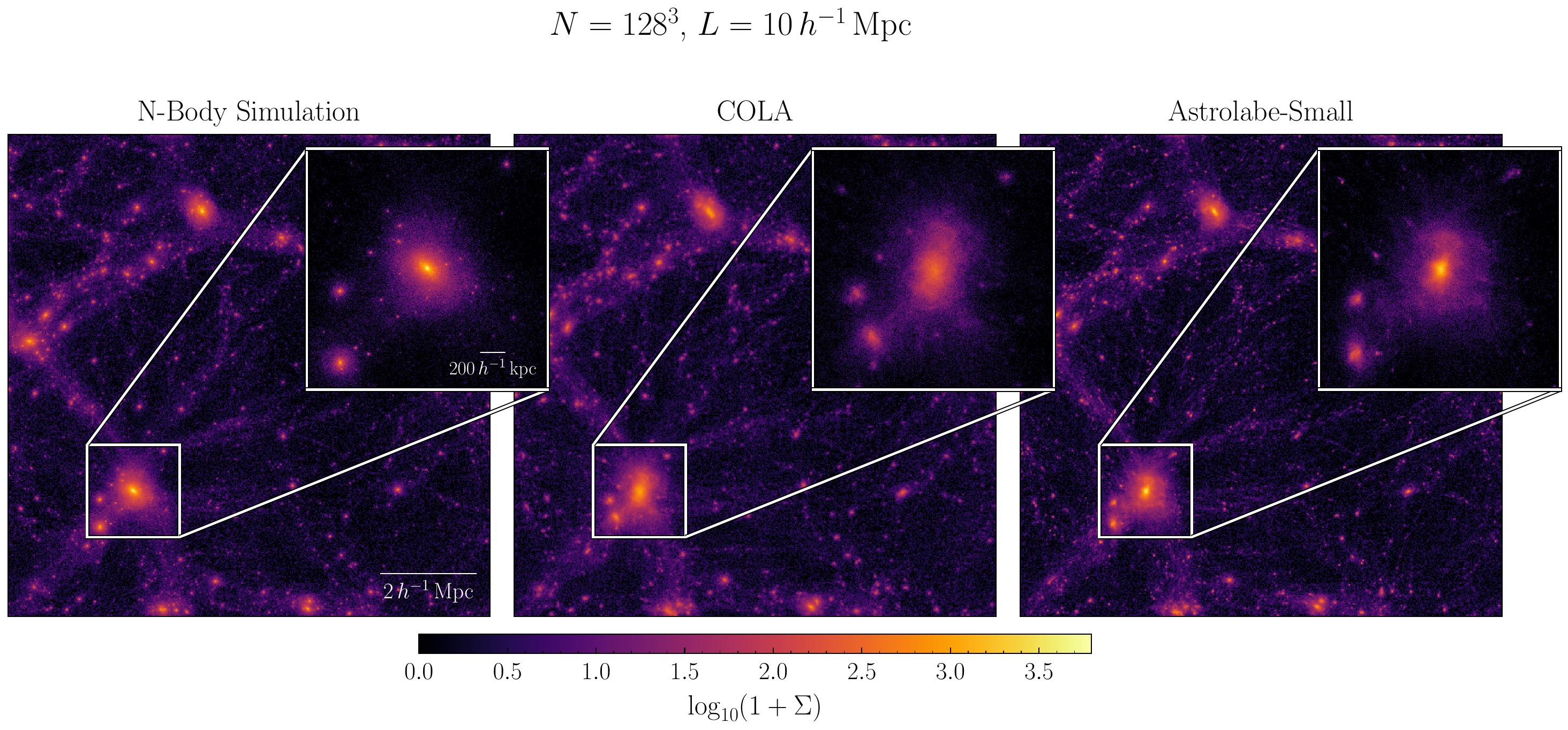}
    \includegraphics[width=.9\linewidth]{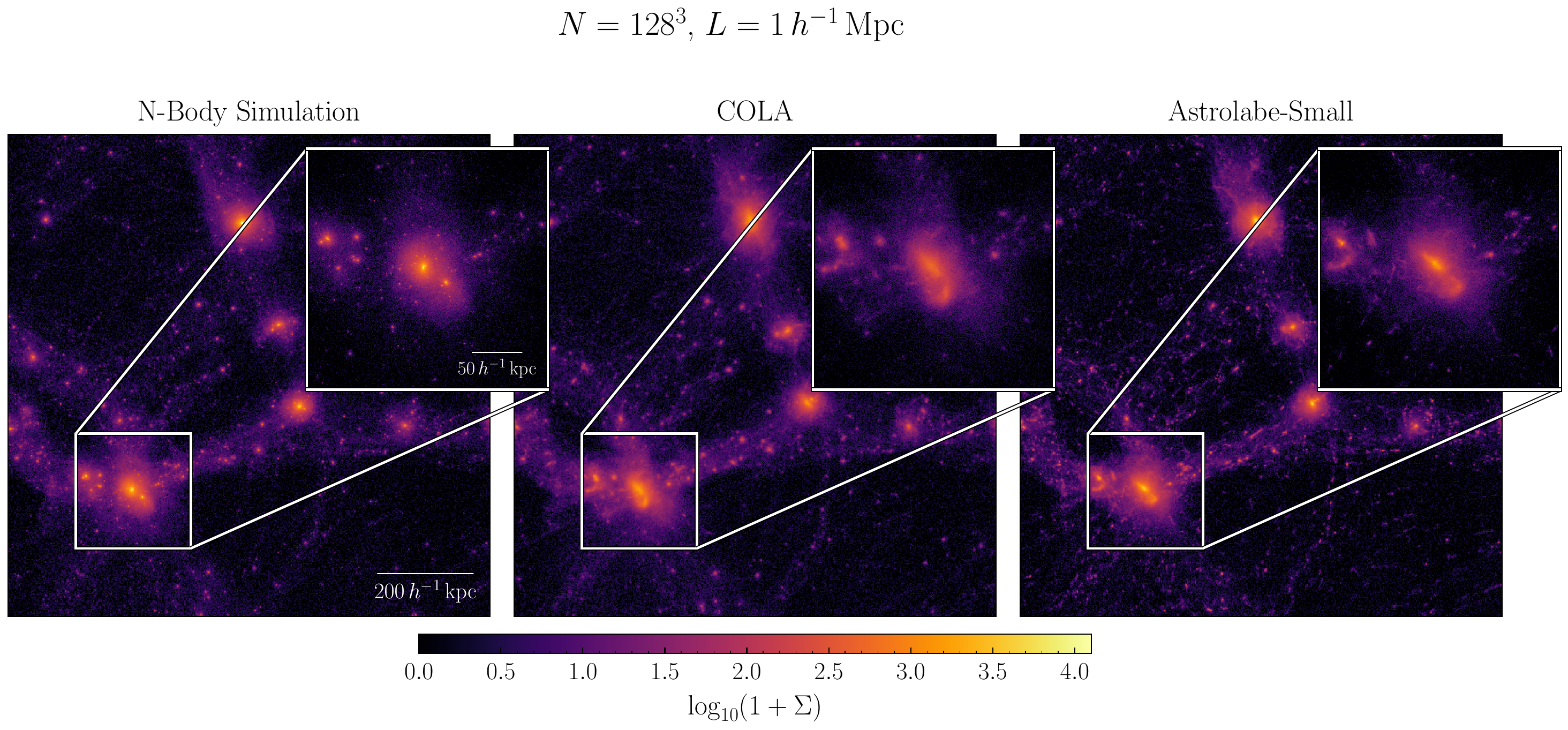}
    \caption{Example $\Lambda$CDM density field for $128^3\approx2\rm M$ particles in a 10 Mpc/$h$ box (\textbf{top row}) or a 1 Mpc/$h$ box (\textbf{bottom row}). \textbf{Left:} A \textsc{GIZMO} \cite{Hopkins:2014qka} \Nbody{} simulation, which is the ground-truth target. \textbf{Middle:} COLA with a mesh of $256^3$ and 40 timesteps. \textbf{Right:} \astrolabe-Small evaluated with 10 ODE steps. In each plot, we show surface density $\Sigma$ in units of particles per pixel on a log scale post-rasterization and zoom into the largest halo.}
    \label{fig:2M-field}
\end{figure}
In this section, we analyze \astrolabe's ability to emulate \Nbody{} simulations. In Sec~\ref{sec:cdm-res}, we evaluate \astrolabe{} on unseen initial conditions with in-distribution $\Lambda$CDM cosmology, box size, and particle count. We follow this in Section \ref{sec:dao-res} with a study on \astrolabe{}'s ability to emulate out-of-distribution non-CDM cosmology with a dark acoustic oscillation (DAO) case study. We then explore in Sec.~\ref{sec:larger-part-res} the ease in which \astrolabe{} can be fine-tuned for larger particle counts and different box sizes, including a 16M particle simulation along with the much larger 150M particle \texttt{TNG50-3-dark} \cite{Pillepich:2017jle} sim. Finally, in Sec.~\ref{sec:compute-res}, we compare the compute savings \astrolabe{} emulation can provide with respect to standard \Nbody{} simulations, where we use the results of an inference ablation in Appendix \ref{app:inference-ablation} to find a ``fast inference'' setting that results in dramatic compute gains. 

\subsection{$\Lambda$CDM (in-distribution) evaluation}
\label{sec:cdm-res}

\begin{figure}[t!]
    \centering
    \includegraphics[width=.83\linewidth]{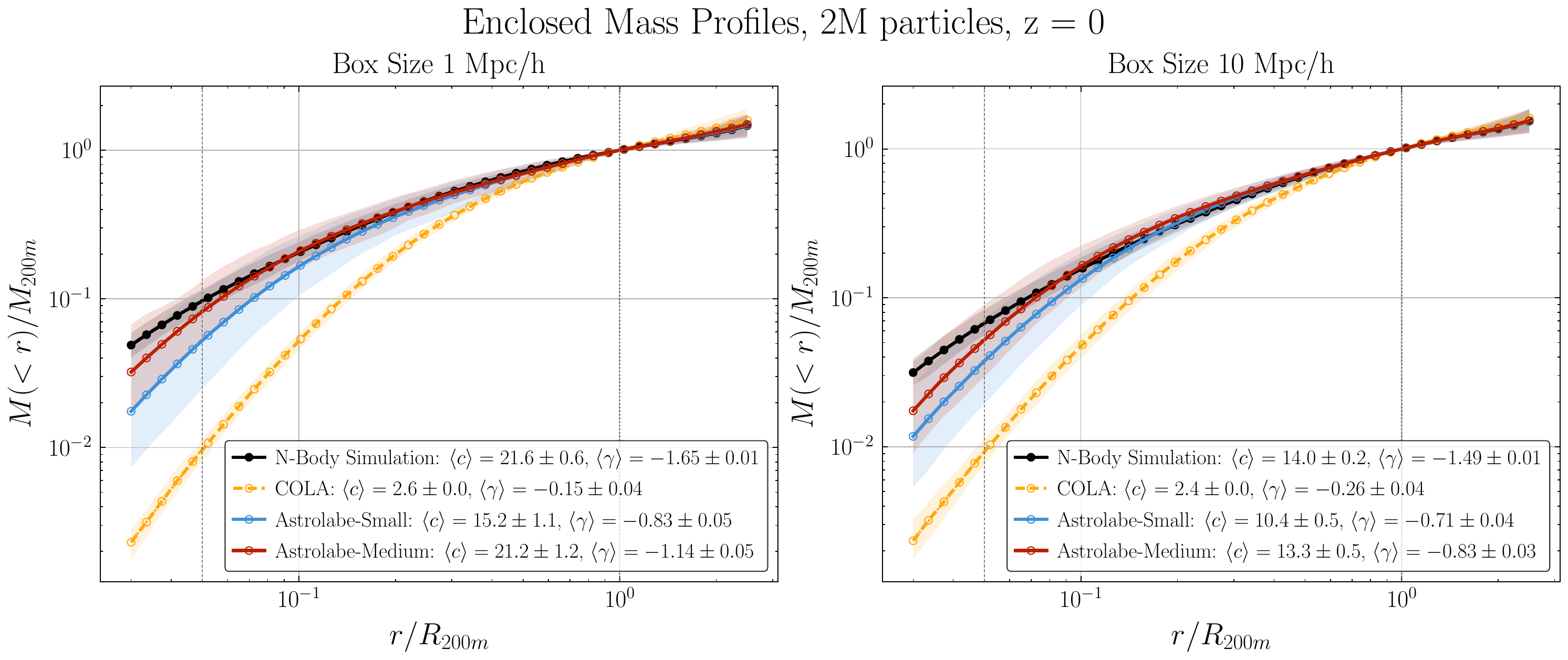}\vspace{1em}
    \includegraphics[width=0.8\linewidth]{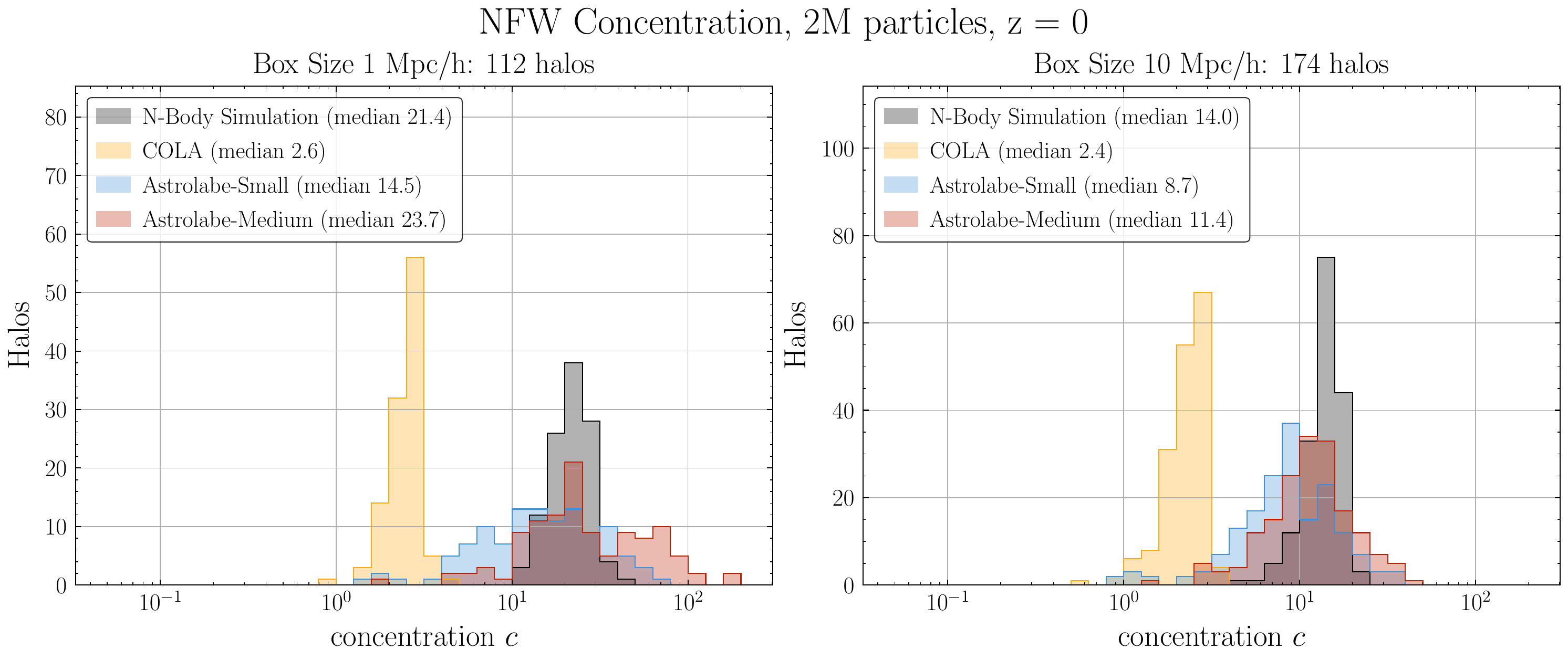}\vspace{1em}
    \caption{Consistency tests for NFW halo density profiles, see text for details. \textbf{Top:} Halo mass profile averaged over all halos with more than 5000 particles for each box size, with concentration $c$ from an NFW fit and inner slope $\gamma$ displayed. \textbf{Bottom:} A histogram of NFW-concentration $c$ over all halos with greater than $5000$ particles in all eight initial conditions in both box sizes. } 
    \label{fig:2M-CDM-conct-compare}
\end{figure}

We evaluate \astrolabe{} on a test set of in-distribution 2M particle simulations with 
$\Lambda$CDM cosmology at the two training box sizes of 1 Mpc/$h$ and 10 Mpc/$h$. The test set is composed of eight different initial conditions for ten different redshifts. We compare truth-level \Nbody{} sims with (1) COLA alone, (2) \astrolabe{}-small, (3) \astrolabe{}-medium, (4) \astrolabe{}-small fast inference, (5) \astrolabe{}-medium fast inference. Fast inference modes are done with fewer ODE integration steps and use a subsampling of particle evaluations which we define and discuss in Section \ref{sec:compute-res} and Appendix \ref{app:inference-ablation}. Figure \ref{fig:2M-field} displays example rasterized field comparisons for 10 Mpc/$h$ and 1 Mpc/$h$ boxes for \astrolabe{}-small, with the largest halo zoomed in. As was demonstrated in Fig.~\ref{fig:2M-field}, the visual accuracy of \astrolabe{} is clear, as is its ability to improve on small-scale structure with respect to COLA; in particular, the halos produced by \astrolabe{} look visually sharper and more resolved near the centers.

To make our observations about small-scale halos more quantitative, we consider three metrics that test the model's ability to produce physical dark matter halos: namely, a halo concentration error, an inner slope error, and a Jeans consistency test. For the concentration error, we compute a best-fite Navarro-Frenk-White (NFW) \cite{Navarro:1996gj} profile within the virial radius $R_{200}$, the radius from the halo center at which the enclosed density is equal to $200\bar\rho_m$~\cite{Navarro:1996gj}, where $\bar\rho_m = \Omega_m\rho_{\rm crit} \approx 86.3 ~\frac{M_\odot/h}{(\rm kpc/h)^3}$ \cite{Planck:2018vyg,ParticleDataGroup:2022pth} is the standard reference density. We do this for the set of halos each method produces that are found with a friends-of-friends halo finder beyond a minimum halo mass, set by the mass of 5,000 particles for the 2M evaluation and 50,000 particles for 16M and \texttt{TNG} evaluation.\footnote{Converted to physical masses, this corresponds to $8.2\times10^7~M_\odot/h$ for 2M 1Mpc, $8.2\times10^{10}~M_\odot/h$ for 2M 10 Mpc, $1\times10^8~M_\odot/h$ for 16M 1 Mpc, $4.7\times10^{11}~M_\odot/h$ for \texttt{TNG 50}, and $7.7\times10^{12}~M_\odot/h$ for \texttt{TNG 100}.}. Specifically, we fit the cumulative mass profile with the functional form from \cite{Navarro:1996gj,Lokas:2000mu},
\begin{equation}
    \frac{M(<r)}{M(<R_{200})} = \frac{\mu(c\cdot \frac{r}{R_{200}})}{\mu(c)}, \qquad \mu(y) = \ln(1+y) - \frac{y}{1+y}.
\end{equation}
Given a per-halo NFW fit, we take the associated concentration parameter $c$ and compute $\langle |\ln \frac{c_{\rm pred}}{c_{\rm truth}} | \rangle$ as the error, propagating initial condition variance if applicable. Figure \ref{fig:2M-CDM-conct-compare} shows the results of the NFW consistency tests. The top panel shows the mean cumulative mass profile averaged over all halos above our mass cutoff, and the bottom panel shows histograms of the best-fit concentration parameter $c$. It is clear that \astrolabe{} significantly improves on COLA in reproducing density profiles and concentrations that are closer to ground truth.

\begin{figure}
    \centering
      \includegraphics[width=0.8\linewidth]{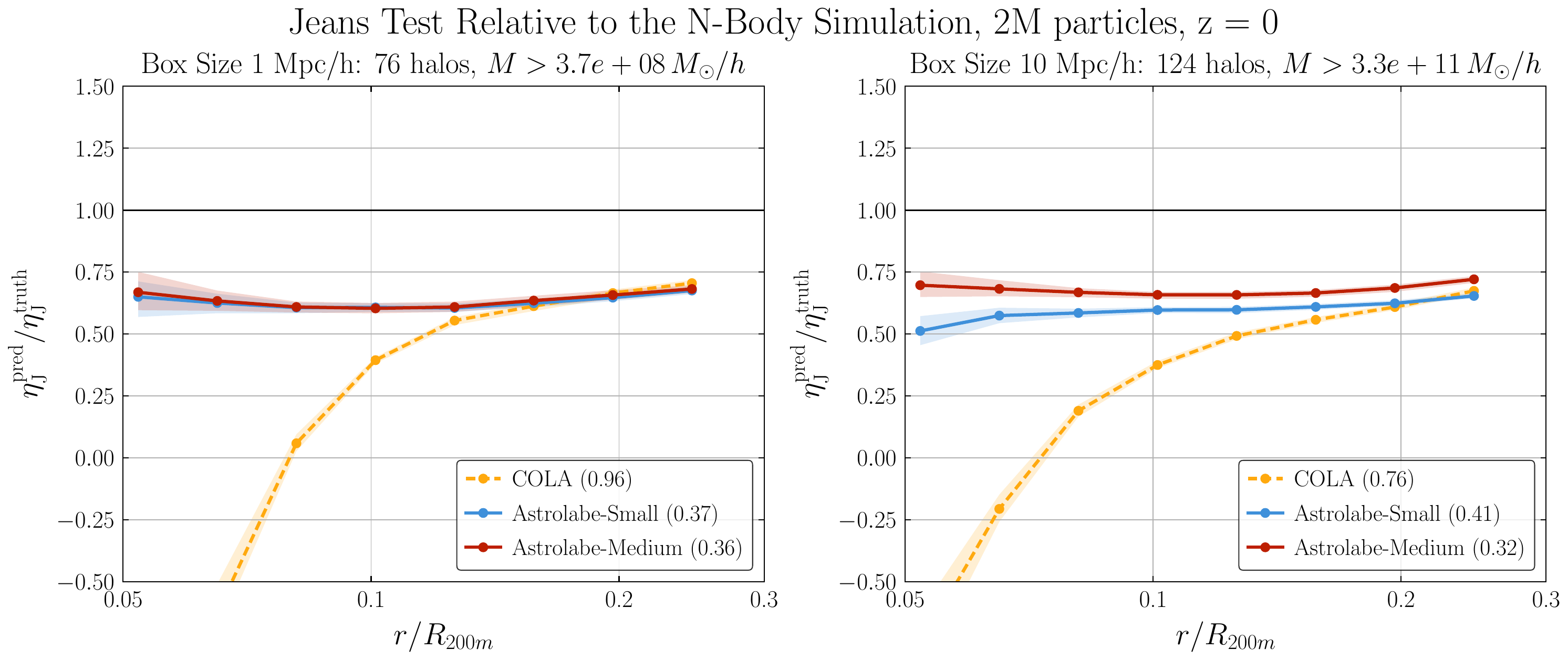}
    \caption{Jeans consistency test for virialized dynamics. We compute $\eta_J=\frac{M(<r)_J}{M(<r)}$ for each model plot the ratio with $\eta_J^{\rm truth}$. The number in parentheses is the Jeans test error $\langle |1 - \eta_{\rm J}^{\rm pred}/\eta_{\rm J}^{\rm truth}|\rangle$ for each simulation.}
    \label{fig:Jeans-test}
\end{figure}

\begin{figure}
    \centering
    \includegraphics[width=\linewidth]{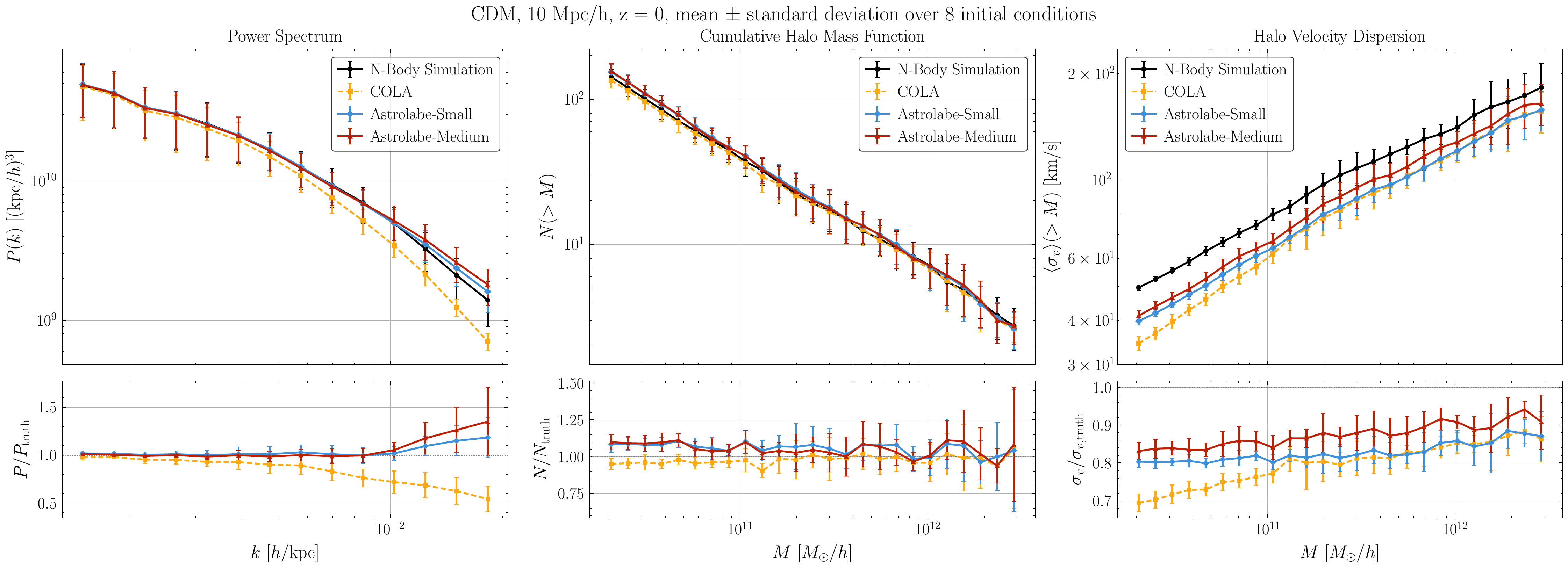}
    \includegraphics[width=\linewidth]{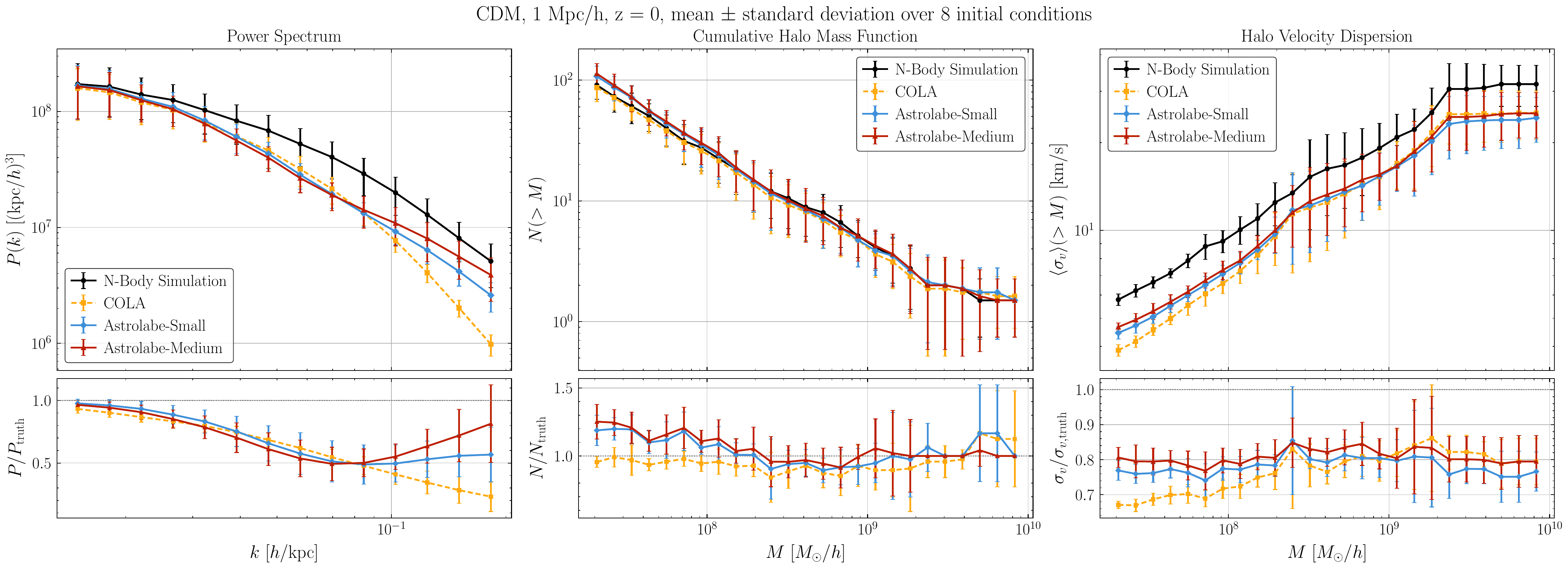}
    \caption{Late-time ($z=0$) summary statistics for a 10 Mpc/$h$ $\Lambda$CDM box (\textbf{top}) or a 1 Mpc/h $\Lambda$CDM box (\textbf{bottom}), with mean and spread computed over eight initial conditions. \textbf{Left:} comoving power spectrum. \textbf{Middle:} cumulative halo mass function. See footnote \ref{ft:halo-finder} for our Halo finder. \textbf{Right:} Halo velocity dispersion as a function of halo mass.}
    \label{fig:z0-curves}
\end{figure}
We further compute an inner slope error, the logarithmic slope of $\rho(r)$ in the central region of the halo well below the virial radius. For each halos with mass greater than the same cutoff used for the concentration error, we fit a power law to the halo mass profile for all radii in $r<0.05 \cdot R_{200}$\footnote{For the smallest halos we consider $r<0.05 \cdot R_{200}$ corresponds to $\sim2-3\times$ the softening length.},
\begin{equation}
    \ln[\frac{M(<r)}{M(<R_{200})}] = a\cdot \ln (\frac{r}{R_{200}}) + b
\end{equation}
and convert the enclosed mass slope $a$ to a density slope via,
\begin{equation}
    \gamma \equiv \frac{d\ln \rho}{d\ln r}\bigg|_{r<0.05\cdot R_{200}} = a -3
\end{equation}
since $M(<r)\propto r^a$ implies $\rho(r)\propto r^{a-3}$. We do this for truth and predicted and we then compute the inner slope error as $\langle |\frac{\gamma_{\rm pred}}{\gamma_{\rm truth}}-1|\rangle$, once again propagating initial condition variance if applicable.

The Jeans consistency metric tests the virialized dynamics of each generated halo. We take the spherical Jeans equation solved for the cumulative mass $M_J(<r)=-\frac{r \sigma_r^2}{G}\left(\frac{d\ln\rho}{d\ln r} + \frac{d\ln\sigma_r^2}{d\ln r}+2\beta\right)$ as in Refs.~\cite{Lokas:2000mu,DynamicsAstrophysicsGalaxies}. We then compute, for both the truth and generated distributions,
\begin{equation}
    \eta_{\rm J}(r) \equiv \frac{M_J(<r)}{M(<r)} = \frac{1}{M(<r)}\bigg[ -\frac{r \sigma_r^2}{G}\left(\frac{d\ln\rho}{d\ln r} + \frac{d\ln\sigma_r^2}{d\ln r} + 2\beta\right)\bigg]
\end{equation}
where $G$ is Newton's constant, $r$ is the radius from the halo center, $M(<r)$ is the measured cumulative mass up to radius $r$, $\rho(r)$ is the density at radius $r$, $\sigma_r^2(r)$ the variance of the radial velocity, and $\beta$ is the anisotropy parameter $1 - \frac{\sigma_t^2}{2\sigma_r^2}$ with $\sigma_t^2$ is the variance of the tangential velocity. We compute $\eta_{\rm J}(r)$ by averaging each term in the equation over the top 33\% of halos by mass. We then compute the Jeans error $\langle |1 - \eta_{\rm J}^{\rm pred}/\eta_{\rm J}^{\rm truth}|\rangle$ as an average over $r/R_{200}$ radial bins. Figure \ref{fig:Jeans-test} shows our Jeans test metric as a function of $r/R_{200}$ in radial bins, which demonstrates significant improvement of \astrolabe{} over COLA for virialized dynamics below the virial radius.

To test consistency with physics across a wide variety of scales, we show in Figure \ref{fig:z0-curves} the power spectrum, halo mass function, and velocity dispersion as a function of halo mass for all eight initial conditions at $z=0$, for both the 10 Mpc/$h$ and 1 Mpc/$h$ boxes. For the halo mass function, and velocity dispersion as a function of halo mass we set our cut off to be halos with $\ge500$ particles found with the same FOF-finder as in footnote \ref{ft:halo-finder}, following \cite{Tinker:2008ff,Warren:2005ey,Watson:2012mt}.

We see from the power spectrum curves that \astrolabe{} successfully recovers much of the small-scale power lost in the COLA approximation, and further performs better at 10 Mpc/$h$ than at 1 Mpc/$h$. This may be due to the increased nonlinear dynamics active in the small box, which makes COLA perform worse, and thus is difficult for \astrolabe{} to correct. We leave investigating and improving this to future work. We also see that \astrolabe{} overproduces small halos, but recovers a more accurate velocity dispersion distribution compared to COLA. 

\begin{figure}
    \centering
    \includegraphics[width=.8\linewidth]{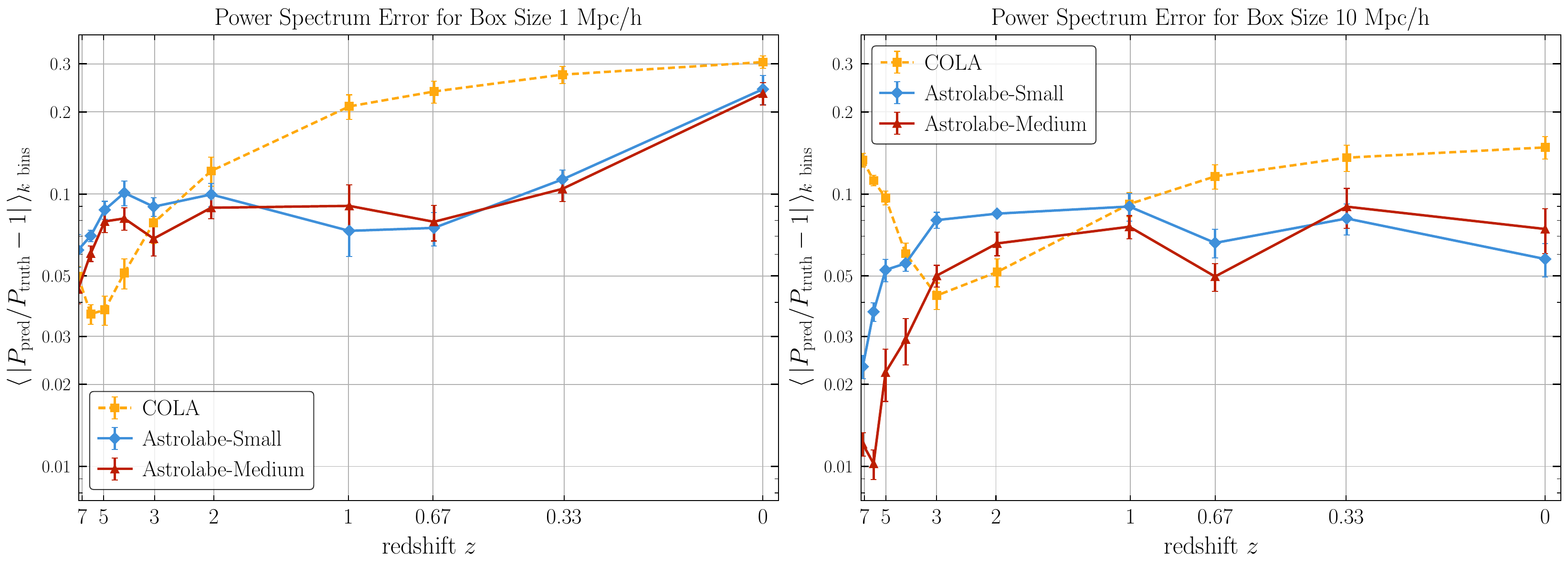}
    \caption{Test set power spectrum error as a function of redshift $z$, for the small-box (\textbf{left}) and large-box (\textbf{right}) simulations. The mean and standard deviation are computed over eight initial conditions.}
    \label{fig:zcurve-err}
\end{figure}

Finally, as with the halo consistency tests, we compute global field errors based on the metrics in Fig.~\ref{fig:z0-curves}. We compute the power spectrum $P(k)$ and velocity dispersion as a function of halo mass $\sigma_v(M)$. We then take the binned curves shown in Figure \ref{fig:z0-curves} and compute $\langle |\frac{P_{\rm pred}(k)}{P_{\rm truth}(k)}|-1\rangle$ and $\langle| \frac{\sigma_v^{\rm pred}(M)}{\sigma_v^{\rm truth}(M)}|-1\rangle$, where the average is taken over all $k$ or mass bins, respectively. After averaging over all eight realizations of the 2M simulation, we compute the sample mean and standard deviation for each redshift $z$, as shown in Fig.~\ref{fig:zcurve-err}. The error metrics for $z = 0$ are given in Tab.~\ref{tab:cdm-error} along with the halo consistency errors. 

These consistency checks make clear the advantages of \astrolabe{} in capturing small-scale structure. The two model sizes differ mainly in the learned velocity distribution and the concentration/inner slope distribution, where \astrolabe{}-medium outperforms the small variant. \astrolabe{} significantly  improves COLA cored-profile, moving toward (but not fully attaining) a cuspy profile. Table~\ref{tab:cdm-error} shows the error metrics for the test set; we give the raw values for all three halo metrics (concentration, inner slope, and Jeans test) in Appendix \ref{app:res}.

\begin{table}[t!]
    \centering
 \scriptsize
\begin{tabular}{llccccc}
\hline
Simulation & Model & Concentration $c$ error & Inner slope $\gamma$ error & $\eta_{\rm J}$ error & $\sigma_v(M)$ error & $P(k)$ error \\
\hline
CDM & COLA & 2.15 $\pm$ 0.01 & 1.46 $\pm$ 0.04 & 0.92 $\pm$ 0.03 & 0.306 $\pm$ 0.005 & 0.305 $\pm$ 0.017 \\
  1 Mpc/$h$ & \astrolabe-Small & 0.67 $\pm$ 0.05 & 0.81 $\pm$ 0.05 & \textbf{0.39 $\pm$ 0.01} & 0.200 $\pm$ 0.005 & 0.242 $\pm$ 0.032 \\
 & \astrolabe-Medium & \textbf{0.59 $\pm$ 0.06} & \textbf{0.58 $\pm$ 0.04} & 0.42 $\pm$ 0.01 & \textbf{0.166 $\pm$ 0.008} & 0.234 $\pm$ 0.023 \\
 & \astrolabe-Small (fast) & 0.60 $\pm$ 0.07 & 0.79 $\pm$ 0.05 & 0.59 $\pm$ 0.01 & 0.250 $\pm$ 0.014 & 0.135 $\pm$ 0.021 \\
 & \astrolabe-Medium (fast) & 0.73 $\pm$ 0.03 & 0.60 $\pm$ 0.02 & 0.61 $\pm$ 0.02 & 0.245 $\pm$ 0.010 & \textbf{0.120 $\pm$ 0.018} \\
\hline
CDM & COLA & 1.80 $\pm$ 0.02 & 1.22 $\pm$ 0.03 & 0.76 $\pm$ 0.02 & 0.270 $\pm$ 0.007 & 0.148 $\pm$ 0.015 \\
  10 Mpc/$h$ & \astrolabe-Small & 0.64 $\pm$ 0.07 & 0.79 $\pm$ 0.02 & 0.40 $\pm$ 0.01 & 0.188 $\pm$ 0.007 & \textbf{0.058 $\pm$ 0.009} \\
 & \astrolabe-Medium & 0.42 $\pm$ 0.02 & 0.69 $\pm$ 0.03 & \textbf{0.34 $\pm$ 0.01} & \textbf{0.150 $\pm$ 0.008} & 0.074 $\pm$ 0.015 \\
 & \astrolabe-Small (fast) & 0.44 $\pm$ 0.04 & 0.81 $\pm$ 0.02 & 0.57 $\pm$ 0.01 & 0.245 $\pm$ 0.007 & 0.153 $\pm$ 0.012 \\
 & \astrolabe-Medium (fast) & \textbf{0.38 $\pm$ 0.02} & \textbf{0.69 $\pm$ 0.04} & 0.56 $\pm$ 0.01 & 0.232 $\pm$ 0.010 & 0.134 $\pm$ 0.020 \\
\hline
\end{tabular}
\caption{\label{tab:cdm-error}Performance metrics for test set at $z=0$ for COLA, \astrolabe-Small and \astrolabe-Medium on our $\Lambda$CDM simulations at both box sizes. We take the mean over eight initial conditions and estimate variance. Bold corresponds to the best performance.}
\end{table}

\subsection{Out-of-distribution generalization: case study with dark acoustic oscillations}
\label{sec:dao-res}

\begin{figure}
    \centering
    \includegraphics[width=.8\linewidth]{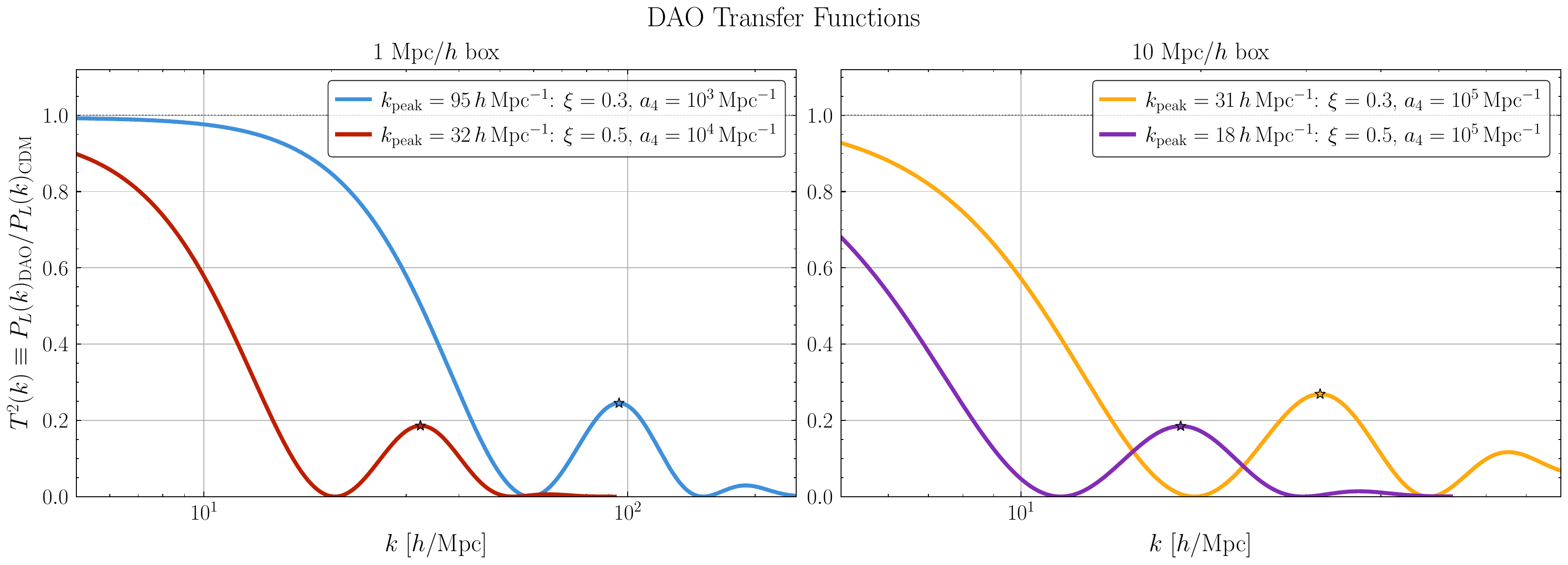}
    \caption{DAO Transfer Functions at $z=99$ for the four configurations we define, two per box size. \textbf{Left:} DAO weak ($k_{\rm peak}=95~h\rm Mpc^{-1}$) and DAO strong ($k_{\rm peak}=32~h\rm Mpc^{-1}$) configuration for the 1 Mpc/$h$ box. \textbf{Right:} DAO weak ($k_{\rm peak}=31~h\rm Mpc^{-1}$) and DAO strong ($k_{\rm peak}=18~h\rm Mpc^{-1}$) configuration for the 10 Mpc/$h$ box.}
    \label{fig:dao-configs}
\end{figure}

In this section, we test \astrolabe's ability to emulate non-$\Lambda$CDM initial conditions, despite the fact that such scenarios are out of the training distribution. If this generalization is reliable, this would be valuable for the utilization of \astrolabe{}, without fine-tuning, for scans over the subset of dark matter theory space that minimally affect cosmological evolution beyond very early redshift. It would also support the notion that \astrolabe{} learns small-scale gravitational dynamics of DM evolution regardless of the specific initial conditions.
\begin{figure}[t]
    \centering
    \includegraphics[width=.9\linewidth]{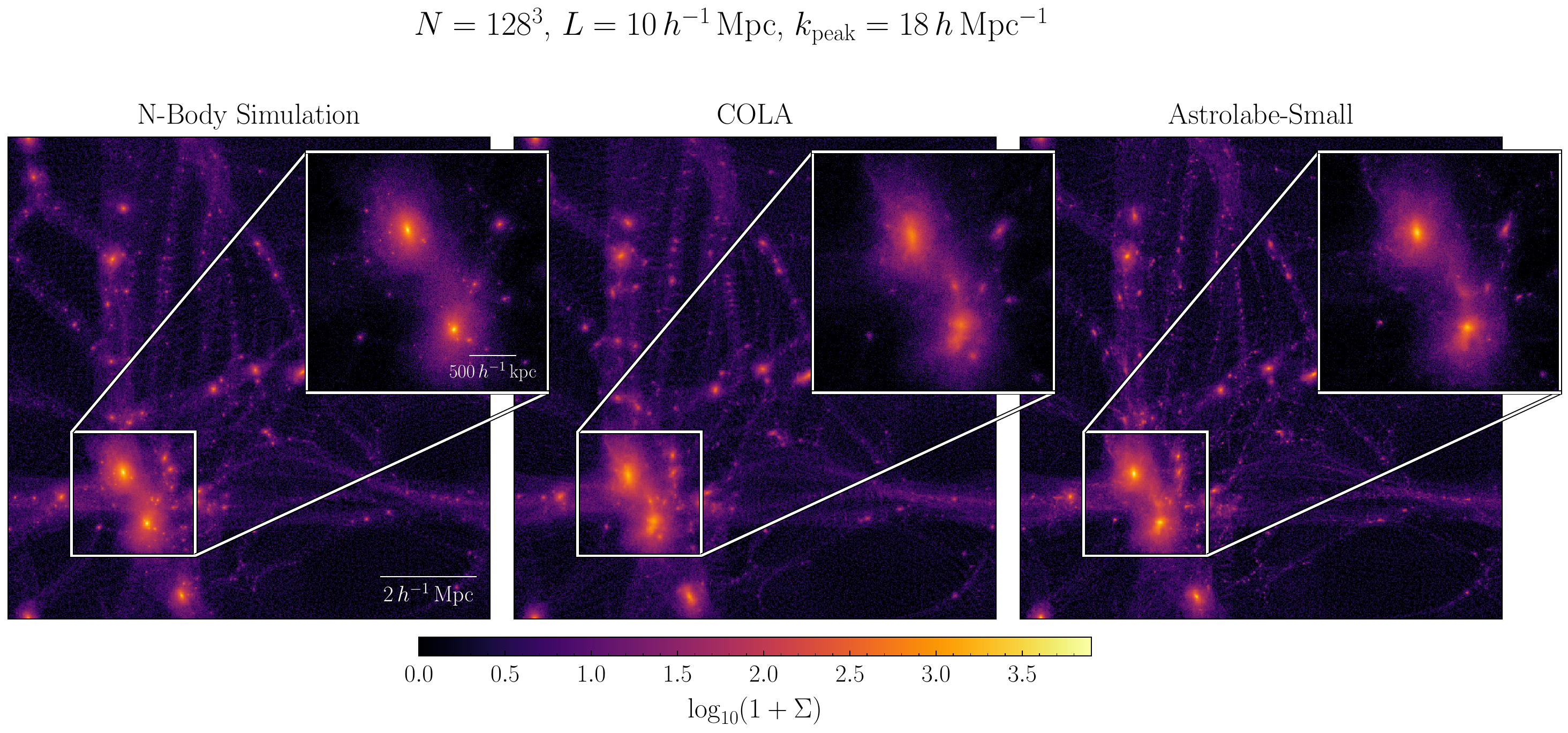}
    \caption{Example DAO initial conditions ($k_{\rm peak} = 18 ~h  \rm Mpc^{-1}$) for $128^3\approx2\rm M$ particles in a 10 Mpc/$h$ box. \textbf{Left:} A \textsc{GIZMO} \cite{Hopkins:2014qka} \Nbody{} simulation with DAO initial conditions. \textbf{Middle:} COLA with a mesh of $256^3$ and 40 timesteps. \textbf{Right:} \astrolabe-Small evaluated with 10 ODE steps. As in Fig.~\ref{fig:2M-field}, we plot density on a log scale post-rasterization and zoom into the largest halo. As in Fig.~\ref{fig:2M-field}, in each plot, we show surface density $\Sigma$ in units of particles per pixel on a log scale post-rasterization and zoom into the largest halo.}
    \label{fig:2M-field-10MPC-DAO}
\end{figure}
To test out-of-distribution generalization, we evaluate the model on initial conditions perturbed by dark acoustic oscillations (DAO) from a dark-radiation bath interacting with the dark candidate. We use the ETHOS parametrization of such a scenario fed to \textsc{CLASS}. We vary the ratio of the dark radiation temperature to the CMB temperature $\xi\equiv\frac{T_{DR}}{T_{CMB}}$ and the leading coefficient on the momentum transfer rate of dark-matter-dark-radiation interactions $a_{4}$ to build four DAO variants, a strong and weak version for the two different box sizes \cite{Cyr-Racine:2015ihg,Blas:2011rf,Archidiacono:2019wdp}, see Figure \ref{fig:dao-configs}. We simulate with \textsc{GIZMO} \cite{Hopkins:2014qka} three different initial conditions for each of the four configurations with 2M particles and evaluate COLA and both model sizes of \astrolabe{}.

\begin{figure}
    \centering
    \includegraphics[width=.8\linewidth]{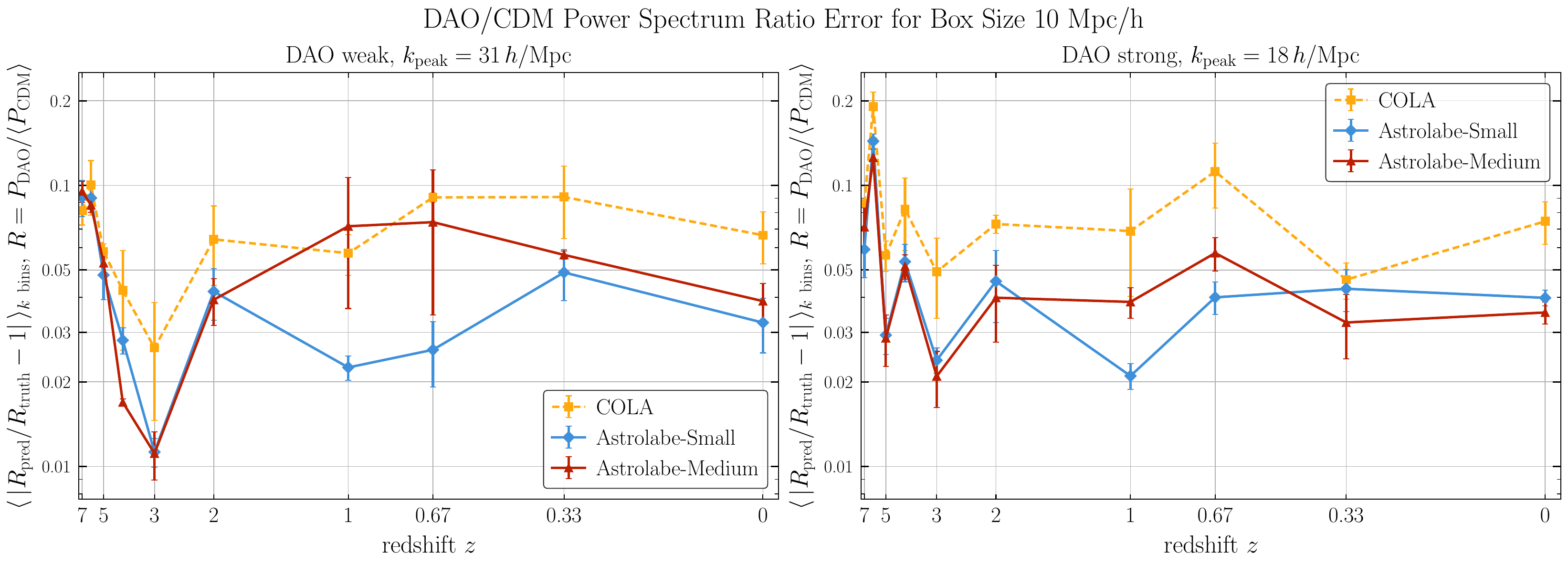}
    \caption{Ratio of DAO to $\Lambda$CDM Power Spectrum Error as compared to \textsc{GIZMO} \cite{Hopkins:2014qka} as a function of redshift $z$, with mean and standard deviation computed over three DAO initial conditions for two DAO configurations in a 10 Mpc/$h$ box. Smaller is better.}
    \label{fig:zcurve-ps-err-dao}
\end{figure}

We plot example fields in Fig.~\ref{fig:2M-field-10MPC-DAO} (analogous to Fig.~\ref{fig:2M-field} for $\Lambda$CDM) and give summarized metrics in Table \ref{tab:dao-error}, analogous to Tab.~\ref{tab:cdm-error}. In Fig.~ \ref{fig:zcurve-ps-err-dao}, we also plot the error in the ratio of DAO to $\Lambda$CDM power spectrum (i.e.\ the transfer function) as compared to \textsc{GIZMO} \cite{Hopkins:2014qka} as a function of redshift $z$. With the exception of the velocity distribution in the 1 Mpc/h box, it is clear \astrolabe{} provides a significant improvement to COLA, just as it did for $\Lambda$CDM. The weakness relative to COLA of the velocities for DAO only for the 1 Mpc/$h$ box is an interesting observation and the only obvious indication of an out-of-distribution failure.

\begin{table}
    \centering
\scriptsize
\begin{tabular}{llccccc}
\hline
Simulation & Model & Concentration $c$ error & Inner slope $\gamma$ error & $\eta_{\rm J}$ error & $\sigma_v(M)$ error & $P(k)$ error \\
\hline
DAO-strong & COLA & 1.44 $\pm$ 0.08 & 1.08 $\pm$ 0.03 & 0.66 $\pm$ 0.05 & 0.162 $\pm$ 0.008 & 0.161 $\pm$ 0.069 \\
1 Mpc/$h$ & \astrolabe-Small & \textbf{0.57 $\pm$ 0.17} & 0.60 $\pm$ 0.07 & 0.33 $\pm$ 0.03 & \textbf{0.107 $\pm$ 0.008} & \textbf{0.156 $\pm$ 0.053} \\
($k_{\rm peak}=32$) & \astrolabe-Medium & 0.74 $\pm$ 0.04 & 0.64 $\pm$ 0.02 & \textbf{0.27 $\pm$ 0.04} & 0.125 $\pm$ 0.009 & 0.166 $\pm$ 0.048 \\
 & \astrolabe-Small (fast) & 0.75 $\pm$ 0.01 & 0.52 $\pm$ 0.08 & 0.47 $\pm$ 0.04 & 0.131 $\pm$ 0.016 & 0.222 $\pm$ 0.031 \\
 & \astrolabe-Medium (fast) & 0.71 $\pm$ 0.08 & \textbf{0.50 $\pm$ 0.10} & 0.42 $\pm$ 0.08 & 0.154 $\pm$ 0.013 & 0.184 $\pm$ 0.040 \\
\hline
DAO-weak & COLA & 1.81 $\pm$ 0.03 & 1.06 $\pm$ 0.03 & 0.91 $\pm$ 0.06 & 0.242 $\pm$ 0.021 & 0.299 $\pm$ 0.040 \\
1 Mpc/$h$ & \astrolabe-Small & 0.55 $\pm$ 0.02 & 0.66 $\pm$ 0.07 & 0.35 $\pm$ 0.02 & 0.146 $\pm$ 0.014 & 0.201 $\pm$ 0.036 \\
($k_{\rm peak}=95$) & \astrolabe-Medium & \textbf{0.42 $\pm$ 0.09} & \textbf{0.60 $\pm$ 0.08} & \textbf{0.27 $\pm$ 0.02} & \textbf{0.111 $\pm$ 0.011} & 0.226 $\pm$ 0.038 \\
 & \astrolabe-Small (fast) & 0.51 $\pm$ 0.05 & 0.71 $\pm$ 0.05 & 0.57 $\pm$ 0.02 & 0.186 $\pm$ 0.005 & \textbf{0.132 $\pm$ 0.007} \\
 & \astrolabe-Medium (fast) & 0.59 $\pm$ 0.07 & 0.67 $\pm$ 0.17 & 0.57 $\pm$ 0.02 & 0.198 $\pm$ 0.015 & 0.158 $\pm$ 0.021 \\
\hline
DAO-strong & COLA & 1.67 $\pm$ 0.04 & 1.10 $\pm$ 0.04 & 0.74 $\pm$ 0.03 & 0.249 $\pm$ 0.008 & 0.156 $\pm$ 0.034 \\
10 Mpc/$h$ & \astrolabe-Small & 0.47 $\pm$ 0.05 & 0.72 $\pm$ 0.01 & 0.42 $\pm$ 0.02 & 0.163 $\pm$ 0.008 & \textbf{0.035 $\pm$ 0.012} \\
($k_{\rm peak}=18$) & \astrolabe-Medium & 0.41 $\pm$ 0.02 & \textbf{0.54 $\pm$ 0.02} & \textbf{0.34 $\pm$ 0.02} & \textbf{0.112 $\pm$ 0.011} & 0.054 $\pm$ 0.013 \\
 & \astrolabe-Small (fast) & 0.49 $\pm$ 0.03 & 0.68 $\pm$ 0.06 & 0.58 $\pm$ 0.01 & 0.229 $\pm$ 0.011 & 0.125 $\pm$ 0.032 \\
 & \astrolabe-Medium (fast) & \textbf{0.39 $\pm$ 0.05} & 0.55 $\pm$ 0.00 & 0.57 $\pm$ 0.01 & 0.221 $\pm$ 0.011 & 0.164 $\pm$ 0.029 \\
\hline
DAO-weak & COLA & 1.76 $\pm$ 0.00 & 1.12 $\pm$ 0.09 & 0.75 $\pm$ 0.02 & 0.283 $\pm$ 0.004 & 0.183 $\pm$ 0.027 \\
10 Mpc/$h$ & \astrolabe-Small & 0.55 $\pm$ 0.05 & 0.80 $\pm$ 0.06 & 0.40 $\pm$ 0.02 & 0.179 $\pm$ 0.005 & 0.035 $\pm$ 0.004 \\
($k_{\rm peak}=31$) & \astrolabe-Medium & 0.42 $\pm$ 0.05 & \textbf{0.59 $\pm$ 0.04} & \textbf{0.32 $\pm$ 0.02} & \textbf{0.157 $\pm$ 0.007} & \textbf{0.031 $\pm$ 0.010} \\
 & \astrolabe-Small (fast) & 0.50 $\pm$ 0.05 & 0.78 $\pm$ 0.06 & 0.58 $\pm$ 0.01 & 0.255 $\pm$ 0.009 & 0.120 $\pm$ 0.027 \\
 & \astrolabe-Medium (fast) & 0.39 $\pm$ 0.03 & 0.64 $\pm$ 0.06 & 0.58 $\pm$ 0.01 & 0.241 $\pm$ 0.009 & 0.104 $\pm$ 0.026 \\
\hline
\end{tabular}

\caption{\label{tab:dao-error} Error on metrics at $z=0$ for COLA, \astrolabe-Small and \astrolabe-Medium on our four DAO simulation configurations, out of the training ($\Lambda$CDM) distribution. The mean and variance are computed over three initial conditions. Bold corresponds to the best performance.}
\end{table}
Figure~\ref{fig:zcurve-ps-err-dao} shows significant improvement in our ability to capture the evolution of the ratio of DAO to $\Lambda$CDM power spectrum throughout the redshift range as compared to COLA. Furthermore, for small-scale halo-observables (concentration, inner slope, and the Jeans test) the improvement is even larger, as for $\Lambda$CDM. Overall, we find this to be compelling evidence that such ML emulators, when anchored to initial conditions and strong physical principles as encoded by COLA evolution, can reasonably generalize to out-of-distribution dark matter models that controllably perturb initial conditions. To emphasize the point, one model trained only on $\Lambda$CDM learns to produce correct $z=0$ fields for completely different initial conditions driven by different dark sector physics that produces a very different (so much so as to be strongly excluded by observations) universe at $z=0$ with \emph{no} extra training. This suggests that \astrolabe{} is learning generalizable features about the physics of gravitational evolution of dark matter, and inspires some confidence that such an emulator may indeed be useful for scanning through dark matter theory space without the need for fine-tuning.

\subsection{Fine-tuning for Larger Particle Count}
\label{sec:larger-part-res}
\begin{figure}
    \centering
    \includegraphics[width=.9\linewidth]{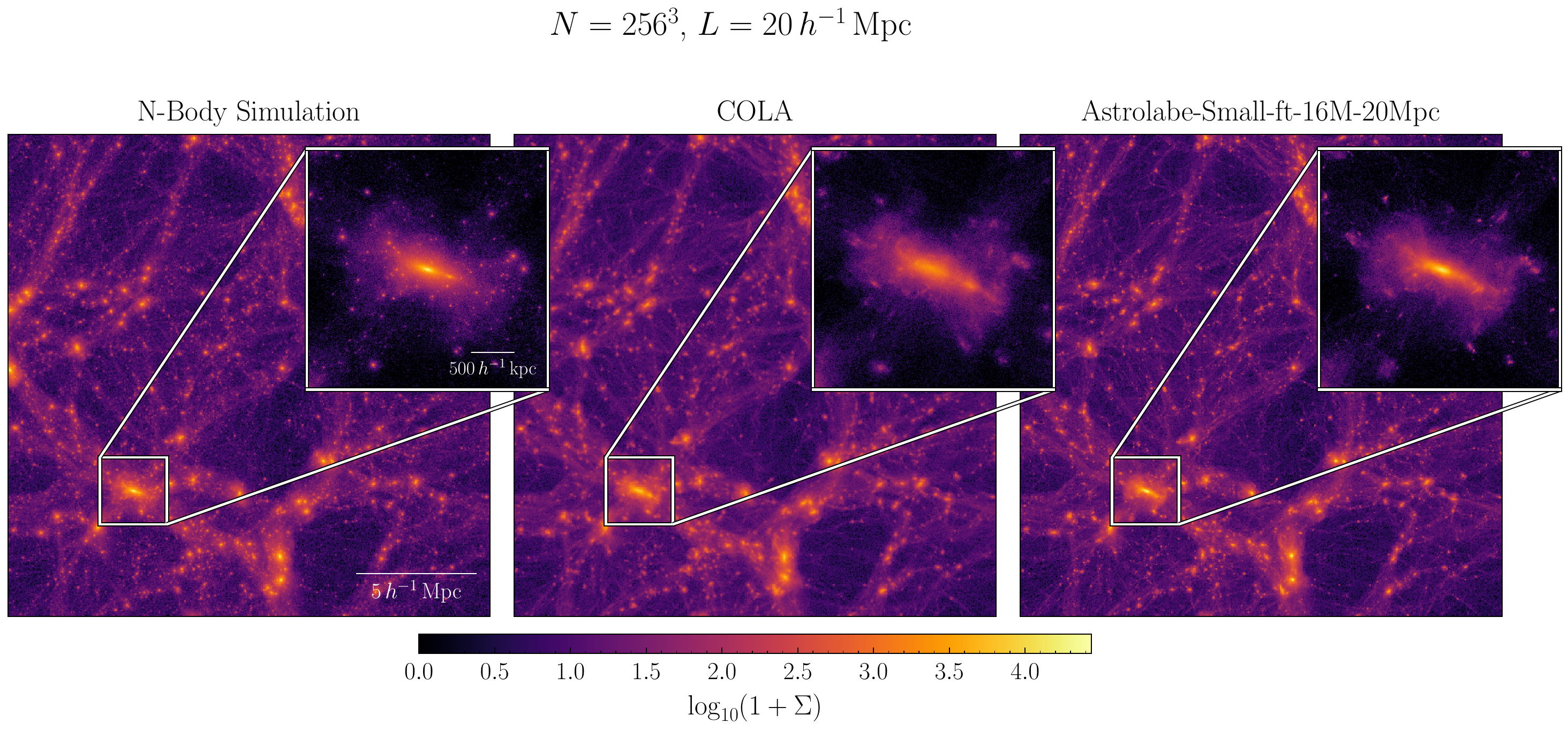}
        \caption{Example $\Lambda$CDM Field for $256^3\approx16\rm M$ particles in a 20 Mpc/h box. \textbf{Left:} A \textsc{GIZMO} \cite{Hopkins:2014qka} \Nbody{} simulation with 16M particles. \textbf{Middle:} COLA with a mesh of $256^3$ and 40 timesteps. \textbf{Right:} The fine-tuned model, \astrolabe-Small-ft-16M, evaluated with 10 ODE steps. As in Fig.~\ref{fig:2M-field}, in each plot, we show surface density $\Sigma$ in units of particles per pixel on a log scale post-rasterization and zoom into the largest halo.}
    \label{fig:16M-field-1MPC}
\end{figure}
\begin{figure}
    \centering
    \includegraphics[width=.72\linewidth]{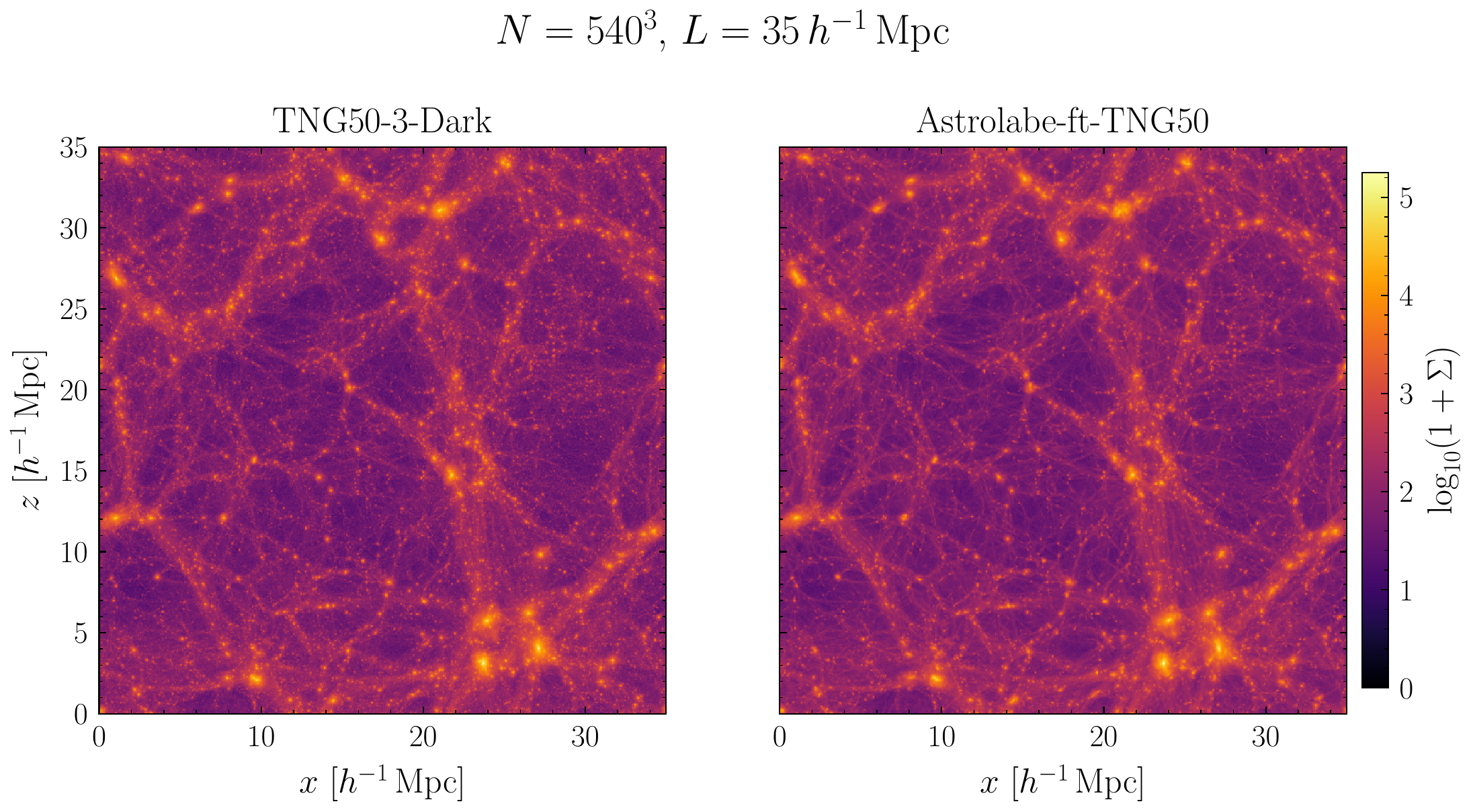}
    \includegraphics[width=.72\linewidth]{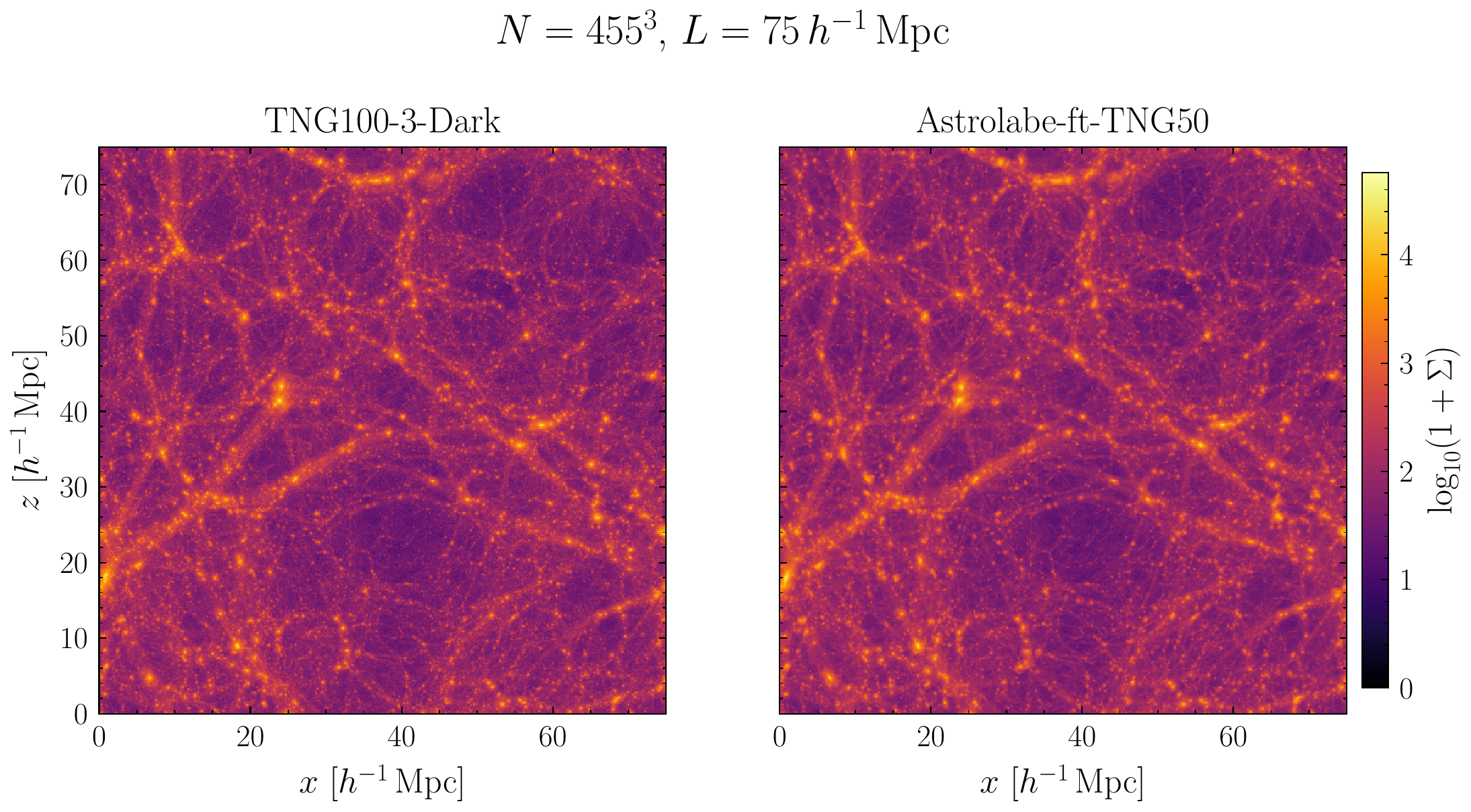}
    \caption{\textbf{Left:} Truth-level simulations at $z = 0$: \texttt{TNG50-3-Dark} (\textbf{top}) and \texttt{TNG100-3-Dark} (\textbf{bottom}). \textbf{Right:} \astrolabe-Small fine-tuned on redshift snapshots of \texttt{TNG50-3-Dark} evaluated with 10 ODE steps from COLA initial conditions with a mesh of $256^3$ and 40 timesteps. As in Fig.~\ref{fig:2M-field}, in each plot, we show surface density $\Sigma$ in units of particles per pixel on a log scale post-rasterization.}
    \label{fig:tng-field}
\end{figure}
Another important direction of generalization to evaluate is larger particle counts and different box sizes. Our hybrid architecture allows us to generate samples from the model at larger particle counts than encountered during training, but improved performance may be obtained by fine-tuning on dedicated large-particle-number simulations, without having to retrain the model from scratch. We focus only on \astrolabe-Small and $\Lambda$CDM initial conditions for this exploration and adopt an early redshift fine-tuning strategy. Specifically, we take our \astrolabe-Small pretrained on 2M samples, and fine-tune on early-redshift snapshots of one simulation with a higher particle count, performing our evaluation at $z=0$ by comparing with a separate simulation with a similar particle count. Our fine-tuning proceeds similar to Ref. \cite{Elsharkawy:2026mev} on 4x4 A100 nodes with the same training hyperparameters as before, starting from the pretrained weights and fine-tuned for 5,000 gradient descent updates with early stopping. For all comparisons, we compare the performance of COLA alone, \astrolabe-Small with no fine-tuning evaluated on the out-of-distribution particle number, and \astrolabe fine-tuned on the larger particle number. 
\begin{figure}
    \centering
    \includegraphics[width=\linewidth]{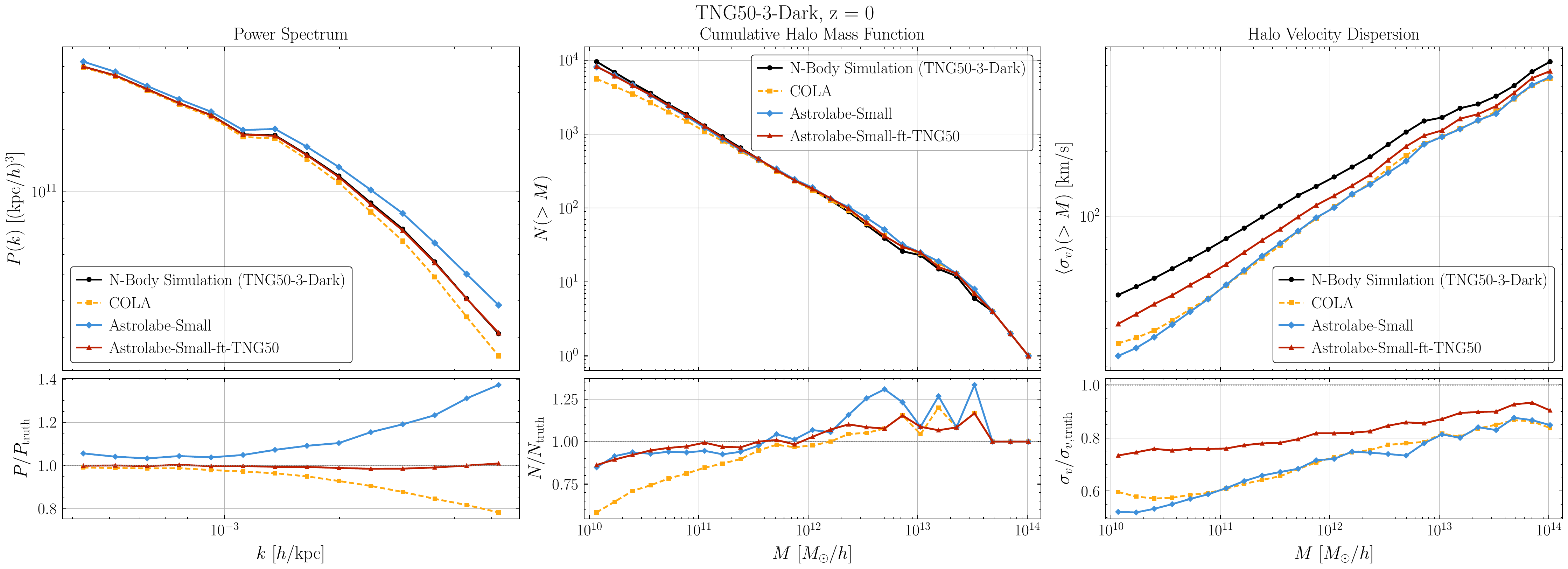}
    \caption{Late-time ($z=0$) summary statistics for \texttt{TNG50-3-Dark}. Mean and spread are computed over halo bins. \textbf{Left:} power spectrum. \textbf{Middle:} cumulative halo mass function. \textbf{Right:} halo velocity dispersion as a function of halo mass.}
    \label{fig:z0-curves-TNG50}
\end{figure}

\begin{table}[t!]
   \centering 
   \scriptsize
\begin{tabular}{llccccc}
\hline
Simulation & Model & Concentration $c$ error & Inner slope $\gamma$ error & $\eta_{\rm J}$ error & $\sigma_v(M)$ error & $P(k)$ error \\
\hline
16M, 1 Mpc/$h$ & COLA & 2.11 & 1.46 & 0.88 & 0.318 & 0.251 \\
(training sim & \astrolabe-Small & 1.33 & 1.08 & 0.49 & 0.223 & 0.242 \\
heldout $z=0$) & \astrolabe-Small-ft-16M & 1.07 & 1.21 & \textbf{0.36} & \textbf{0.220} & \textbf{0.069} \\
 & \astrolabe-Small-ft-16M (fast) & \textbf{0.71} & \textbf{1.05} & 0.56 & 0.240 & 0.100 \\
\hline
16M, 1 Mpc/$h$ & COLA & 2.16 & 1.36 & 1.11 & 0.350 & 0.232 \\
(eval sim) & \astrolabe-Small & \textbf{1.19} & \textbf{1.17} & 0.50 & 0.268 & 0.204 \\
 & \astrolabe-Small-ft-16M & 1.73 & 1.38 & \textbf{0.22} & \textbf{0.265} & 0.164 \\
 & \astrolabe-Small-ft-16M (fast) & 1.33 & 1.32 & 0.50 & 0.281 & \textbf{0.126} \\
\hline
16M, 20 Mpc/$h$ & COLA & 1.73 & 1.17 & 0.73 & 0.330 & 0.088 \\
(training sim & \astrolabe-Small & 0.73 & 1.05 & 0.58 & 0.321 & 0.127 \\
heldout $z=0$) & \astrolabe-Small-ft-16M-20Mpc & 0.46 & 0.87 & \textbf{0.44} & \textbf{0.232} & \textbf{0.035} \\
 & \astrolabe-Small-ft-16M-20Mpc (fast) & \textbf{0.37} & \textbf{0.82} & 0.55 & 0.286 & 0.073 \\
\hline
16M, 20 Mpc/$h$ & COLA & 1.70 & 1.01 & 0.81 & 0.272 & 0.165 \\
(eval sim) & \astrolabe-Small & 1.24 & 1.17 & 0.72 & 0.258 & 0.043 \\
 & \astrolabe-Small-ft-16M-20Mpc & 0.58 & \textbf{0.81} & \textbf{0.50} & \textbf{0.213} & \textbf{0.033} \\
 & \astrolabe-Small-ft-16M-20Mpc (fast) & \textbf{0.44} & 0.81 & 0.57 & 0.232 & 0.092 \\
\hline
\end{tabular}

\caption{\label{tab:16m-error}Error on metrics for COLA, \astrolabe-Small and the fine-tuned model \astrolabe-Small-ft-16M for both the training simulation (evaluated on the held-out $z=0$ snapshot) and the held-out evaluation simulation. The $c$ error is the mean over halos with $\geq 50{,}000$ particles on the same halos for all models. Bold corresponds to the best performance.}
\end{table}

\begin{table}
\centering
\scriptsize
\begin{tabular}{llccccc}
\hline
Simulation & Model & Concentration $c$ error & Inner slope $\gamma$ error & $\eta_{\rm J}$ error & $\sigma_v(M)$ error & $P(k)$ error \\
\hline
TNG50-3-Dark & COLA & 2.42 & \textbf{1.09} & 0.73 & 0.354 & 0.059 \\
 & \astrolabe-Small & 2.75 & 1.42 & 0.43 & 0.363 & 0.116 \\
 & \astrolabe-Small-ft-TNG50 & 1.74 & \textbf{1.24} & \textbf{0.36} & \textbf{0.218} & \textbf{0.005} \\
 & \astrolabe-Small-ft-TNG50 (fast) & \textbf{1.22} & 1.29 & 0.56 & 0.261 & 0.025 \\
\hline
TNG100-3-Dark & COLA & 2.26 & \textbf{1.01} & 0.65 & 0.369 & 0.041 \\
 & \astrolabe-Small & 2.38 & 1.36 & 0.71 & 0.398 & 0.058 \\
 & \astrolabe-Small-ft-TNG50 & 1.81 & 1.24 & \textbf{0.42} & \textbf{0.242} & \textbf{0.008} \\
 & \astrolabe-Small-ft-TNG50 (fast) & \textbf{1.38} & \textbf{1.19} & 0.57 & 0.284 & 0.024 \\
\hline
\end{tabular}

    \caption{\label{tab:tng-error}Error on metrics for COLA, \astrolabe-Small, \astrolabe-Small-ft-TNG (\astrolabe{}-Small fine-tuned on \texttt{TNG50}) on \texttt{TNG50-3-Dark} and \texttt{TNG100-3-Dark} at $z=0$. The $c$ error is the mean over halos with $\geq 50{,}000$ particles on the same halos for all models. Bold corresponds to the best performance.}
\end{table}
We first take as our ground truth a \textsc{GIZMO} 16M particle simulation with box size 1 Mpc/$h$ and a box size of 20 Mpc/$h$. The 1 Mpc/$h$ box tests performance under increased resolution for the same box size, while the 20 Mpc/$h$ box maintains the same interparticle spacing as in 2M 10Mpc/$h$ and thus tests generalization to different box sizes. We fine-tune on many snapshots with redshifts in $z\in[40,0.11]$. We show an example field in Figure \ref{fig:16M-field-1MPC} and summarize our error metrics in Table \ref{tab:16m-error}. 

We see that fine-tuning is very effective in improving \astrolabe's performance for both box sizes for the held-out $z=0$ snapshot with the same initial conditions. For the 20 Mpc/$h$ box, fine-tuning also leads to significant improvement on unseen 16M initial conditions, thus implying generalization. On the other hand, for the 1 Mpc/$h$ box, fine-tuning only marginally benefits \astrolabe's performance on the held-out initial condition. The difference between the two box sizes hints that fine-tuning is capable of adapting to different box sizes, while shrinking interparticle spacing (that is, increasing resolution) may require a full re-training. This out-of-distribution failure mode is consistent with the model learning local physics. Due to computational expense, we only test on one held-out 16M particle sim, so our conclusions are necessarily limited, but we intend to further explore this direction in future work.


\label{sec:tng-res}


 We follow this with a test on \texttt{TNG50-3-dark}, an \Nbody{} simulation containing $540^3 \approx 150$M particles in a 35 Mpc/$h$ box, simulated with different software and different boundary conditions. We fine-tune on a coarser grid of redshifts given by $z = 2.0, 1.0, 0.5, 0.24$ and we evaluate on a held-out snapshot of \texttt{TNG50-3-dark} at $z=0$. We also evaluate on \texttt{TNG100-3-dark} at $z=0$, a 75 Mpc/$h$ box with fewer particles, $N = 455^3 \approx 100$M, which we use as a proxy for a held-out evaluation simulation despite the mismatch in particle number.

We plot example fields in Figure \ref{fig:tng-field} and display metrics in Table \ref{tab:tng-error}. We further plot $z=0$ curves for power spectrum, halo mass function, and velocity dispersion as a function of halo mass in Figure \ref{fig:z0-curves-TNG50}. It is clear that fine-tuning is both beneficial and necessary. \astrolabe{} does worse than COLA without fine-tuning, but with fine-tuning the model improves significantly on COLA, even on the technically out-of-distribution \texttt{TNG-100-dark-3}. We note that to emulate the \texttt{TNG50-3-dark} fields, we had to evaluate \astrolabe{} on 150M individual particles using 10 Euler steps to integrate the flow matching ODE. The fact that this is feasible given our computing resources highlights \astrolabe{}'s scalability to large particle numbers.

\subsection{Comparison of compute costs}
\label{sec:compute-res}

\begin{figure}
    \centering
    \includegraphics[width=\linewidth]{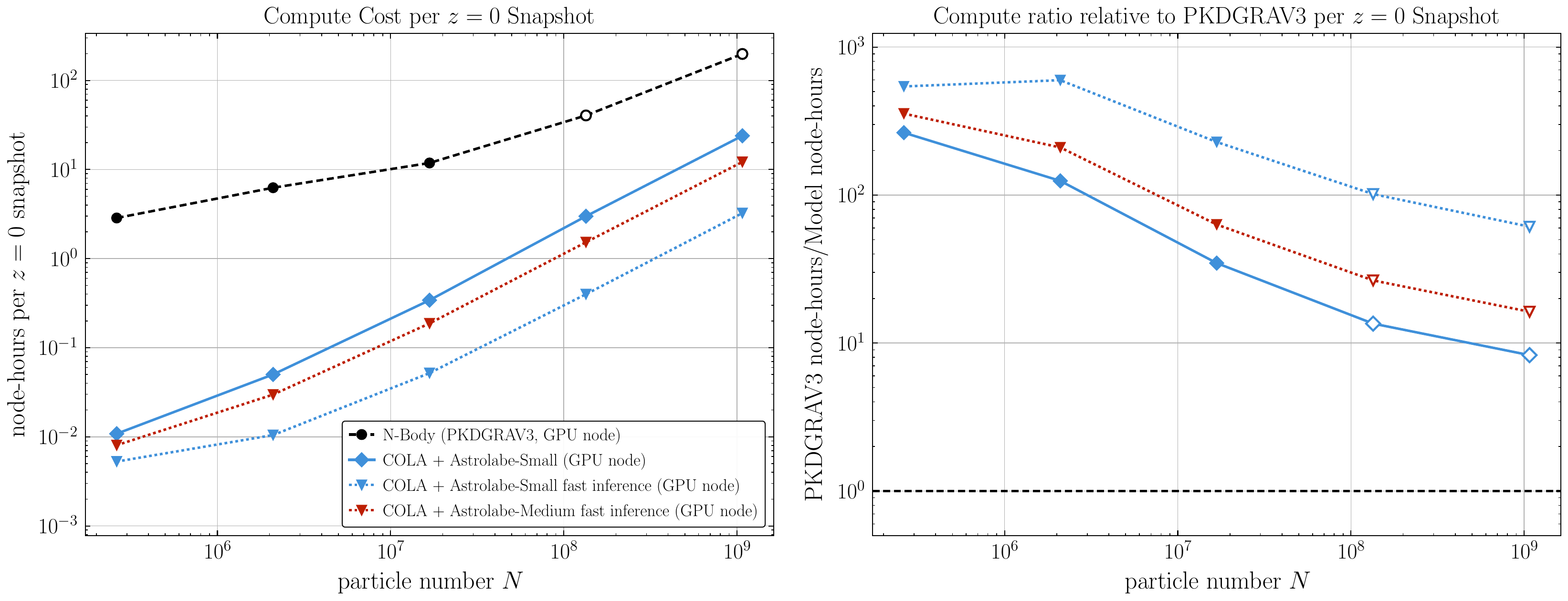}
    \caption{\textbf{Left:} Node-hours required for generating a $N$-particle snapshot at $z=0$. \textbf{Right:} Compute multiplier for \textsc{PKDGRAV3} compared to COLA+\astrolabe{} for generating $N$ particle $z=0$ snapshot. COLA+\astrolabe{} runs on 4xA100 40GB GPU nodes. \astrolabe-Small fast inference uses 5 Euler steps with a subsampling factor of 4. See text for how we approximate \textsc{PKDGRAV3} timing for large particle counts.}
    \label{fig:Comput-Cost-node-Hours}
\end{figure}

Here we study the compute savings which generative emulators could achieve compared to GPU-based \textsc{PKDGRAV3} \cite{Potter:2016ttn} \Nbody{} simulations. In Figure \ref{fig:Comput-Cost-node-Hours} we plot, as a function of particle number $N$, the number of node-hours needed to generate a $z=0$ snapshot of a given simulation. We keep simulation resolution the same as $128^3\approx2$M for $1\rm Mpc$ and increase box size accordingly as $N$ increases. Thus, we probe the same level of non-linear physics for each $N$ and do not penalize \textsc{PKDGRAV3} for the extra timestep it would take for a larger $N$ at the same box size.\footnote{Specifically, we test $N\in[\sim262k,\sim2M,\sim16M,\sim134M,\sim1B]$ corresponding to box sizes of $L \in [0.5, 1, 2, 4, 8]\ h^{-1} ~\mathrm{Mpc}$.}

We also plot the \astrolabe{} ratio of compute as compared to \textsc{PKDGRAV3} as a function of particle number $N$.  We plot this curve for \astrolabe{}, and for \textsc{PKDGRAV3}, timed on 4x4xA100 40GB GPU nodes. Open markers represent approximate extrapolations for \textsc{PKDGRAV3} so that the simulation is computationally tractable. We use
\begin{equation}
    T_{\rm total} = \sum_{k=1}^{N_{\rm step}} t_k + A \sum_{k=N_{\rm step}+1}^{100} p_k + T_{\rm setup}
\end{equation}
where we time the cost of $N_{\rm step}=2$ steps, $p_k$ is the time per step of the 2M run at the same resolution used as reference, and $A$ is the approximate scaling computed by comparing the 2M run per time step to the steps we compute past the first step given by $A = \frac{\sum_{k=2}^{n} t_k}{\sum_{k=2}^{n} p_k}$. Finally, $T_{\rm setup}$ is the measured initial cost time before the simulation starts.

In addition, we plot the fast-inference mode for \astrolabe{} based on the ablation studied in Appendix \ref{app:inference-ablation}. To speed up inference, we take fewer Euler steps during the flow matching integration and further subsample particles, evaluating only a subset and copying the flow velocity to nearest neighbors. As can be seen in Figure \ref{fig:Comput-Cost-node-Hours} this significantly reduces compute costs while still producing realistic fields, an example of which can be seen in Figure \ref{fig:fast-inference}. We chose 5 Euler steps and a subsampling factor of 4 based on the ablation.

\astrolabe{}-small achieves a $\sim10-250\times$ speedup over \textsc{PKDGRAV3} depending on particle number, while \astrolabe{}-small-fast-inference achieves $\sim60-500\times$ speedup. In other words, with \astrolabe{}-small-fast-inference, we may be able to perform 200 times the number of billion-particle simulations as one could with \textsc{PKDGRAV3} on the same compute budget. We note that \textsc{PKDGRAV3} does not saturate all $16$ GPUs at lower particle numbers, and thus becomes more efficient relative to \astrolabe{} as particle number increases, as we can see in the right panel of Figure \ref{fig:Comput-Cost-node-Hours}. We expect this effect to plateau as $N$ grows larger.

\section{Conclusion and Outlook}
\label{sec:conclusion}

In this paper we have introduced \astrolabe{}, a fast emulator of \Nbody{} simulations that acts on top of COmoving Lagrangian Acceleration (COLA), built with the purpose of enabling large scans through dark matter theory space and other cosmological parameters that is computationally intractable for traditional \Nbody{} simulations. \astrolabe{} acts directly on the 6-dimensional phase space of \Nbody{} simulation particles instead of voxel summaries, allowing for a better inductive bias, and the removal of any voxel resolution/dynamic range limitations. This is made tractable with a field-particle hybrid architecture, where each particle attends to voxel summaries of its surroundings. This approach further allows for simple scaling of training and inference to large $N$, which we demonstrate by emulating the 150M particle \texttt{TNG50-3-dark} simulation \cite{Nelson:2018uso}.

Our use of physical flow matching for \astrolabe{} training and inference imposes the important physical priors of physical initial conditions and conservation laws directly on the model architecture. We find that \astrolabe{} is capable of producing realistic fields and halos, with accurate power spectra, velocity dispersions, and halo mass profiles at many redshifts, box sizes, and particle numbers. Moreover, \astrolabe{} takes a fraction of the compute cost of traditional \Nbody{} simulations; we find an compute advantage of one or more orders of magnitude over \textsc{PKDGRAV3} depending on particle number and \astrolabe{} inference mode. We have also demonstrated the ability of \astrolabe{}, following fine-tuning, to generalize across box sizes and boundary conditions not seen in training. 

We plan to further explore the exciting possibilities enabled by \astrolabe{} in future work. A key direction for future exploration is to train an \astrolabe{}-style emulator for baryonic physics, using physical flow matching to correlate the baryon field to the dark matter field at each redshift. If such emulation is achieved at the same or similar cost to \astrolabe{} for DMO simulations, the compute savings of the emulator compared to the full $N$-body plus hydrodynamical simulations will only grow. We hope that progress of this type could eventually improve our ability to robustly understand the effects of different dark matter models at large and small scales. The architecture we introduce may enable large scans through dark matter theory space that are currently computationally intractable for traditional \Nbody{} simulations, and thus lead to a better understanding of how rich dark sector physics could affect the structure of our universe.

\section*{Interactive Visualization, Code, and Data Availability}

The code and dataset used for this study are to be released soon. A blog post on the method with 3-dimensional interactive visualizations and films of the \astrolabe{} and \Nbody{} fields can be found in \url{https://ibrahimelsharkawy.com/blog/astrolabe-flow-matching.html}.

\section*{Acknowledgments}

We thank Yann LeCun for an in-depth discussion on world models which inspired our approach, and Norman Murray and Ting Li for helpful discussions on relevant astrophysical observables for validating our approach. We are especially grateful to Sandip Roy for helpful comments on a draft version of this manuscript and especially grateful to Wahid Bhimji for an abundance of support including computing infrastructure. I.E. thanks Vinicius Mikuni, Joschka Birk, and Benjamin Nachman for helpful insight into Point-Edge Transformers. This research used the resources of the National Energy Research Scientific Computing Center (NERSC), a U.S. Department of Energy Office of Science User Facility located at Lawrence Berkeley National Laboratory, using NERSC award \#ERCAP0041413.
I.E is supported in part by the Connaught International Scholarship at the University of Toronto and the Natural Sciences and Engineering Research Council of Canada (NSERC) Canada Graduate Research Scholarship. Y.K. and I.E. acknowledge the support of an NSERC Discovery Grant. I.E. thanks Marat Freytsis and Anthropic for access to Claude Code, which assisted with aspects of building the models used in this work.
The work of DC is in part supported by Discovery Grants from the Natural Sciences and Engineering Research Council of Canada, the Canada Research Chair program, the Ontario Early Researcher Award, and the University of Toronto McLean Award. NW is supported by NSF
under grant PHY-2610067, the BSF under grant 2022287 and the Simons Foundation.

\section*{AI Disclosure}
In our analysis and the preparation of this manuscript, Claude was used in
the following way: to assist with building coding segments in regards to the model and evaluation infrastructure, to increase the efficiency of our implementation, to help monitor and queue compute jobs, and to optimize plots, all of which were human-verified. Claude was also widely used in building the 3D interactive figures in \url{https://ibrahimelsharkawy.com/blog/astrolabe-flow-matching.html}, and was used for the manuscript by assisting with many literature searches and providing feedback on draft versions of the
paper to help identify inconsistencies and editing mistakes.

\clearpage

\appendix

\section{Inference Ablation}
\label{app:inference-ablation}
\begin{figure}
    \centering
    \includegraphics[width=.6\linewidth]{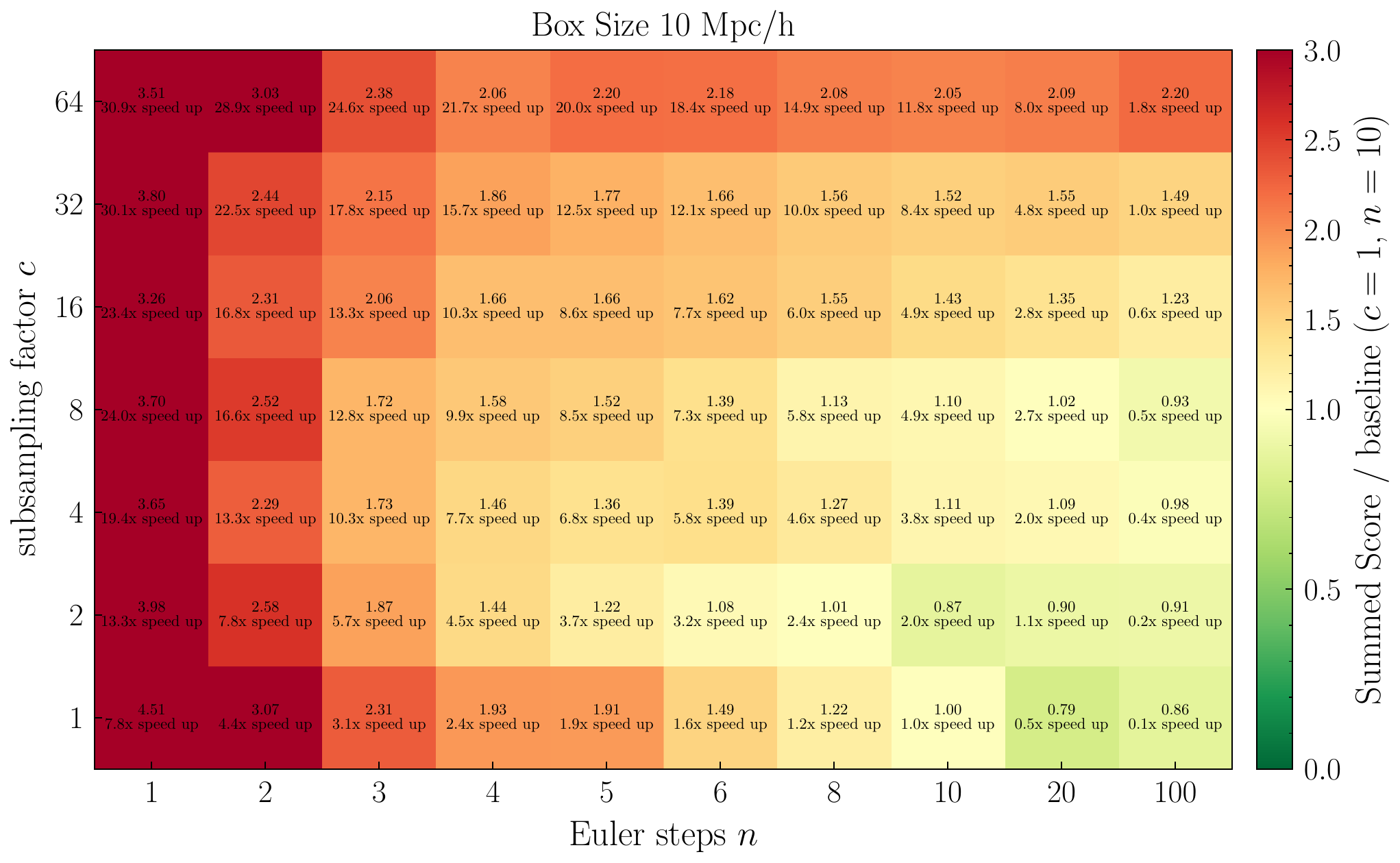}
    \includegraphics[width=.6\linewidth]{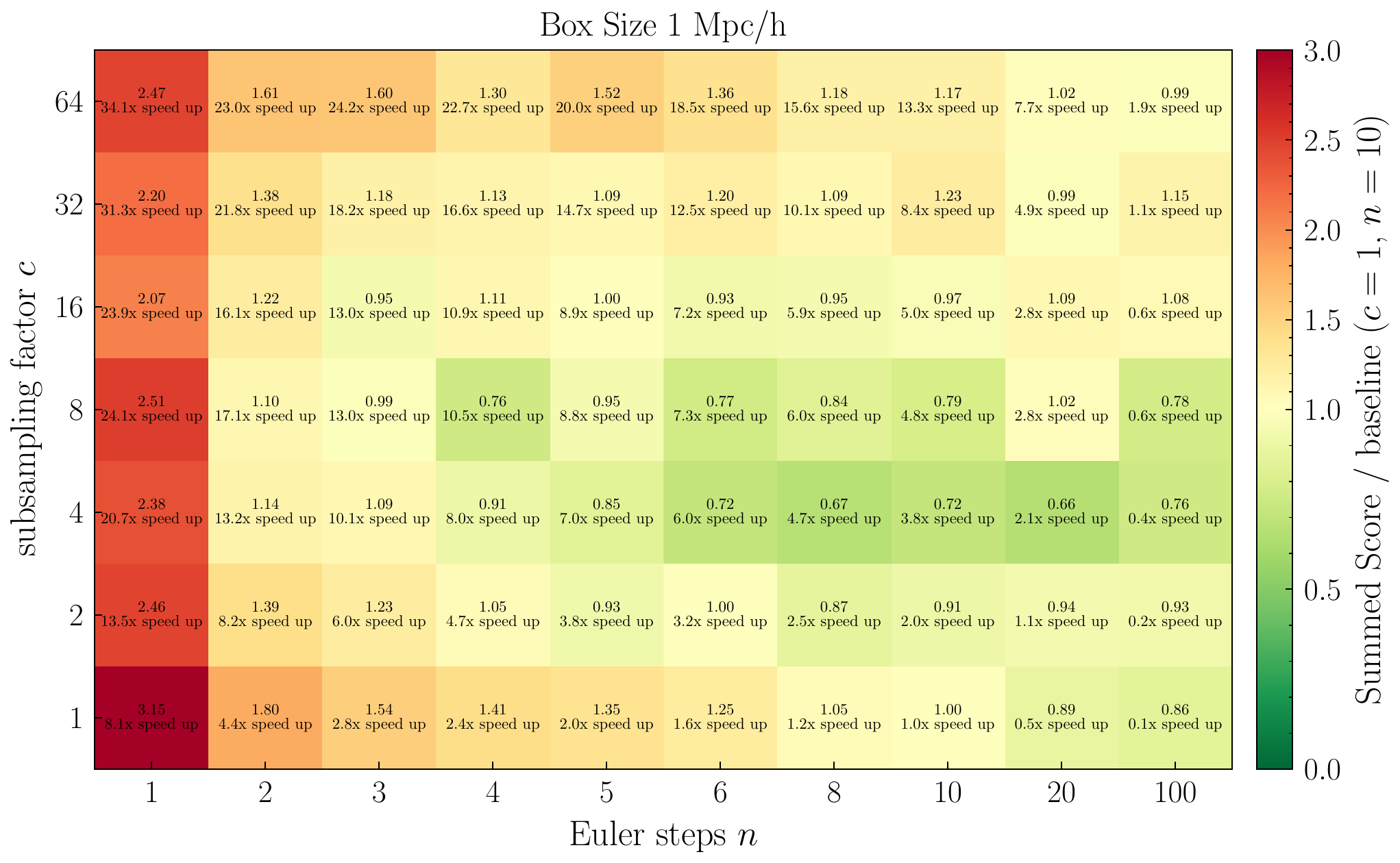}
    \caption{Ablation varying the number of Euler steps $n$ and the subsampling factor $c$ over a grid. Lower is better. For fast inference, we chose $n=5$ and $c=4$.}
    \label{fig:Comput-Cost-grid-scan}
\end{figure}

In this appendix we explore methods to speed up inference. Our standard inference setup requires $n =10$ steps through our Euler integration scheme and evaluates the network for a flow velocity for all particles in the simulation. Here we scan through a 2D grid varying both $n$ and a sub-sampling factor $c$. For a given $c$, we evaluate the flow velocity of only $N_{\rm sim}/c$ randomly-chosen particles. Particles that are not evaluated are assigned the same flow velocity as their nearest evaluated neighbor in 3-dimensional position space.\footnote{We have attempted doing the same in full 6-dimensional phase space, which would also incorporate closeness in velocity space,  but we find less success than the pure position-space distance. We leave a study on improving such subsampling for future work.} We do this for one initial condition per box size and for \astrolabe{}-small. After generating all fields, we evaluate each point with a combined error metric of all error metrics in Section \ref{sec:results} standardized by its variation across the scan, 
\begin{equation}
    E = \frac{1}{\sigma_{e_{\rm PS}}}e_{\rm PS} + \frac{1}{\sigma_{e_{\sigma_v}}}e_{\sigma_v} + \frac{1}{\sigma_{e_{c}}}e_{c} + \frac{1}{\sigma_{e_{\gamma}}}e_{\gamma}+\frac{1}{\sigma_{e_{\eta_J}}}e_{\eta_J}.
    \label{eq:combined-error}
\end{equation}
We then divide by $E_{\rm baseline}$, given by the error of our standard $n10$ steps and no subsampling configuration. The ablation results are plotted in Figure \ref{fig:Comput-Cost-grid-scan} as a 2D heat map, where smaller values are better. From these, we choose $n=5$ and $c=4$ for fast inference as a reasonable setting for both box sizes that increases speed significantly with a controlled loss of performance. 

\section{Hash Encoding of Particle Positions}
\label{app:hash}
\begin{figure}
    \centering
    \includegraphics[width=.25\linewidth]{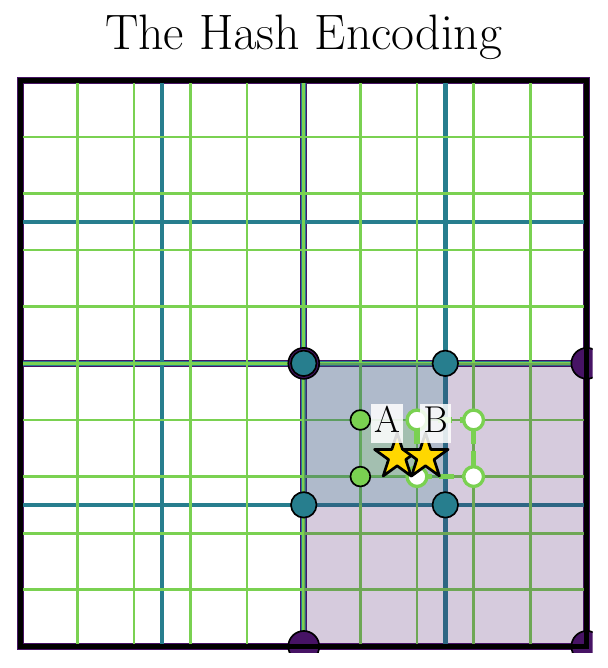}
    \caption{Diagrammatic Representation of the Hash Grids.}
    \label{fig:HashEncoder}
\end{figure}

The network is given each particle's and each field cell's comoving position in units relative to the box size (that is $\mathbf{r} \in [0,1)^3$). Thus, in a 1 Mpc box, two particles that differ in position by a kpc would differ in network input at the $10^{-3}$ level. Neural networks are known to struggle with capturing small-scale correlations \cite{rahaman2019spectralbiasneuralnetworks} and thus benefit from a feature basis that changes relative input scale. One common choice is a Fourier basis, and another common to physical applications is the multi-resolution hash encoder \cite{takikawa2021neuralgeometricleveldetail,Huang_2024,muller2022instant}.   

In it, we build $L=12$ grids of different resolutions from $32^3$ to $2767^3$ cells, finer than the particle mesh we train on. Each grid cell is represented in a per-grid hash table that contains a learned feature vector of dimension $T_D =2$ per grid cell. When a particle position is encoded, we find which cell it belongs to for all $L=12$ grids and concatenate the associated learned feature vector of all cells, forming a positional encoding vector of dimension $H_{\rm hash} = 24$, which is then passed through a small MLP to reach the network's full hidden dimension. Figure \ref{fig:HashEncoder} gives a diagrammatic representation of the grids.

\section{Model and Training Hyperparameters}
\label{app:hparams}
Table~\ref{tab:hparams_sizes} shows the hyperparameters of \astrolabe{} that vary by model size, while Table~\ref{tab:hparams_shared} shows hyperparameters of \astrolabe{} that are shared by both model sizes. 

\begin{table}[h]
\centering
\small
\begin{tabular}{l cc}
\toprule
 & \textbf{small} & \textbf{medium} \\
\midrule
Hidden dimension $d$                         & 128  & 512 \\
Transformer blocks                           & 8    & 12 \\
Attention heads                              & 8    & 16 \\
Voxels per token cell per side             & 8    & 10 \\
Token convolution channels                   & 32   & 48 \\
Particle Batch Size            & 8096 & 4096 \\
\bottomrule
\end{tabular}
\caption{\label{tab:hparams_sizes}\astrolabe-Small and \astrolabe-Medium hyperparameters. Parameters shared between the two sizes are given in Table~\ref{tab:hparams_shared}.}
\end{table}

\begin{table}[h]
\centering
\scriptsize
\renewcommand{\arraystretch}{1.15}
\begin{tabular}{l c @{\hspace{1.5em}} l c}
\toprule
\multicolumn{4}{c}{\textbf{Shared hyperparameters}} \\
\midrule
\multicolumn{2}{l}{\emph{Architecture}} & \multicolumn{2}{l}{\emph{Optimization}} \\
Particle input channels $(\vec x_{\rm COLA}, \vec v_{\rm COLA}, \vec x_{\rm IC}, \vec v_{\rm IC})$ & 12 & Optimizer & Ranger \\
Token grids $g$                     & 2, 4, 8, 16, 32 & Learning rate              & $5\times10^{-4}$ \\
Token encoder layers                & 2               & $(\beta_1, \beta_2)$       & $(0.95, 0.98)$ \\
Output head layers                  & 2               & Weight decay               & $10^{-4}$ \\
Global summary tokens               & 4               & Warm-up steps              & 1000 \\
Nearest neighbors $K$               & 256             & Schedule                   & cosine \\
Pair radial basis functions         & 32              & Gradient clipping (norm)   & 1.0 \\
Power-spectrum tokens               & 16              & Precision                  & bf16 \\
Distance-gated grids                & 8, 16, 32       & Halo-weighted loss weight  & 10 \\
Gating radius (fraction of the box) & 0.3, 0.2, 0.1   & Inference Euler steps      & 10 \\
\midrule
\multicolumn{2}{l}{\emph{Position encoding}} & \multicolumn{2}{l}{\emph{Conditioning and regularization}} \\
Hash levels                         & 12              & Conditioning                    & $a$, $N$, $L$, initial conditions \\
Hash features per level             & 2               & Conditioning dropout            & 0.1 \\
Fourier output features             & 64              & Initial-condition dropout       & 0.1 \\
                                    &                 & Token-grid dropout (per grid)   & 0, 0, 0, 0.1, 0.15 \\
                                    &                 & Power-spectrum token dropout    & 0.1 \\
                                    &                 & Attention dropout               & 0.1 \\
                                    &                 & MLP dropout                     & 0.1 \\
\bottomrule
\end{tabular}
\caption{\label{tab:hparams_shared}Hyperparameters shared between the two model sizes.}
\end{table}

\section{More Halo Results}
\label{app:res}

Here we give specific mean concentration, inner slope and Jeans test values for each of the model and \Nbody{} simulation cases we consider. Table \ref{tab:16m-raw} displays the raw values for the 1 Mpc 16M particle box, and Table \ref{tab:tng-raw} does so for the two TNG cases we consider. Table \ref{tab:cdm-raw} gives these values for both box sizes of 2M $\Lambda$CDM initial conditions, while Table \ref{tab:dao-raw} gives these values for all four box size and DAO cases for 2M DAO initial conditions.

\begin{table}
    \centering
\begin{tabular}{llccc}
\hline
Simulation & Model & Concentration $c$ & Inner slope $\gamma$ & $\eta_{\rm J}$ \\
\hline
16M, 1 Mpc/$h$ & \Nbody{} & 20.2 $\pm$ 2.0 & $-1.77 \pm 0.01$ & 0.99 $\pm$ 0.01 \\
(training sim & COLA & 2.6 $\pm$ 0.1 & $-0.28 \pm 0.02$ & 0.16 $\pm$ 0.03 \\
heldout $z=0$) & \astrolabe-Small & 6.0 $\pm$ 0.6 & $-0.57 \pm 0.08$ & 0.52 $\pm$ 0.03 \\
 & \astrolabe-Small-ft-16M & 8.1 $\pm$ 1.6 & $-0.58 \pm 0.09$ & 0.64 $\pm$ 0.04 \\
 & \astrolabe-Small-ft-16M (fast) & 12.2 $\pm$ 2.4 & $-0.71 \pm 0.05$ & 0.44 $\pm$ 0.02 \\
\hline
16M, 1 Mpc/$h$ & \Nbody{} & 22.1 $\pm$ 1.5 & $-1.74 \pm 0.02$ & 0.97 $\pm$ 0.01 \\
(eval sim) & COLA & 2.7 $\pm$ 0.1 & $-0.37 \pm 0.04$ & 0.14 $\pm$ 0.03 \\
 & \astrolabe-Small & 7.0 $\pm$ 0.7 & $-0.54 \pm 0.03$ & 0.53 $\pm$ 0.04 \\
 & \astrolabe-Small-ft-16M & 4.3 $\pm$ 1.2 & $-0.30 \pm 0.07$ & 0.77 $\pm$ 0.11 \\
 & \astrolabe-Small-ft-16M (fast) & 7.5 $\pm$ 1.1 & $-0.49 \pm 0.05$ & 0.50 $\pm$ 0.03 \\
\hline
\end{tabular}
\caption{\label{tab:16m-raw}Concentration $c$, inner slope $\gamma$ and Jeans test $\eta_{\rm J}$ for \Nbody{} simulations, \astrolabe{} and COLA for the $\Lambda$CDM 16M particle case in a 1 Mpc box size. Mean and variance are computed over all halos with $\geq 50,000$ particles in the two initial conditions.}
\end{table}

\begin{table}[t!]
    \centering
\begin{tabular}{llccc}
\hline
Simulation & Model & Concentration $c$ & Inner slope $\gamma$ & $\eta_{\rm J}$ \\
\hline
TNG50-3-Dark & \Nbody{} & 11.8 $\pm$ 0.3 & $-1.60 \pm 0.01$ & 1.02 $\pm$ 0.01 \\
 & COLA & 1.2 $\pm$ 0.1 & $-0.46 \pm 0.03$ & 0.28 $\pm$ 0.02 \\
 & \astrolabe-Small & 0.9 $\pm$ 0.1 & $-0.18 \pm 0.03$ & 0.57 $\pm$ 0.04 \\
 & \astrolabe-Small-ft-TNG50 & 2.2 $\pm$ 0.2 & $-0.35 \pm 0.04$ & 0.65 $\pm$ 0.04 \\
 & \astrolabe-Small-ft-TNG50 (fast) & 3.5 $\pm$ 0.2 & $-0.31 \pm 0.03$ & 0.45 $\pm$ 0.02 \\
\hline
TNG100-3-Dark & \Nbody{} & 9.2 $\pm$ 0.3 & $-1.52 \pm 0.01$ & 1.04 $\pm$ 0.01 \\
 & COLA & 1.1 $\pm$ 0.1 & $-0.53 \pm 0.06$ & 0.37 $\pm$ 0.04 \\
 & \astrolabe-Small & 1.0 $\pm$ 0.1 & $-0.16 \pm 0.05$ & 0.31 $\pm$ 0.05 \\
 & \astrolabe-Small-ft-TNG50 & 2.0 $\pm$ 0.2 & $-0.26 \pm 0.01$ & 0.60 $\pm$ 0.05 \\
 & \astrolabe-Small-ft-TNG50 (fast) & 2.7 $\pm$ 0.2 & $-0.34 \pm 0.03$ & 0.44 $\pm$ 0.03 \\
\hline
\end{tabular}
\caption{\label{tab:tng-raw}Concentration $c$, inner slope $\gamma$ and Jeans test $\eta_{\rm J}$ for \Nbody{} simulations, \astrolabe{} and COLA for \texttt{TNG50-3-dark} and \texttt{TNG100-3-dark} simulations. Mean and variance are computed over all halos with $\geq 50,000$ particles in the two simulations.}
\end{table}

\begin{table}
    \centering
\begin{tabular}{llccc}
\hline
Simulation & Model & Concentration $c$ & Inner slope $\gamma$ & $\eta_{\rm J}$ \\
\hline
CDM & \Nbody{} & 21.9 $\pm$ 0.5 & $-1.64 \pm 0.02$ & 0.99 $\pm$ 0.01 \\
  1 Mpc/$h$ & COLA & 2.6 $\pm$ 0.0 & $-0.20 \pm 0.03$ & 0.19 $\pm$ 0.02 \\
 & \astrolabe-Small & 13.8 $\pm$ 1.4 & $-0.79 \pm 0.06$ & 0.61 $\pm$ 0.01 \\
 & \astrolabe-Medium & 25.9 $\pm$ 2.5 & $-1.19 \pm 0.10$ & 0.57 $\pm$ 0.01 \\
 & \astrolabe-Small (fast) & 20.3 $\pm$ 1.6 & $-0.89 \pm 0.02$ & 0.40 $\pm$ 0.02 \\
 & \astrolabe-Medium (fast) & 35.6 $\pm$ 2.9 & $-1.07 \pm 0.03$ & 0.38 $\pm$ 0.02 \\
\hline
CDM & \Nbody{} & 14.1 $\pm$ 0.3 & $-1.51 \pm 0.01$ & 1.01 $\pm$ 0.01 \\
  10 Mpc/$h$ & COLA & 2.3 $\pm$ 0.1 & $-0.36 \pm 0.02$ & 0.27 $\pm$ 0.02 \\
 & \astrolabe-Small & 8.5 $\pm$ 0.5 & $-0.67 \pm 0.03$ & 0.61 $\pm$ 0.01 \\
 & \astrolabe-Medium & 11.7 $\pm$ 0.3 & $-0.84 \pm 0.03$ & 0.67 $\pm$ 0.01 \\
 & \astrolabe-Small (fast) & 14.1 $\pm$ 0.9 & $-0.72 \pm 0.02$ & 0.44 $\pm$ 0.01 \\
 & \astrolabe-Medium (fast) & 15.7 $\pm$ 0.5 & $-0.82 \pm 0.04$ & 0.45 $\pm$ 0.01 \\
\hline
\end{tabular}
\caption{\label{tab:cdm-raw}Concentration $c$, inner slope $\gamma$ and Jeans test $\eta_{\rm J}$ for \Nbody{} simulations, \astrolabe{} and COLA for the $\Lambda$CDM 2M particle case in both box sizes. Mean and variance are computed over all halos with $\geq 5,000$ particles in all eight initial conditions.}
\end{table}

\begin{table}
    \centering
\begin{tabular}{llccc}
\hline
Simulation & Model & Concentration $c$ & Inner slope $\gamma$ & $\eta_{\rm J}$ \\
\hline
DAO-strong & \Nbody{} & 8.7 $\pm$ 0.4 & $-1.35 \pm 0.05$ & 1.01 $\pm$ 0.03 \\
1 Mpc/$h$ & COLA & 2.3 $\pm$ 0.2 & $-0.30 \pm 0.09$ & 0.37 $\pm$ 0.04 \\
($k_{\rm peak}=32$) & \astrolabe-Small & 8.2 $\pm$ 1.4 & $-0.72 \pm 0.08$ & 0.68 $\pm$ 0.03 \\
 & \astrolabe-Medium & 11.1 $\pm$ 2.0 & $-1.08 \pm 0.15$ & 0.75 $\pm$ 0.04 \\
 & \astrolabe-Small (fast) & 13.2 $\pm$ 3.1 & $-0.74 \pm 0.04$ & 0.54 $\pm$ 0.04 \\
 & \astrolabe-Medium (fast) & 14.7 $\pm$ 2.7 & $-0.88 \pm 0.09$ & 0.59 $\pm$ 0.08 \\
\hline
DAO-weak & \Nbody{} & 15.7 $\pm$ 1.3 & $-1.50 \pm 0.01$ & 0.99 $\pm$ 0.02 \\
1 Mpc/$h$ & COLA & 2.7 $\pm$ 0.0 & $-0.34 \pm 0.01$ & 0.21 $\pm$ 0.02 \\
($k_{\rm peak}=95$) & \astrolabe-Small & 12.3 $\pm$ 1.1 & $-0.79 \pm 0.10$ & 0.64 $\pm$ 0.02 \\
 & \astrolabe-Medium & 16.6 $\pm$ 0.8 & $-1.00 \pm 0.11$ & 0.71 $\pm$ 0.02 \\
 & \astrolabe-Small (fast) & 16.6 $\pm$ 2.4 & $-0.78 \pm 0.06$ & 0.42 $\pm$ 0.01 \\
 & \astrolabe-Medium (fast) & 22.1 $\pm$ 4.4 & $-0.98 \pm 0.07$ & 0.41 $\pm$ 0.02 \\
\hline
DAO-strong & \Nbody{} & 13.1 $\pm$ 0.6 & $-1.43 \pm 0.02$ & 1.02 $\pm$ 0.01 \\
10 Mpc/$h$ & COLA & 2.6 $\pm$ 0.0 & $-0.29 \pm 0.06$ & 0.29 $\pm$ 0.02 \\
($k_{\rm peak}=18$) & \astrolabe-Small & 10.0 $\pm$ 0.4 & $-0.71 \pm 0.04$ & 0.59 $\pm$ 0.02 \\
 & \astrolabe-Medium & 13.2 $\pm$ 1.2 & $-0.94 \pm 0.10$ & 0.68 $\pm$ 0.02 \\
 & \astrolabe-Small (fast) & 14.4 $\pm$ 1.4 & $-0.76 \pm 0.04$ & 0.43 $\pm$ 0.01 \\
 & \astrolabe-Medium (fast) & 16.5 $\pm$ 0.6 & $-0.90 \pm 0.05$ & 0.44 $\pm$ 0.01 \\
\hline
DAO-weak & \Nbody{} & 13.9 $\pm$ 0.8 & $-1.41 \pm 0.03$ & 1.02 $\pm$ 0.01 \\
10 Mpc/$h$ & COLA & 2.5 $\pm$ 0.1 & $-0.27 \pm 0.07$ & 0.27 $\pm$ 0.02 \\
($k_{\rm peak}=31$) & \astrolabe-Small & 10.2 $\pm$ 0.4 & $-0.68 \pm 0.04$ & 0.61 $\pm$ 0.02 \\
 & \astrolabe-Medium & 13.0 $\pm$ 0.8 & $-0.84 \pm 0.01$ & 0.70 $\pm$ 0.02 \\
 & \astrolabe-Small (fast) & 12.1 $\pm$ 0.6 & $-0.65 \pm 0.03$ & 0.43 $\pm$ 0.01 \\
 & \astrolabe-Medium (fast) & 16.7 $\pm$ 0.7 & $-0.77 \pm 0.03$ & 0.43 $\pm$ 0.01 \\
\hline
\end{tabular}
\caption{\label{tab:dao-raw}Concentration $c$, inner slope $\gamma$ and Jeans test $\eta_{\rm J}$ for \Nbody{} simulations, \astrolabe{} and COLA for the DAO 2M particle case in both box sizes and DAO strengths. Mean and variance are computed over all halos with $\geq 5,000$ particles in all three initial conditions.}
\end{table}

\clearpage
\bibliography{bib}

@article{Heitmann:2013bra,
    author = "Heitmann, Katrin and Lawrence, Earl and Kwan, Juliana and Habib, Salman and Higdon, David",
    title = "{The Coyote Universe Extended: Precision Emulation of the Matter Power Spectrum}",
    eprint = "1304.7849",
    archivePrefix = "arXiv",
    primaryClass = "astro-ph.CO",
    reportNumber = "ANL-HEP-PR-13-10",
    doi = "10.1088/0004-637X/780/1/111",
    journal = "Astrophys. J.",
    volume = "780",
    pages = "111",
    year = "2014"
}

@article{Tassev:2013pn,
    author = "Tassev, Svetlin and Zaldarriaga, Matias and Eisenstein, Daniel",
    title = "{Solving Large Scale Structure in Ten Easy Steps with COLA}",
    eprint = "1301.0322",
    archivePrefix = "arXiv",
    primaryClass = "astro-ph.CO",
    doi = "10.1088/1475-7516/2013/06/036",
    journal = "JCAP",
    volume = "06",
    pages = "036",
    year = "2013"
}

@article{Euclid:2018mlb,
    author = "Knabenhans, Mischa and others",
    collaboration = "Euclid",
    title = "{Euclid preparation: II. The EuclidEmulator -- A tool to compute the cosmology dependence of the nonlinear matter power spectrum}",
    eprint = "1809.04695",
    archivePrefix = "arXiv",
    primaryClass = "astro-ph.CO",
    doi = "10.1093/mnras/stz197",
    journal = "Mon. Not. Roy. Astron. Soc.",
    volume = "484",
    pages = "5509--5529",
    year = "2019"
}

@article{Nishimichi:2018etk,
    author = "Nishimichi, Takahiro and others",
    title = "{Dark Quest. I. Fast and Accurate Emulation of Halo Clustering Statistics and Its Application to Galaxy Clustering}",
    eprint = "1811.09504",
    archivePrefix = "arXiv",
    primaryClass = "astro-ph.CO",
    reportNumber = "YITP-19-73",
    doi = "10.3847/1538-4357/ab3719",
    journal = "Astrophys. J.",
    volume = "884",
    pages = "29",
    year = "2019"
}

@article{Blas:2011rf,
    author = "Blas, Diego and Lesgourgues, Julien and Tram, Thomas",
    title = "{The Cosmic Linear Anisotropy Solving System (CLASS) II: Approximation schemes}",
    eprint = "1104.2933",
    archivePrefix = "arXiv",
    primaryClass = "astro-ph.CO",
    reportNumber = "CERN-PH-TH-2011-082, LAPTH-010-11",
    doi = "10.1088/1475-7516/2011/07/034",
    journal = "JCAP",
    volume = "07",
    pages = "034",
    year = "2011"
}

@article{Cyr-Racine:2015ihg,
    author = "Cyr-Racine, Francis-Yan and Sigurdson, Kris and Zavala, Jesus and Bringmann, Torsten and Vogelsberger, Mark and Pfrommer, Christoph",
    title = "{ETHOS{\textemdash}an effective theory of structure formation: From dark particle physics to the matter distribution of the Universe}",
    eprint = "1512.05344",
    archivePrefix = "arXiv",
    primaryClass = "astro-ph.CO",
    doi = "10.1103/PhysRevD.93.123527",
    journal = "Phys. Rev. D",
    volume = "93",
    number = "12",
    pages = "123527",
    year = "2016"
}

@article{Hopkins:2014qka,
    author = "Hopkins, Philip F.",
    title = "{A new class of accurate, mesh-free hydrodynamic simulation methods}",
    eprint = "1409.7395",
    archivePrefix = "arXiv",
    primaryClass = "astro-ph.CO",
    doi = "10.1093/mnras/stv195",
    journal = "Mon. Not. Roy. Astron. Soc.",
    volume = "450",
    number = "1",
    pages = "53--110",
    year = "2015"
}

@article{Ding:2023uro,
    author = "Ding, Jiacheng and Li, Shaohong and Zheng, Yi and Luo, Xiaolin and Zhang, Le and Li, Xiao-Dong",
    title = "{Fast Generation of Mock Galaxy Catalogs with COLA}",
    eprint = "2311.00981",
    archivePrefix = "arXiv",
    primaryClass = "astro-ph.CO",
    doi = "10.3847/1538-4365/ad0c5b",
    journal = "Astrophys. J. Suppl.",
    volume = "270",
    number = "2",
    pages = "25",
    year = "2024"
}

@article{Potter:2016ttn,
    author = "Potter, Douglas and Stadel, Joachim and Teyssier, Romain",
    title = "{PKDGRAV3: beyond trillion particle cosmological simulations for the next era of galaxy surveys}",
    eprint = "1609.08621",
    archivePrefix = "arXiv",
    primaryClass = "astro-ph.IM",
    doi = "10.1186/s40668-017-0021-1",
    journal = "Comput. Astrophys. Cosmol.",
    volume = "4",
    number = "1",
    pages = "2",
    year = "2017"
}

@article{Archidiacono:2019wdp,
    author = "Archidiacono, Maria and Hooper, Deanna C. and Murgia, Riccardo and Bohr, Sebastian and Lesgourgues, Julien and Viel, Matteo",
    title = "{Constraining Dark Matter-Dark Radiation interactions with CMB, BAO, and Lyman-$\alpha$}",
    eprint = "1907.01496",
    archivePrefix = "arXiv",
    primaryClass = "astro-ph.CO",
    doi = "10.1088/1475-7516/2019/10/055",
    journal = "JCAP",
    volume = "10",
    pages = "055",
    year = "2019"
}

@misc{lipman2023flowmatchinggenerativemodeling,
      title={Flow Matching for Generative Modeling}, 
      author={Yaron Lipman and Ricky T. Q. Chen and Heli Ben-Hamu and Maximilian Nickel and Matt Le},
      year={2023},
      eprint={2210.02747},
      archivePrefix={arXiv},
      primaryClass={cs.LG},
      url={https://arxiv.org/abs/2210.02747}, 
}

@misc{liu2022flowstraightfastlearning,
      title={Flow Straight and Fast: Learning to Generate and Transfer Data with Rectified Flow}, 
      author={Xingchao Liu and Chengyue Gong and Qiang Liu},
      year={2022},
      eprint={2209.03003},
      archivePrefix={arXiv},
      primaryClass={cs.LG},
      url={https://arxiv.org/abs/2209.03003}, 
}

@misc{albergo2023buildingnormalizingflowsstochastic,
      title={Building Normalizing Flows with Stochastic Interpolants}, 
      author={Michael S. Albergo and Eric Vanden-Eijnden},
      year={2023},
      eprint={2209.15571},
      archivePrefix={arXiv},
      primaryClass={cs.LG},
      url={https://arxiv.org/abs/2209.15571}, 
}

@article{Huang_2024,
   title={Efficient physics-informed neural networks using hash encoding},
   volume={501},
   ISSN={0021-9991},
   url={http://dx.doi.org/10.1016/j.jcp.2024.112760},
   DOI={10.1016/j.jcp.2024.112760},
   journal={Journal of Computational Physics},
   publisher={Elsevier BV},
   author={Huang, Xinquan and Alkhalifah, Tariq},
   year={2024},
   month=Mar, pages={112760} }

@misc{takikawa2021neuralgeometricleveldetail,
      title={Neural Geometric Level of Detail: Real-time Rendering with Implicit 3D Shapes}, 
      author={Towaki Takikawa and Joey Litalien and Kangxue Yin and Karsten Kreis and Charles Loop and Derek Nowrouzezahrai and Alec Jacobson and Morgan McGuire and Sanja Fidler},
      year={2021},
      eprint={2101.10994},
      archivePrefix={arXiv},
      primaryClass={cs.CV},
      url={https://arxiv.org/abs/2101.10994}, 
}

@misc{mildenhall2020nerfrepresentingscenesneural,
      title={NeRF: Representing Scenes as Neural Radiance Fields for View Synthesis}, 
      author={Ben Mildenhall and Pratul P. Srinivasan and Matthew Tancik and Jonathan T. Barron and Ravi Ramamoorthi and Ren Ng},
      year={2020},
      eprint={2003.08934},
      archivePrefix={arXiv},
      primaryClass={cs.CV},
      url={https://arxiv.org/abs/2003.08934}, 
}

@misc{tancik2020fourierfeaturesletnetworks,
      title={Fourier Features Let Networks Learn High Frequency Functions in Low Dimensional Domains}, 
      author={Matthew Tancik and Pratul P. Srinivasan and Ben Mildenhall and Sara Fridovich-Keil and Nithin Raghavan and Utkarsh Singhal and Ravi Ramamoorthi and Jonathan T. Barron and Ren Ng},
      year={2020},
      eprint={2006.10739},
      archivePrefix={arXiv},
      primaryClass={cs.CV},
      url={https://arxiv.org/abs/2006.10739}, 
}

@misc{sitzmann2020implicitneuralrepresentationsperiodic,
      title={Implicit Neural Representations with Periodic Activation Functions}, 
      author={Vincent Sitzmann and Julien N. P. Martel and Alexander W. Bergman and David B. Lindell and Gordon Wetzstein},
      year={2020},
      eprint={2006.09661},
      archivePrefix={arXiv},
      primaryClass={cs.CV},
      url={https://arxiv.org/abs/2006.09661}, 
}

@misc{rahaman2019spectralbiasneuralnetworks,
      title={On the Spectral Bias of Neural Networks}, 
      author={Nasim Rahaman and Aristide Baratin and Devansh Arpit and Felix Draxler and Min Lin and Fred A. Hamprecht and Yoshua Bengio and Aaron Courville},
      year={2019},
      eprint={1806.08734},
      archivePrefix={arXiv},
      primaryClass={stat.ML},
      url={https://arxiv.org/abs/1806.08734}, 
}

@misc{song2021scorebasedgenerativemodelingstochastic,
      title={Score-Based Generative Modeling through Stochastic Differential Equations}, 
      author={Yang Song and Jascha Sohl-Dickstein and Diederik P. Kingma and Abhishek Kumar and Stefano Ermon and Ben Poole},
      year={2021},
      eprint={2011.13456},
      archivePrefix={arXiv},
      primaryClass={cs.LG},
      url={https://arxiv.org/abs/2011.13456}, 
}

@misc{ho2020denoisingdiffusionprobabilisticmodels,
      title={Denoising Diffusion Probabilistic Models}, 
      author={Jonathan Ho and Ajay Jain and Pieter Abbeel},
      year={2020},
      eprint={2006.11239},
      archivePrefix={arXiv},
      primaryClass={cs.LG},
      url={https://arxiv.org/abs/2006.11239}, 
}

@misc{karras2022elucidatingdesignspacediffusionbased,
      title={Elucidating the Design Space of Diffusion-Based Generative Models}, 
      author={Tero Karras and Miika Aittala and Timo Aila and Samuli Laine},
      year={2022},
      eprint={2206.00364},
      archivePrefix={arXiv},
      primaryClass={cs.CV},
      url={https://arxiv.org/abs/2206.00364}, 
}

@misc{rezende2016variationalinferencenormalizingflows,
      title={Variational Inference with Normalizing Flows}, 
      author={Danilo Jimenez Rezende and Shakir Mohamed},
      year={2016},
      eprint={1505.05770},
      archivePrefix={arXiv},
      primaryClass={stat.ML},
      url={https://arxiv.org/abs/1505.05770}, 
}

@misc{dinh2015nicenonlinearindependentcomponents,
      title={NICE: Non-linear Independent Components Estimation}, 
      author={Laurent Dinh and David Krueger and Yoshua Bengio},
      year={2015},
      eprint={1410.8516},
      archivePrefix={arXiv},
      primaryClass={cs.LG},
      url={https://arxiv.org/abs/1410.8516}, 
}

@misc{dinh2017densityestimationusingreal,
      title={Density estimation using Real NVP}, 
      author={Laurent Dinh and Jascha Sohl-Dickstein and Samy Bengio},
      year={2017},
      eprint={1605.08803},
      archivePrefix={arXiv},
      primaryClass={cs.LG},
      url={https://arxiv.org/abs/1605.08803}, 
}

@inproceedings{Elsharkawy:2025six,
    author = "Elsharkawy, Ibrahim and Kahn, Yonatan",
    title = "{Contrastive Normalizing Flows for Uncertainty-Aware Parameter Estimation}",
    eprint = "2505.08709",
    archivePrefix = "arXiv",
    primaryClass = "physics.data-an",
    month = "5",
    year = "2025"
}

@article{Angulo:2020vky,
    author = {Angulo, Raul E. and Zennaro, Matteo and Contreras, Sergio and Aric{\`o}, Giovanni and Pellejero-Iba{\~n}ez, Marcos and St{\"u}cker, Jens},
    title = "{The BACCO simulation project: exploiting the full power of large-scale structure for cosmology}",
    eprint = "2004.06245",
    archivePrefix = "arXiv",
    primaryClass = "astro-ph.CO",
    doi = "10.1093/mnras/stab2018",
    journal = "Mon. Not. Roy. Astron. Soc.",
    volume = "507",
    number = "4",
    pages = "5869--5881",
    year = "2021"
}

@article{Bird:2018efe,
    author = "Bird, Simeon and Rogers, Keir K. and Peiris, Hiranya V. and Verde, Licia and Font-Ribera, Andreu and Pontzen, Andrew",
    title = "{An Emulator for the Lyman-alpha Forest}",
    eprint = "1812.04654",
    archivePrefix = "arXiv",
    primaryClass = "astro-ph.CO",
    doi = "10.1088/1475-7516/2019/02/050",
    journal = "JCAP",
    volume = "02",
    pages = "050",
    year = "2019"
}

@misc{Bogorad:2026oxa,
    author = "Bogorad, Zachary and Elsharkawy, Ibrahim and Kahn, Yonatan and Larkoski, Andrew J. and Levi, Noam",
    title = "{Generative models on phase space}",
    eprint = "2604.02415",
    archivePrefix = "arXiv",
    primaryClass = "hep-ph",
    reportNumber = "FERMILAB-PUB-26-0177-T",
    month = "4",
    year = "2026"
}

@article{Pedersen:2020kaw,
    author = "Pedersen, Christian and Font-Ribera, Andreu and Rogers, Keir K. and McDonald, Patrick and Peiris, Hiranya V. and Pontzen, Andrew and Slosar, An{\v{z}}e",
    title = "{An emulator for the Lyman-$\alpha$ forest in beyond-$\Lambda$CDM cosmologies}",
    eprint = "2011.15127",
    archivePrefix = "arXiv",
    primaryClass = "astro-ph.CO",
    doi = "10.1088/1475-7516/2021/05/033",
    journal = "JCAP",
    volume = "05",
    pages = "033",
    year = "2021"
}

@article{CAMELS:2020cof,
    author = "Villaescusa-Navarro, Francisco and others",
    collaboration = "CAMELS",
    title = "{The CAMELS project: Cosmology and Astrophysics with MachinE Learning Simulations}",
    eprint = "2010.00619",
    archivePrefix = "arXiv",
    primaryClass = "astro-ph.CO",
    doi = "10.3847/1538-4357/abf7ba",
    journal = "Astrophys. J.",
    volume = "915",
    pages = "71",
    year = "2021"
}

@article{Roy:2023zar,
    author = "Roy, Sandip and Shen, Xuejian and Lisanti, Mariangela and Curtin, David and Murray, Norman and Hopkins, Philip F.",
    title = "{Simulating Atomic Dark Matter in Milky Way Analogs}",
    eprint = "2304.09878",
    archivePrefix = "arXiv",
    primaryClass = "astro-ph.GA",
    doi = "10.3847/2041-8213/ace2c8",
    journal = "Astrophys. J. Lett.",
    volume = "954",
    number = "2",
    pages = "L40",
    year = "2023"
}

@article{Rogers:2020ltq,
    author = "Rogers, Keir K. and Peiris, Hiranya V.",
    title = "{Strong Bound on Canonical Ultralight Axion Dark Matter from the Lyman-Alpha Forest}",
    eprint = "2007.12705",
    archivePrefix = "arXiv",
    primaryClass = "astro-ph.CO",
    doi = "10.1103/PhysRevLett.126.071302",
    journal = "Phys. Rev. Lett.",
    volume = "126",
    number = "7",
    pages = "071302",
    year = "2021"
}

@misc{Yuan:2026zlr,
    author = "Yuan, Zhihan and Gemmell, Caleb and Rogers, Keir K. and Barron, Jared and Roy, Sandip and Curtin, David and Murray, Norman",
    title = "{Strongest constraints on dark acoustic oscillations from the Lyman-alpha forest}",
    eprint = "2606.28482",
    archivePrefix = "arXiv",
    primaryClass = "astro-ph.CO",
    month = "6",
    year = "2026"
}

@article{Rose:2024xcb,
    author = "Rose, Jonah C. and others",
    title = "{Introducing the DREAMS Project: DaRk mattEr and Astrophysics with Machine Learning and Simulations}",
    eprint = "2405.00766",
    archivePrefix = "arXiv",
    primaryClass = "astro-ph.GA",
    doi = "10.3847/1538-4357/adb8e5",
    journal = "Astrophys. J.",
    volume = "982",
    number = "2",
    pages = "68",
    year = "2025"
}

@article{He:2018ggn,
    author = "He, Siyu and Li, Yin and Feng, Yu and Ho, Shirley and Ravanbakhsh, Siamak and Chen, Wei and P{\'o}czos, Barnab{\'a}s",
    title = "{Learning to Predict the Cosmological Structure Formation}",
    eprint = "1811.06533",
    archivePrefix = "arXiv",
    primaryClass = "astro-ph.CO",
    doi = "10.1073/pnas.1821458116",
    journal = "Proc. Nat. Acad. Sci.",
    volume = "116",
    number = "28",
    pages = "13825--13832",
    year = "2019"
}

@article{Rodriguez:2018mjb,
    author = "Rodriguez, Andres C. and Kacprzak, Tomasz and Lucchi, Aurelien and Amara, Adam and Sgier, Raphael and Fluri, Janis and Hofmann, Thomas and R{\'e}fr{\'e}gier, Alexandre",
    title = "{Fast cosmic web simulations with generative adversarial networks}",
    eprint = "1801.09070",
    archivePrefix = "arXiv",
    primaryClass = "astro-ph.CO",
    doi = "10.1186/s40668-018-0026-4",
    journal = "Comput. Astrophys. Cosmol.",
    volume = "5",
    number = "1",
    pages = "4",
    year = "2018"
}

@article{Perraudin:2019bxl,
    author = {Perraudin, Nathana{\"e}l and Srivastava, Ankit and Lucchi, Aurelien and Kacprzak, Tomasz and Hofmann, Thomas and R{\'e}fr{\'e}gier, Alexandre},
    title = "{Cosmological N-body simulations: a challenge for scalable generative models}",
    eprint = "1908.05519",
    archivePrefix = "arXiv",
    primaryClass = "physics.comp-ph",
    doi = "10.1186/s40668-019-0032-1",
    journal = "Comput. Astrophys. Cosmol.",
    volume = "6",
    number = "1",
    pages = "5",
    year = "2019"
}

@article{Ramanah:2020vyl,
    author = "Ramanah, Doogesh Kodi and Charnock, Tom and Villaescusa-Navarro, Francisco and Wandelt, Benjamin D.",
    title = "{Super-resolution emulator of cosmological simulations using deep physical models}",
    eprint = "2001.05519",
    archivePrefix = "arXiv",
    primaryClass = "astro-ph.CO",
    doi = "10.1093/mnras/staa1428",
    journal = "Mon. Not. Roy. Astron. Soc.",
    volume = "495",
    pages = "4227",
    year = "2020"
}

@misc{Li:2020vor,
    author = "Li, Yin and Ni, Yueying and Croft, Rupert A. C. and Di Matteo, Tiziana and Bird, Simeon and Feng, Yu",
    title = "{AI-assisted super-resolution cosmological simulations}",
    eprint = "2010.06608",
    archivePrefix = "arXiv",
    primaryClass = "astro-ph.CO",
    doi = "10.1073/pnas.2022038118",
    month = "10",
    year = "2020"
}

@article{Cuesta-Lazaro:2023zuk,
    author = "Cuesta-Lazaro, Carolina and Mishra-Sharma, Siddharth",
    title = "{Point cloud approach to generative modeling for galaxy surveys at the field level}",
    eprint = "2311.17141",
    archivePrefix = "arXiv",
    primaryClass = "astro-ph.CO",
    reportNumber = "MIT-CTP/5651",
    doi = "10.1103/PhysRevD.109.123531",
    journal = "Phys. Rev. D",
    volume = "109",
    number = "12",
    pages = "123531",
    year = "2024"
}

@article{Nguyen:2024ndo,
    author = "Nguyen, Tri and others",
    title = "{How DREAMS Are Made: Emulating Satellite Galaxy and Subhalo Populations with Diffusion Models and Point Clouds}",
    eprint = "2409.02980",
    archivePrefix = "arXiv",
    primaryClass = "astro-ph.GA",
    doi = "10.3847/1538-4357/ae28dc",
    journal = "Astrophys. J.",
    volume = "997",
    number = "2",
    pages = "336",
    year = "2026"
}

@misc{Huang:2025dmm,
    author = "Huang, Ningyuan and Stiskalek, Richard and Lee, Jun-Young and Bayer, Adrian E. and Margossian, Charles C. and Jespersen, Christian Kragh and Perez, Lucia A. and Saul, Lawrence K. and Villaescusa-Navarro, Francisco",
    title = "{CosmoBench: A Multiscale, Multiview, Multitask Cosmology Benchmark for Geometric Deep Learning}",
    eprint = "2507.03707",
    archivePrefix = "arXiv",
    primaryClass = "cs.LG",
    month = "7",
    year = "2025"
}

@misc{Mikuni:2025ocp,
    author = "Mikuni, Vinicius and Elsharkawy, Ibrahim and Nachman, Benjamin",
    title = "{OmniCosmos: Transferring Particle Physics Knowledge Across the Cosmos}",
    eprint = "2512.24422",
    archivePrefix = "arXiv",
    primaryClass = "astro-ph.CO",
    month = "12",
    year = "2025"
}

@book{Evans:1998,
  author    = {Evans, James},
  title     = {The History and Practice of Ancient Astronomy},
  publisher = {Oxford University Press},
  address   = {New York},
  year      = {1998},
  isbn      = {9780195095395},
  url       = {https://global.oup.com/academic/product/the-history-and-practice-of-ancient-astronomy-9780195095395}
}

@book{King:2004,
  author    = {King, David A.},
  title     = {In Synchrony with the Heavens: Studies in Astronomical
               Timekeeping and Instrumentation in Medieval Islamic
               Civilization. Volume 1: The Call of the Muezzin},
  publisher = {Brill},
  address   = {Leiden and Boston},
  year      = {2004},
  url       = {https://philpapers.org/rec/KINISW}
}

@book{King:2005,
  author    = {King, David A.},
  title     = {In Synchrony with the Heavens: Studies in Astronomical
               Timekeeping and Instrumentation in Medieval Islamic
               Civilization. Volume 2: Instruments of Mass Calculation},
  publisher = {Brill},
  address   = {Leiden and Boston},
  year      = {2005},
  isbn      = {9789004141889},
  series    = {Islamic Philosophy, Theology and Science. Texts and Studies},
  volume    = {55},
  url       = {https://brill.com/display/title/11037}
}

@book{Charette:2003,
  author    = {Charette, Fran\c{c}ois},
  title     = {Mathematical Instrumentation in Fourteenth-Century Egypt and
               Syria: The Illustrated Treatise of Najm al-D\={\i}n
               al-Mi\d{s}r\={\i}},
  publisher = {Brill},
  address   = {Leiden and Boston},
  year      = {2003},
  isbn      = {9004130152},
  doi       = {10.1163/9789047402176},
  series    = {Islamic Philosophy, Theology and Science. Texts and Studies},
  volume    = {51},
  url       = {https://brill.com/view/title/8302}
}

@article{Calvo:2006,
  author  = {Calvo, Emilia and Puig, Roser},
  title   = {The Universal Plate Revisited},
  journal = {Suhayl: Journal for the History of the Exact and Natural
             Sciences in Islamic Civilisation},
  volume  = {6},
  year    = {2006},
  note    = {Open access via RACO},
  url     = {https://raco.cat/index.php/Suhayl}
}

@book{Hockey:2007,
  editor    = {Hockey, Thomas and Trimble, Virginia and Williams, Thomas R.
               and Bracher, Katherine and Jarrell, Richard A. and
               March{\'e} II, Jordan D. and Ragep, F. Jamil},
  title     = {The Biographical Encyclopedia of Astronomers},
  publisher = {Springer},
  address   = {New York},
  year      = {2007},
  note      = {s.v. ``Faz\={a}r\={\i}'', ``\d{S}\={u}f\={\i}'',
               ``Zarq\={a}l\={\i}''},
  url       = {https://link.springer.com/referencework/10.1007/978-0-387-30400-7}
}

@misc{Elsharkawy:2026kwp,
    author = "Elsharkawy, Ibrahim and Mikuni, Vinicius and Bhimji, Wahid and Nachman, Benjamin",
    title = "{OmniMol: Transferring Particle Physics Knowledge to Molecular Dynamics with Point-Edge Transformers}",
    eprint = "2601.10791",
    archivePrefix = "arXiv",
    primaryClass = "physics.chem-ph",
    month = "1",
    year = "2026"
}

@article{Bhimji:2025isp,
    author = "Bhimji, Wahid and Harris, Chris and Mikuni, Vinicius and Nachman, Benjamin",
    title = "{Foundation model framework for all tasks involving jet physics}",
    eprint = "2510.24066",
    archivePrefix = "arXiv",
    primaryClass = "hep-ph",
    doi = "10.1103/knmd-f5jm",
    journal = "Phys. Rev. D",
    volume = "113",
    number = "3",
    pages = "032020",
    year = "2026"
}

@book{Morrison:2007,
  author    = {Morrison, James E.},
  title     = {The Astrolabe},
  publisher = {Janus},
  address   = {Rehoboth Beach, DE},
  year      = {2007}
}

@misc{mardani2024residualcorrectivediffusionmodeling,
      title={Residual Corrective Diffusion Modeling for Km-scale Atmospheric Downscaling}, 
      author={Morteza Mardani and Noah Brenowitz and Yair Cohen and Jaideep Pathak and Chieh-Yu Chen and Cheng-Chin Liu and Arash Vahdat and Mohammad Amin Nabian and Tao Ge and Akshay Subramaniam and Karthik Kashinath and Jan Kautz and Mike Pritchard},
      year={2024},
      eprint={2309.15214},
      archivePrefix={arXiv},
      primaryClass={cs.LG},
      url={https://arxiv.org/abs/2309.15214}, 
}

@misc{pathak2024kilometerscaleconvectionallowingmodel,
      title={Kilometer-Scale Convection Allowing Model Emulation using Generative Diffusion Modeling}, 
      author={Jaideep Pathak and Yair Cohen and Piyush Garg and Peter Harrington and Noah Brenowitz and Dale Durran and Morteza Mardani and Arash Vahdat and Shaoming Xu and Karthik Kashinath and Michael Pritchard},
      year={2024},
      eprint={2408.10958},
      archivePrefix={arXiv},
      primaryClass={physics.ao-ph},
      url={https://arxiv.org/abs/2408.10958}, 
}

@misc{sundar2025taudiffhighlyefficientkilometerscale,
      title={TAUDiff: Highly efficient kilometer-scale downscaling using generative diffusion models}, 
      author={Rahul Sundar and Yucong Hu and Nishant Parashar and Antoine Blanchard and Boyko Dodov},
      year={2025},
      eprint={2412.13627},
      archivePrefix={arXiv},
      primaryClass={cs.LG},
      url={https://arxiv.org/abs/2412.13627}, 
}

@misc{kim2026mindresidualgapprobabilistic,
      title={Mind the Residual Gap: Probabilistic Downscaling under Real-World Bias}, 
      author={Yujin Kim and Nidhi Soma and Sarah Dean},
      year={2026},
      eprint={2606.30821},
      archivePrefix={arXiv},
      primaryClass={cs.LG},
      url={https://arxiv.org/abs/2606.30821}, 
}

@article{Bar_Sinai_2019,
   title={Learning data-driven discretizations for partial differential equations},
   volume={116},
   ISSN={1091-6490},
   url={http://dx.doi.org/10.1073/pnas.1814058116},
   DOI={10.1073/pnas.1814058116},
   number={31},
   journal={Proceedings of the National Academy of Sciences},
   publisher={National Academy of Sciences},
   author={Bar-Sinai, Yohai and Hoyer, Stephan and Hickey, Jason and Brenner, Michael P.},
   year={2019},
   month=jul, pages={15344–15349} }

@misc{um2021solverinthelooplearningdifferentiablephysics,
      title={Solver-in-the-Loop: Learning from Differentiable Physics to Interact with Iterative PDE-Solvers}, 
      author={Kiwon Um and Robert Brand and Yun and Fei and Philipp Holl and Nils Thuerey},
      year={2021},
      eprint={2007.00016},
      archivePrefix={arXiv},
      primaryClass={physics.comp-ph},
      url={https://arxiv.org/abs/2007.00016}, 
}

@article{Kochkov_2024,
   title={Neural general circulation models for weather and climate},
   volume={632},
   ISSN={1476-4687},
   url={http://dx.doi.org/10.1038/s41586-024-07744-y},
   DOI={10.1038/s41586-024-07744-y},
   number={8027},
   journal={Nature},
   publisher={Springer Science and Business Media LLC},
   author={Kochkov, Dmitrii and Yuval, Janni and Langmore, Ian and Norgaard, Peter and Smith, Jamie and Mooers, Griffin and Klöwer, Milan and Lottes, James and Rasp, Stephan and Düben, Peter and Hatfield, Sam and Battaglia, Peter and Sanchez-Gonzalez, Alvaro and Willson, Matthew and Brenner, Michael P. and Hoyer, Stephan},
   year={2024},
   month=jul, pages={1060–1066} }

@article{Kochkov_2021,
   title={Machine learning–accelerated computational fluid dynamics},
   volume={118},
   ISSN={1091-6490},
   url={http://dx.doi.org/10.1073/pnas.2101784118},
   DOI={10.1073/pnas.2101784118},
   number={21},
   journal={Proceedings of the National Academy of Sciences},
   publisher={National Academy of Sciences},
   author={Kochkov, Dmitrii and Smith, Jamie A. and Alieva, Ayya and Wang, Qing and Brenner, Michael P. and Hoyer, Stephan},
   year={2021},
   month=May }

@inproceedings{NEURIPS2021_0987b8b3,
 author = {Pezeshki, Mohammad and Kaba, Oumar and Bengio, Yoshua and Courville, Aaron and Precup, Doina and Lajoie, Guillaume},
 booktitle = {Advances in Neural Information Processing Systems},
 editor = {M. Ranzato and A. Beygelzimer and Y. Dauphin and P.S. Liang and J. Wortman Vaughan},
 pages = {1256--1272},
 publisher = {Curran Associates, Inc.},
 title = {Gradient Starvation: A Learning Proclivity in Neural Networks},
 url = {https://proceedings.neurips.cc/paper_files/paper/2021/file/0987b8b338d6c90bbedd8631bc499221-Paper.pdf},
 volume = {34},
 year = {2021}
}

@article{Dai:2018vvv,
    author = "Dai, Biwei and Feng, Yu and Seljak, Uros",
    title = "{A gradient based method for modeling baryons and matter in halos of fast simulations}",
    eprint = "1804.00671",
    archivePrefix = "arXiv",
    primaryClass = "astro-ph.CO",
    doi = "10.1088/1475-7516/2018/11/009",
    journal = "JCAP",
    volume = "11",
    pages = "009",
    year = "2018"
}

@article{Bartlett:2024cra,
    author = "Bartlett, Deaglan J. and Chiarenza, Marco and Doeser, Ludvig and Leclercq, Florent",
    title = "{COmoving Computer Acceleration (COCA): N-body simulations in an emulated frame of reference}",
    eprint = "2409.02154",
    archivePrefix = "arXiv",
    primaryClass = "astro-ph.IM",
    doi = "10.1051/0004-6361/202452217",
    journal = "Astron. Astrophys.",
    volume = "694",
    pages = "A287",
    year = "2025"
}

@misc{liu2021varianceadaptivelearningrate,
      title={On the Variance of the Adaptive Learning Rate and Beyond}, 
      author={Liyuan Liu and Haoming Jiang and Pengcheng He and Weizhu Chen and Xiaodong Liu and Jianfeng Gao and Jiawei Han},
      year={2021},
      eprint={1908.03265},
      archivePrefix={arXiv},
      primaryClass={cs.LG},
      url={https://arxiv.org/abs/1908.03265}, 
}

@misc{zhang2019lookaheadoptimizerksteps,
      title={Lookahead Optimizer: k steps forward, 1 step back}, 
      author={Michael R. Zhang and James Lucas and Geoffrey Hinton and Jimmy Ba},
      year={2019},
      eprint={1907.08610},
      archivePrefix={arXiv},
      primaryClass={cs.LG},
      url={https://arxiv.org/abs/1907.08610}, 
}

@misc{wright2021ranger21synergisticdeeplearning,
      title={Ranger21: a synergistic deep learning optimizer}, 
      author={Less Wright and Nestor Demeure},
      year={2021},
      eprint={2106.13731},
      archivePrefix={arXiv},
      primaryClass={cs.LG},
      url={https://arxiv.org/abs/2106.13731}, 
}

@article{Hahn:2011uy,
    author = "Hahn, Oliver and Abel, Tom",
    title = "{Multi-scale initial conditions for cosmological simulations}",
    eprint = "1103.6031",
    archivePrefix = "arXiv",
    primaryClass = "astro-ph.CO",
    reportNumber = "SLAC-PUB-14463",
    doi = "10.1111/j.1365-2966.2011.18820.x",
    journal = "Mon. Not. Roy. Astron. Soc.",
    volume = "415",
    pages = "2101--2121",
    year = "2011"
}

@article{Navarro:1996gj,
    author = "Navarro, Julio F. and Frenk, Carlos S. and White, Simon D. M.",
    title = "{A Universal density profile from hierarchical clustering}",
    eprint = "astro-ph/9611107",
    archivePrefix = "arXiv",
    doi = "10.1086/304888",
    journal = "Astrophys. J.",
    volume = "490",
    pages = "493--508",
    year = "1997"
}

@article{muller2022instant,
  title={Instant neural graphics primitives with a multiresolution hash encoding},
  author={M{\"u}ller, Thomas and Evans, Alex and Schied, Christoph and Keller, Alexander},
  journal={ACM transactions on graphics (TOG)},
  volume={41},
  number={4},
  pages={1--15},
  year={2022},
  publisher={ACM New York, NY, USA}
}

@article{Springel:2020plp,
    author = {Springel, Volker and Pakmor, R{\"u}diger and Zier, Oliver and Reinecke, Martin},
    title = "{Simulating cosmic structure formation with the gadget-4 code}",
    eprint = "2010.03567",
    archivePrefix = "arXiv",
    primaryClass = "astro-ph.IM",
    doi = "10.1093/mnras/stab1855",
    journal = "Mon. Not. Roy. Astron. Soc.",
    volume = "506",
    number = "2",
    pages = "2871--2949",
    year = "2021"
}

@article{Springel:2009aa,
    author = "Springel, Volker",
    title = "{E pur si muove: Galiliean-invariant cosmological hydrodynamical simulations on a moving mesh}",
    eprint = "0901.4107",
    archivePrefix = "arXiv",
    primaryClass = "astro-ph.CO",
    doi = "10.1111/j.1365-2966.2009.15715.x",
    journal = "Mon. Not. Roy. Astron. Soc.",
    volume = "401",
    pages = "791",
    year = "2010"
}

@article{Pillepich:2017jle,
    author = "Pillepich, Annalisa and others",
    title = "{Simulating Galaxy Formation with the IllustrisTNG Model}",
    eprint = "1703.02970",
    archivePrefix = "arXiv",
    primaryClass = "astro-ph.GA",
    doi = "10.1093/mnras/stx2656",
    journal = "Mon. Not. Roy. Astron. Soc.",
    volume = "473",
    number = "3",
    pages = "4077--4106",
    year = "2018"
}

@article{Springel:2017tpz,
    author = "Springel, Volker and others",
    title = "{First results from the IllustrisTNG simulations: matter and galaxy clustering}",
    eprint = "1707.03397",
    archivePrefix = "arXiv",
    primaryClass = "astro-ph.GA",
    doi = "10.1093/mnras/stx3304",
    journal = "Mon. Not. Roy. Astron. Soc.",
    volume = "475",
    number = "1",
    pages = "676--698",
    year = "2018"
}

@article{Nelson:2018uso,
    author = "Nelson, Dylan and others",
    title = "{The IllustrisTNG simulations: public data release}",
    eprint = "1812.05609",
    archivePrefix = "arXiv",
    primaryClass = "astro-ph.GA",
    doi = "10.1186/s40668-019-0028-x",
    journal = "Comput. Astrophys. Cosmol.",
    volume = "6",
    number = "1",
    pages = "2",
    year = "2019"
}

@article{Schaye:2014tpa,
    author = "Schaye, Joop and others",
    title = "{The EAGLE project: Simulating the evolution and assembly of galaxies and their environments}",
    eprint = "1407.7040",
    archivePrefix = "arXiv",
    primaryClass = "astro-ph.GA",
    doi = "10.1093/mnras/stu2058",
    journal = "Mon. Not. Roy. Astron. Soc.",
    volume = "446",
    pages = "521--554",
    year = "2015"
}

@article{Hopkins:2017ycn,
    author = "Hopkins, Philip F and others",
    title = "{FIRE-2 Simulations: Physics versus Numerics in Galaxy Formation}",
    eprint = "1702.06148",
    archivePrefix = "arXiv",
    primaryClass = "astro-ph.GA",
    doi = "10.1093/mnras/sty1690",
    journal = "Mon. Not. Roy. Astron. Soc.",
    volume = "480",
    number = "1",
    pages = "800--863",
    year = "2018"
}

@article{Villaescusa-Navarro:2019bje,
    author = "Villaescusa-Navarro, Francisco and others",
    title = "{The Quijote simulations}",
    eprint = "1909.05273",
    archivePrefix = "arXiv",
    primaryClass = "astro-ph.CO",
    doi = "10.3847/1538-4365/ab9d82",
    journal = "Astrophys. J. Suppl.",
    volume = "250",
    number = "1",
    pages = "2",
    year = "2020"
}

@article{dosovitskiy2020image,
  title={An image is worth 16x16 words: Transformers for image recognition at scale},
  author={Dosovitskiy, Alexey},
  journal={arXiv preprint arXiv:2010.11929},
  year={2020}
}

@article{vaswani2017attention,
  title={Attention is all you need},
  author={Vaswani, Ashish and Shazeer, Noam and Parmar, Niki and Uszkoreit, Jakob and Jones, Llion and Gomez, Aidan N and Kaiser, {\L}ukasz and Polosukhin, Illia},
  journal={Advances in neural information processing systems},
  volume={30},
  year={2017}
}

@article{More:2011dc,
    author = "More, Surhud and Kravtsov, Andrey and Dalal, Neal and Gottlober, Stefan",
    title = "{The overdensity and masses of the friends-of-friends halos and universality of the halo mass function}",
    eprint = "1103.0005",
    archivePrefix = "arXiv",
    primaryClass = "astro-ph.CO",
    doi = "10.1088/0067-0049/195/1/4",
    journal = "Astrophys. J. Suppl.",
    volume = "195",
    pages = "4",
    year = "2011"
}

@article{Davis:1985rj,
    author = "Davis, Marc and Efstathiou, George and Frenk, Carlos S. and White, Simon D. M.",
    editor = "Srednicki, M. A.",
    title = "{The Evolution of Large Scale Structure in a Universe Dominated by Cold Dark Matter}",
    reportNumber = "NSF-ITP-84-129",
    doi = "10.1086/163168",
    journal = "Astrophys. J.",
    volume = "292",
    pages = "371--394",
    year = "1985"
}

@book{DynamicsAstrophysicsGalaxies,
    title={Dynamics and Astrophysics of Galaxies},
    author={{Bovy}, J.},
    year={2026},
    publisher={Princeton University Press},
    address= {Princeton, NJ}
}

@article{Lokas:2000mu,
    author = "Lokas, Ewa L. and Mamon, Gary A.",
    title = "{Properties of spherical galaxies and clusters with an nfw density profile}",
    eprint = "astro-ph/0002395",
    archivePrefix = "arXiv",
    doi = "10.1046/j.1365-8711.2001.04007.x",
    journal = "Mon. Not. Roy. Astron. Soc.",
    volume = "321",
    pages = "155",
    year = "2001"
}

@article{Planck:2018vyg,
    author = "Aghanim, N. and others",
    collaboration = "Planck",
    title = "{Planck 2018 results. VI. Cosmological parameters}",
    eprint = "1807.06209",
    archivePrefix = "arXiv",
    primaryClass = "astro-ph.CO",
    doi = "10.1051/0004-6361/201833910",
    journal = "Astron. Astrophys.",
    volume = "641",
    pages = "A6",
    year = "2020",
    note = "[Erratum: Astron.Astrophys. 652, C4 (2021)]"
}

@article{ParticleDataGroup:2022pth,
    author = "Workman, R. L. and others",
    collaboration = "Particle Data Group",
    title = "{Review of Particle Physics}",
    doi = "10.1093/ptep/ptac097",
    journal = "PTEP",
    volume = "2022",
    pages = "083C01",
    year = "2022"
}

@misc{lai2025principlesdiffusionmodels,
      title={The Principles of Diffusion Models}, 
      author={Chieh-Hsin Lai and Yang Song and Dongjun Kim and Yuki Mitsufuji and Stefano Ermon},
      year={2025},
      eprint={2510.21890},
      archivePrefix={arXiv},
      primaryClass={cs.LG},
      url={https://arxiv.org/abs/2510.21890}, 
}

@misc{Elsharkawy:2026dte,
    author = "Elsharkawy, Ibrahim and Knapp-Perez, Victoria and Bhimji, Wahid and Ghosh, Aishik",
    title = "{Searching for BSM Experimental Signatures with Large Lagrangian Models}",
    eprint = "2609.38309",
    archivePrefix = "arXiv",
    primaryClass = "hep-ph",
    month = "9",
    year = "2026"
}

@misc{Elsharkawy:2026mev,
    author = "Elsharkawy, Ibrahim and Birk, Joschka and Mikuni, Vinicius and Bhimji, Wahid and Kasieczka, Gregor and Nachman, Benjamin",
    title = "{Pre-Training for Simulation-Based Science: A Study on Jet Foundation Model Training Objectives}",
    eprint = "2606.14870",
    archivePrefix = "arXiv",
    primaryClass = "hep-ph",
    month = "6",
    year = "2026"
}

@misc{Dai:2026jme,
    author = "Dai, Biwei and others",
    title = "{FAIR Universe Weak Lensing ML Uncertainty Challenge: Handling Uncertainties and Distribution Shifts for Precision Cosmology}",
    eprint = "2604.14451",
    archivePrefix = "arXiv",
    primaryClass = "astro-ph.CO",
    month = "4",
    year = "2026"
}

@article{zhang2026h,
  title={H-JEPA: End-to-End Learning of Hierarchical World Models for Visual Planning},
  author={Zhang, Wancong and Terver, Basile and Rabbat, Michael and LeCun, Yann and Balestriero, Randall},
  journal={arXiv preprint arXiv:2610.06805},
  year={2026}
}

@article{albergo2022building,
  title={Building normalizing flows with stochastic interpolants},
  author={Albergo, Michael S and Vanden-Eijnden, Eric},
  journal={arXiv preprint arXiv:2209.15571},
  year={2022}
}

@article{albergo2025stochastic,
  title={Stochastic interpolants: A unifying framework for flows and diffusions},
  author={Albergo, Michael and Boffi, Nicholas M and Vanden-Eijnden, Eric},
  journal={Journal of Machine Learning Research},
  volume={26},
  number={209},
  pages={1--80},
  year={2025}
}

@article{Warren:2005ey,
    author = "Warren, Michael S. and Abazajian, Kevork and Holz, Daniel E. and Teodoro, Luis",
    title = "{Precision determination of the mass function of dark matter halos}",
    eprint = "astro-ph/0506395",
    archivePrefix = "arXiv",
    reportNumber = "LA-UR-05-4013",
    doi = "10.1086/504962",
    journal = "Astrophys. J.",
    volume = "646",
    pages = "881--885",
    year = "2006"
}

@article{Tinker:2008ff,
    author = "Tinker, Jeremy L. and Kravtsov, Andrey V. and Klypin, Anatoly and Abazajian, Kevork and Warren, Michael S. and Yepes, Gustavo and Gottlober, Stefan and Holz, Daniel E.",
    title = "{Toward a halo mass function for precision cosmology: The Limits of universality}",
    eprint = "0803.2706",
    archivePrefix = "arXiv",
    primaryClass = "astro-ph",
    doi = "10.1086/591439",
    journal = "Astrophys. J.",
    volume = "688",
    pages = "709--728",
    year = "2008"
}

@article{Watson:2012mt,
    author = "Watson, William A. and Iliev, Ilian T. and D'Aloisio, Anson and Knebe, Alexander and Shapiro, Paul R. and Yepes, Gustavo",
    title = "{The halo mass function through the cosmic ages}",
    eprint = "1212.0095",
    archivePrefix = "arXiv",
    primaryClass = "astro-ph.CO",
    doi = "10.1093/mnras/stt791",
    journal = "Mon. Not. Roy. Astron. Soc.",
    volume = "433",
    pages = "1230",
    year = "2013"
}
\bibliographystyle{apsrev4-2}

\end{document}